\documentclass[11pt, notitlepage, tightenlines,aps, prd, reprint,superscriptaddress,nofootinbib]{revtex4-1}
\usepackage[utf8]{inputenc}
\usepackage{placeins}
\usepackage{mathtools}
\usepackage{amsfonts}
\usepackage{amssymb,amsmath}
\usepackage{color}
\usepackage{xcolor}
\definecolor{lcolor}{rgb}{0.5,0,0}
\definecolor{citcolor}{rgb}{0,0.3,0.0}
\usepackage[breaklinks,colorlinks,urlcolor=blue,citecolor=citcolor,linkcolor=lcolor]{hyperref}

\usepackage{graphicx}
\usepackage{subfigure}
\graphicspath{ {image/} }
\usepackage{multirow}
\newcommand*\diff{\mathop{}\!\mathrm{d}}
\usepackage{braket}
\usepackage{bm}
\usepackage{txfonts}
\usepackage{mathrsfs} 
\usepackage{slashed}
\usepackage{cleveref}
\usepackage{hepunits}

\def\p{{\boldsymbol p}}
\def\q{{\boldsymbol q}}
\def\k{{\boldsymbol k}}
\def\x{{\boldsymbol x}}
\def\y{{\boldsymbol y}}
\def\r{{\boldsymbol r}}
\def\R{{\boldsymbol R}}

\def\l{{\boldsymbol l}}
\def\v{{\boldsymbol v}}

\DeclareMathOperator{\Tr}{Tr}
\newcommand{\GeV}{{{\,}\textrm{GeV}}}

\newcommand{\expconfig}[1]{{\langle #1 \rangle_m}}
\begin{document}

\bibliographystyle{apsrev4-1}

\title{On the momentum broadening of in-medium jet evolution using a light-front Hamiltonian approach}
\author{Meijian Li}
\email{meijian.li@usc.es}
\affiliation{Department of Physics, P.O. Box 35, FI-40014 University of Jyv\"{a}skyl\"{a},
Finland}
\affiliation{
Helsinki Institute of Physics, P.O. Box 64, FI-00014 University of Helsinki,
Finland
}
\affiliation{Instituto Galego de Fisica de Altas Enerxias (IGFAE), Universidade de Santiago de Compostela, E-15782 Galicia, Spain}

\author{Tuomas Lappi}
\email{tuomas.v.v.lappi@jyu.fi}
\affiliation{Department of Physics, P.O. Box 35, FI-40014 University of Jyv\"{a}skyl\"{a},
Finland}
\affiliation{
Helsinki Institute of Physics, P.O. Box 64, FI-00014 University of Helsinki,
Finland
}

\author{Xingbo Zhao}
\email{xbzhao@impcas.ac.cn}
\affiliation{Institute of Modern Physics, Chinese Academy of Sciences, Lanzhou 730000, China}
\affiliation{University of Chinese Academy of Sciences, Beijing 100049, China}

\author{Carlos A. Salgado}
\email{carlos.salgado@usc.es}
\affiliation{Instituto Galego de Fisica de Altas Enerxias (IGFAE), Universidade de Santiago de Compostela, E-15782 Galicia, Spain}

\begin{abstract}
  Following the non-perturbative light-front Hamiltonian formalism developed in our preceding work~\cite{Li:2021zaw}, we investigate the momentum broadening of a quark jet inside a SU(3) colored medium. 
  We perform the numerical simulation of the real-time jet evolution in Fock spaces of a single quark, a quark-gluon state, and coupled quark- and quark-gluon states
  at various jet momenta $p^+$ and medium densities.
  With the obtained jet light-front wavefunction, we extract the jet transverse momentum distribution, the quenching parameter, and the gluon emission rate. 
  We analyze the dependence of momentum broadening on $p^+$, medium density, color configuration, spatial correlation, and medium-induced gluon emission. 
  For comparison, we also derive analytically the expectation value of the transverse momentum of a quark-gluon state in any color configuration and in an arbitrary spatial distribution in the eikonal limit.
  This work can help understand jet momentum broadening in the non-eikonal regime.
\end{abstract}
\maketitle

\section{Introduction}\label{sec:introduction}
A central goal of ultrarelativistic heavy ion collisions, such as those performed at the Relativistic Heavy-Ion Collider (RHIC) to the Large Hadron Collider (LHC), is to recreate droplets of matter in the early Universe, the quark-gluon plasma, and learn about its properties~\cite{Busza:2018rrf}. 
In heavy-ion collisions, energetic quarks and gluons are produced at early stages, propagating through the dense and hot medium. The initial particle is transformed into a cone-shaped beam of hadrons, a jet. In theoretical studies, a jet initiated by a high energy quark(gluon) is often referred to as the quark(gluon) jet. 
Experimentally, the energy and motion of the original particle are estimated by measuring the hadrons in the jet. The jets are suppressed and modified compared to those in proton-proton collisions, a phenomenon known as jet quenching, observed at RHIC~\cite{BRAHMS:2004adc, PHOBOS:2004zne, STAR:2005gfr,
PHENIX:2004vcz} and LHC~\cite{ATLAS:2010isq, ALICE:2010yje, CMS:2011iwn}. 
Similar processes happen in deeply inelastic scattering (DIS), where  jets lose energy when traversing the cold nuclear matter formed from the large nucleus.

Studies of jet quenching give us information on how the medium responds when traversed by a high-energy quark or gluon jet, and how the jet is modified by the medium~\cite{Baier:2002tc, Casalderrey-Solana:2007knd, dEnterria:2009xfs, Majumder:2010qh, Qin:2015srf, Cao:2020wlm, Apolinario:2022vzg}.
Several perturbative QCD (pQCD)-based studies have been carried out to calculate jet energy loss through multiple scatterings and gluon radiations. In the Baier-Dokshitzer-Mueller-Peigne-Schiff and Zakharov (BDMPS-Z) approach, the medium is modeled as a collection of static scattering centers and soft gluon radiations are induced through multiple scatterings~\cite{Baier:1996kr,Baier:1996sk,Zakharov:1997uu}. 
Gyulassy-Levai-Vitev (GLV) and Wiedemann \cite{Gyulassy:1999zd,Gyulassy:2000fs,Wiedemann:2000za} developed a systematic expansion of the calculation in terms of the number of scatterings. 
In the Arnold, Moore and Yaffe (AMY) \cite{Arnold:2001ba,Arnold:2002ja} approach, the hard thermal loop framework is employed and the medium is treated as in a thermal equilibrium state. 
In the higher-twist (HT) approach, the twist-expansion is used in a collinear factorization formalism and the medium is characterized by matrix elements of gauge field operators \cite{Guo:2000nz,Wang:2001ifa}.
In the SCETG formalism \cite{Ovanesyan:2011kn,Ovanesyan:2011xy}, the
standard Soft Collinear Effective Theory (SCET) Lagrangian is modified to include Glauber modes of gluon field for parton interactions.

In preceding works, we have developed a computational method of simulating the evolution of a quark jet inside a classical color background field, first in the $\ket{q}$ Fock sector~\cite{Li:2020uhl}, then in the $\ket{q}+\ket{qg}$ Fock sector~\cite{Li:2021zaw}. 
This method is known as the time-dependent Basis Light-Front Quantization (tBLFQ)~\cite{1stBLFQ}, a light-front Hamiltonian formalism. 
Unlike the aforementioned pQCD-based approaches, the evolution process is calculated on the amplitude rather than the probability level. 
This method enables us to relax approximations usually made in high-energy collisions, such as the eikonal and the collinear radiation approximations~\cite{Apolinario:2022vzg}.

The tBLFQ method has also been applied to various problems in quantum electrodynamics~\cite{Zhao:2013cma, Chen:2017uuq, Hu:2019hjx, Lei:2022nsk}, and its Quantum Mechanics counterpart---the time-dependent Basis Function (tBF) approach to nuclear structure and scattering~\cite{Du:2018tce, Yin:2022zii}.
The advantages of the light-front Hamiltonian formalism are also used in a recent study in small-$x$ physics to study spin-related observables ~\cite{Li:2023tlw}. 
It is noteworthy that the tBLFQ (and the related tBF) approach is well-suited to be implemented as quantum simulations on a quantum device. Recent developments can be found in Refs.~\cite{Du:2020glq, Barata:2022wim, Yao:2022eqm}. The calculation in this work, though performed on classical computers, also provides a precursor for the future implementation of quantum simulation, with the anticipation of a quantum speedup~\cite{Arute:2019zxq, Lau:2022fky}.

In this work, we present a study on the momentum broadening of in-medium jet evolution using tBLFQ. 
We aim to enhance the understanding of the mechanism in jet momentum broadening: the momentum exchange of the jet constituents with the medium and the effects from gluon radiation.
The new development in this work answers the following key questions:
\begin{enumerate}
    \item \textit{How to extract and interpret the momentum broadening from the jet's wavefunction formulated on a discrete momentum basis space?}
We have shown that in tBLFQ, the quark jet is described by an evolving light-front wavefunction, a superposition of different momenta, color, and helicity modes in the $\ket{q}+\ket{qg}$ space, and we examine various observables and quantities from it~\cite{Li:2021zaw}. 
To complement and further develop the study on the transverse momentum space, we extract the time-dependent transverse momentum square $\braket{\vec p_\perp^2 (x^+)}$ and the quenching parameter $\hat q$ (defined in the next section) from the jet state.
Importantly, one must understand the dependence on the physical and the basis parameters in simulations on a finite basis space, especially the infrared and ultraviolet cutoffs, which we will examine and elaborate on. 
\item \textit{What is the analytical expectation of the momentum broadening of the quark-gluon state at finite $N_c$, in the eikonal limit, given the state arbitrarily distributed in color and transverse space?}
The analytical expression of $\braket{\vec p_\perp^2 (x^+)}$, in terms of medium strength, momentum cutoffs, and evolution time, is known for the single particle state (the quark or gluon jet), derived with Wilson lines in the McLerran-Venugopalan (MV) model~\cite{McLerran:1993ni, McLerran:1993ka, McLerran:1994vd}. 
But the result for the quark-gluon state is absent. We fill this gap by presenting the full derivation using the four-point $\bar q\bar g qg$ Wilson line correlators.
This result can be useful to other studies on jet quenching with quark-gluon components, for example, the production of the quark-gluon dijet in high-energy collisions, especially when one needs to look at color-differential cross sections \cite{Nikolaev:2003zf,Dominguez:2012ad,Apolinario:2014csa,Blaizot:2013vha}. 
Here, it also serves as a benchmark for checking the numerical simulations in the eikonal limit.

\item  \textit{What is the effect from the medium at finite $p^+$ and finite $N_\eta$ (number of uncorrelated medium layers, defined in the next section)?} The picture of jet quenching becomes very complicated when the eikonal approximation gets relaxed: at finite $p^+$, there is a  diffusion in transverse coordinate space resulting from the kinetic energy part of the Hamiltonian, and there exists a continuous gluon emission/absorption throughout the evolution even when only allowing one dynamical gluon at the same time. 
In addition, to be more realistic, we also let the number of uncorrelated medium layers in the MV model, namely $N_\eta$,  be finite. 
We will analyze those effects using the evolved jet wavefunction obtained from numerical simulations. We observe a suppression on $\hat q$ at finite $p^+$ and finite $N_\eta$. The medium enhances the gluon emission compared to the vacuum, but slows down the total momentum broadening of the quark jet state compared to a quark-gluon state. 
\end{enumerate}

The layout of this paper is as follows. We first introduce the method in Sec.~\ref{sec:method}. We then present and discuss the analytical results in the eikonal limit in Sec.~\ref{sec:eikonal} and the numerical results of the full non-perturbative calculation in Sec.~\ref{sec:numerical}. We conclude with a discussion of future steps beyond this work in Sec.~\ref{sec:conclusions}.

\section{Methodology}\label{sec:method}
In Ref.~\cite{Li:2021zaw}, we have developed the formalism of using the tBLFQ approach to simulate the evolution of a quark jet inside a classical color background field in the $\ket{q}+\ket{qg}$ Fock sector. 
Here, we briefly review the basics of this formalism and reformulate the physical quantities in dimensionless variables.

\subsection{Jet evolution in tBLFQ}
The light-front Hamiltonian consists of three parts, $P^-(x^+)=P_{KE}^- + V_{qg}+V_{\mathcal A}(x^+)$, which are the kinetic energy term, the gluon emission/absorption term, and the interaction term with a background field (the medium), respectively. 
The background field $\mathcal A$ describing the medium is given by the MV model~\cite{McLerran:1993ni, McLerran:1993ka, McLerran:1994vd}. 
We refer to Ref.~\cite{Li:2021zaw} for the derivation and the full expression of $P^-$.
The evolution of the state is treated by decomposing the time-evolution operator into many small steps of the light-front time $x^+$, then solved in the time sequence numerically,
\begin{align}\label{eq:ShrodingerEqSol}
  \begin{split}
    \ket{\psi;x^+}=&\mathcal{T}_+\exp\left[-\frac{i}{2}\int_0^{x^+}\diff z^+ P^-(z^+)\right]\ket{\psi;0}\\
    =&\lim_{n\to\infty} \prod^n_{k=1}\mathcal{T}_+ \exp\left[-\frac{i}{2}\int_{x_{k-1}^+}^{x_k^+}\diff z^+P^-(z^+)\right]\ket{\psi;0}
    \;,
  \end{split}
\end{align}
in which $x_k^+=k\delta x^+ (k=0,1,2,\ldots,n)$ with $\delta x^+ \equiv x^+/n$.
The numerical method for this specific problem is optimized in Ref.~\cite{Li:2021zaw}. That is, within each small time step, we treat $P^-_{KE}$ and $V_{\mathcal A}$ as time-constant and carry out matrix exponentiation in the momentum and coordinate space, respectively; the operation with $V_{qg}$ uses the fourth-order Runge-Kutta method in momentum space.

The formulated basis space consists of a square lattice with periodic boundary conditions in the transverse dimensions $\vec x_\perp$, ranging in $[-L_\perp, L_\perp]$ with $2N_\perp$ sites, and a loop with (anti-)periodic boundary condition in the $x^-$ direction, of length $2L$, for the gluon(quark).
The transverse lattice introduces a pair of infrared~(IR) and ultraviolet~(UV) cutoffs in the transverse momentum space $\vec p_\perp$, $\lambda_{IR}=d_p=\pi/L_\perp$ and $\lambda_{UV}=\pi/a_\perp$, with $a_\perp=L_\perp/N_\perp$ as the lattice spacing. 
Therefore the simulations are performed at fixed IR and UV cutoffs, which means setting upper and lower bounds on the $\int p_\perp$ integral in the corresponding analytical calculations.
We will see later that the physical IR regulator $m_g$ will play the role of the IR cutoff instead of $\lambda_{IR}$, but the UV cutoff is still the lattice-dependent $\lambda_{UV}$. To relate to a physical process, one needs to match such cutoffs to realistic momentum scales. For example, the study in Ref.~\cite{Boguslavski:2023alu} uses the Landau matching condition to choose cutoff models that depend on the jet energy and plasma temperature. In this work, we study the dependence of the jet observables at a range of cutoffs. 

The longitudinal momentum $p^+$ is quantized in units of $2\pi/L$, and the gluon(quark) is allowed to take a positive (half-)integer number in this unit.
For the total momentum of the quark jet state, $K$ is a half-integer (note that $K\ge 1.5$ in order to accommodate multiple $p^+$ configurations for the $\ket{qg}$ sector), with $p^+= K 2\pi/L$ and $p^+=p^+_Q=p^+_q+p^+_g$
\footnote{We use the subscripts ``$Q$'' and ``$q$'' to distinguish between the quark in the $\ket{q}$ sector and that in the $\ket{qg}$ sector.}.
Then the longitudinal momentum fraction of the gluon, $z\equiv p^+_g/p^+$, has a resolution of $1/K$.

The total evolution time, which is also the thickness of the medium, is $x^+=[0, L_\eta]$. 
The $x^+$ dimension is discretized into small time steps of $\delta x^+$ for numerically simulating the time evolution, as seen in Eq.~\eqref{eq:ShrodingerEqSol}. Meanwhile, the medium along $x^+$ is discretized into a number of $N_\eta$ uncorrelated layers such that each layer has a duration of $\tau\equiv L_\eta/N_\eta$.
This layer structure is to numerically simulate the stochastic feature of the sources along $x^+$ that generate the medium~\cite{Lappi:2007ku}, characterized in continuum by the correlation relation
\begin{equation}\label{eq:chgcor}
  \expconfig{\rho_a(\vec{x}_\perp,x^+)\rho_b(\vec{y}_\perp,y^+)}=g^2\tilde\mu^2\delta_{ab}\delta^2(\vec{x}_\perp-\vec{y}_\perp)\delta(x^+-y^+)\; .
\end{equation}
This continuum relation corresponds to the limit $N_\eta\to \infty$($\tau\to 0$), but we can keep $\tau$ general and thus introduce a finite longitudinal correlation length into our description of the medium.
The average over the medium configurations is indicated by $\expconfig{\ldots}$. 
The medium color field $\mathcal A$ is calculated from the sources by solving the reduced Yang-Mills equation with an IR regulator $m_g$,
 \begin{align}\label{eq:poisson}
  (m_g^2-\nabla^2_\perp )  \mathcal{A}^-_a(\vec{x}_\perp,x^+)=\rho_a(\vec{x}_\perp,x^+)\;.
 \end{align}
The details on how to simulate the medium in the basis space are discussed in Appendix \ref{app:medium}.

In this Hamiltonian formalism, the jet as a quantum state is described by its light-front wavefunction at different time instances.
The wavefunction reads as a column vector of coefficients in the above-formulated basis space. For a given observable $\hat O$, one can directly evaluate its expectation value as $\braket{O(x^+)}=\braket{ \psi;x^+|\hat O|\psi;x^+} $, using the jet wavefunction solved from a single simulation. Then, by taking the average of $\braket{O(x^+)}$'s from multiple simulations, each with an independently sampled $\rho$, we arrive at the configuration averaged $\expconfig{O(x^+)} $.

\subsection{Parameter dependence}
To have a qualitative understanding of how the physical process should depend on the parameters of the setup, here we consider the integrated Hamiltonian, $P^-$ summed over the spatial dimensions $(\vec x_\perp, x^-)$ and accumulated over a time duration of $\Delta x^+$.
Let us examine the three terms individually.

Firstly, for the kinetic energy term, $P_{KE}^-$, its effect as in the evolution operator mainly depends on three dimensionless quantities,
\begin{align}\label{eq:S_KE}
  e^{-\frac{i}{2} P_{KE}^- \Delta x^+} \sim f\left[\frac{2L \Delta x^+}{(2 a_\perp)^2},\frac{m_q a_\perp}{\pi},\frac{1}{K} \right]\;.
\end{align}
The dependence on the first quantity in the product form is a result of the longitudinal boost invariance. In the view of the full evolution process, viz., $\Delta x^+\to L_\eta$, this action stays the same by scaling the $p^+$ momentum (via inversely scaling $L$) of the incoming quark and its evolution time $L_\eta$ equally. 
The dependence on the second quantity reflects the contribution from the quark mass as compared to the largest transverse momentum mode $\lambda_{UV}$. 
The third quantity reflects the resolution in probing the longitudinal momentum fraction of the particle in the $\ket{qg}$ sector.

Secondly, the gluon emission/absorption term, $V_{qg}$, accumulated over time, depends also on the previously introduced quantities, and in addition, the coupling constant,
\begin{align}
  e^{-\frac{i}{2}V_{qg}\Delta x^+}  \sim f\left[g, \frac{2L \Delta x^+}{(2 a_\perp)^2},\frac{m_q a_\perp}{\pi},\frac{1}{K} \right]\;.
\end{align}
The dependence on the third quantity reflects the ratio between the quark-spin-flip and non-flip transition widths, and we refer to Figure 4 of Ref.~\cite{Li:2021zaw} as an illustration.
The fourth quantity indicates the softest gluon being emitted/absorbed.

Thirdly, there is an effect from the medium interaction term, $V_{\mathcal A}$. Here we consider a time duration $\Delta x^+$ that is multiples of the layer thickness $\tau$, in order to be in the situation where one is properly sensitive to several sources that are uncorrelated in the $x^+$ direction. 
The medium-interaction action depends on two dimensionless variables, 
\begin{align}
  e^{-\frac{i}{2}V_{\mathcal A}\Delta x^+}  \sim f\left[\frac{g^2\tilde{\mu}a_\perp  }{\sqrt{\tau}} \Delta x^+, \frac{m_g a_\perp}{\pi} \right]\;.
\end{align}
Here the combination $g^2\tilde{\mu}a_\perp  /\sqrt{\tau}$ must appear together in this way since $g^2$, $\tilde{\mu}$ and $\tau$ in fact only appear in the calculation in this combination, see Eq.~\eqref{eq:chgcor_dis}.
Taking $\Delta x^+\to \tau$ for one layer, and then adding together $N_\eta$ layers of squared color charge density  (added at the level of squares because the interaction is a diffusion-type process in transverse momentum), the first argument leads to the  emergence of the saturation scale  
\begin{equation}
\label{eq:bareqs}
Q_s^2=C_F (g^2\tilde{\mu})^2L_\eta/(2\pi^2)
\end{equation}
on the lattice, $Q_s a_\perp$. 
The second quantity is the ratio between the smallest and the largest transverse momentum that can be transferred by the medium.

In total, the full process combining all three terms would depend on the above-introduced quantities. We summarize those quantities and address potential constraints in setting up simulation parameters:
\begin{itemize}
  \item The \textit{coupling constant} $g$, and $g=1$ in this work. 
  \item The \textit{free action} $S_{free}\equiv L_\eta \lambda_{UV}^2 /p^+$. 
  Since the largest transverse momentum a particle can acquire is $\lambda_{UV}$, we can interpret $p^+/\lambda_{UV}^2$ as the coherence length of quantum diffusion and gluon emission/absorption. 
  Thus $S_{free}$ characterizes the length scale in $x^+$ at which the jet becomes non-eikonal, cf.~the non-eikonal parameter defined in Refs.~\cite{Altinoluk:2014oxa, Agostini:2022ctk, Agostini:2022oge}.
  \item The \textit{quark mass} in the unit of the lattice UV cutoff, $m_q a_\perp/\pi$. When its value is small, $m_q a_\perp/\pi\approx 0$, the quark-spin-flip gluon emission/absorption would be suppressed. If its value is too large, the kinetic energy term cannot resolve different transverse momentum modes.
  \item The \textit{saturation scale} in the unit of the lattice UV cutoff, $Q_s a_\perp/\pi$. It should be that $Q_s a_\perp/\pi\ll 1$; otherwise, the medium momentum transfer cannot be appropriately accessed on the lattice.
  \item The \textit{medium IR regulator} in the unit of the lattice UV cutoff, $m_g a_\perp/\pi$. This ``gluon mass'' $m_g$ is often introduced as the infrared screening scale of the medium, similar to the Debye mass in Gyulassy-Wang and other model potentials~\cite{Braaten:1989mz, Gyulassy:1993hr, Mehtar-Tani:2013pia, Barata:2020rdn}.
  Note the mass of the dynamical gluon in the $\ket{qg}$ sector of the Fock space is always zero.
  An eligible value of $m_g$ should be covered by the momentum range on the transverse lattice $[\lambda_{IR},\lambda_{UV}]$, at the same time smaller than the saturation scale $Q_s$. This constraint reads
  \begin{align}\label{eq:range_coverage}
    \frac{1}{N_\perp}\ll \frac{m_g a_\perp}{\pi}\ll \frac{Q_s a_\perp}{\pi}\;.
  \end{align}
  The continuum limit, $a_\perp\to 0$ is taken by letting  $N_\perp\to\infty$ so that $Q_s a_\perp/\pi\to 0$ while $Q_s L_\perp/\pi$ remains constant.
  \item The \textit{z resolution} and \textit{cutoff}, $1/K$. In the continuum limit, $K\to\infty$.
\end{itemize}
In running the numerical simulation, choosing parameters that satisfy the aforementioned conditions helps ensure that the physics of interest is captured on the discrete basis space. We refer to Appendix C of Ref.~\cite{Barata:2022wim} for a more detailed explanation of this topic.

\section{ $\hat q$ in the eikonal limit, using Wilson line correlators}\label{sec:eikonal}
In studying the phenomenon of jet momentum broadening inside a medium, a characteristic quantity to examine is the quenching parameter $\hat q$ \cite{Baier:2002tc, Kovner:2003zj,Liu:2006ug, Majumder:2012sh, Benzke:2012sz}, defined as 
\begin{align}\label{eq:dpperp}
  \hat q 
  = \frac{\Delta\braket{p^2_\perp ( x^+)}}{\Delta x^+} 
  \;.
\end{align}
It characterizes the mean square momentum transfer to the jet per unit length in the medium. 

In the eikonal limit, when one can express the propagation of a parton through a medium in terms of Wilson lines, $\hat q$ can be derived analytically using the Wilson line correlators. In this section, we derive it first for a single particle state, then for a quark-gluon state.

The derivation for the single particle state has already been developed in the Wilson line formalism (e.g., ref.~\cite{Casalderrey-Solana:2007knd}). The purpose of our revisiting this problem is two-folded. First, we perform the derivation in the context of the formulated basis space in Sec.~\ref{sec:method}, so that it could help interpret and verify in the eikonal limit the results obtained from the numerical simulations. Second, the derivation for the single particle prepares the necessary ingredients for the more complicated case of the quark-gluon state.

The existing derivations for the quark-gluon state usually use specific truncations and approximations. One considers the quark-gluon state as initially split from a single quark state. Thus the two-particle state resides in the triplet subspace. One often also takes the large $N_c$ limit, such that some correlations can be neglected, e.g., Ref.~\cite{Apolinario:2014csa}. Here, we carry out the calculation in its full color space and keep $N_c$ finite ($N_c=3$).

\subsection{The single particle state}\label{sec:eikonal_q}
The expectation value of the transverse momentum square can be calculated directly knowing the state vector, $\braket{p_\perp^2 (x^+)}
=\bra{\psi;x^+}\hat p_\perp^2 \ket{\psi;x^+}$. 
In the eikonal limit of $p^+=\infty$, only the $V_{\mathcal A}$ term survives in the Hamiltonian, and the evolution operator reduces to the Wilson line. 

For a quark, the Wilson line in the fundamental representation reads,
\begin{align}
  U_F(0,x^+; \vec x_\perp)\equiv\mathcal{T}_+\exp\bigg(
  -i g\int_{0}^{x^+}\diff z^+\mathcal{A}_a^-(\vec x_\perp, z^+)T^a
  \bigg)\;,
\end{align}
in which $T^a$ is the SU(3) generator in the fundamental representation, i.e., the Gell-Mann matrices. Replacing $T^a$ by the generators in the adjoint representation, $t^a$, one gets the adjoint Wilson line for the gluon, $U_A(0,x^+; \vec x_\perp)$. 
The Wilson line in the above expression is a matrix in the corresponding color space.

Then, the color-$\beta$ component of the evolved state that is initially in color state $\alpha$ reads 
 \begin{align}
  \ket{\psi;x^+}_{Eik}
  =\int_{\x}  \tilde \phi (\vec x_\perp)
  U_F(0,x^+;\vec x_\perp)_{\beta\alpha}
  \ket{\vec x_\perp}
  \;,
 \end{align}
in which the initial state is written as a wavefunction in the coordinate basis, $\ket{\psi;0}=\int_{\x} \tilde \phi (\vec x_\perp)\ket{\vec x_\perp}$ with the normalization $\int_{\x} \braket{\vec x_\perp|\vec x_\perp}=1$.
\footnote{Here and throughout the paper we use the shorthand notation $\int_{\p}\equiv \int \diff^2 p_\perp/(2\pi)^2 $ and $\int_{\r}\equiv \int\diff^2 r_\perp $.}

In calculating $\braket{p_\perp^2 (x^+)}$, the initial color space is averaged over, and the final space is summed over, 
therefore
\begin{align}\label{eq:psq_eik}
  \begin{split}
    & \braket{p_\perp^2 (x^+)}_{Eik}
    =
    \int_{\p}
    \vec p_\perp^2
    \int_{\x,\y} \tilde \phi^* (\vec x_\perp)
     \tilde \phi (\vec y_\perp)
    e^{-i \vec p_\perp\cdot (\vec x_\perp-\vec y_\perp)}\\
    &\qquad\times \sum_{\beta=1}^{N_c}
    \frac{1}{N_c}\sum_{\alpha=1}^{N_c}
    \expconfig{U_F^\dagger(0,x^+;\vec x_\perp)_{\alpha\beta}
    U_F(0,x^+;\vec y_\perp)_{\beta \alpha}}
    \;.
  \end{split}
\end{align}
In the second line, the hermitian conjugate of the Wilson line can be viewed as the S-matrix of an antiquark. Consequently, we recognize the second line of the equation as the forward scattering amplitude of an effective quark-antiquark dipole and write it as
\begin{align}\label{eq:SF}
  \begin{split}
   S_F(0,x^+&;|\vec x_\perp-\vec y_\perp|)
    = \frac{1}{N_c}\Tr\expconfig{U_F^\dagger(0,x^+;\vec x_\perp) 
    U_F(0,x^+;\vec y_\perp)}\\
    =&
    \exp\bigg[
    -C_F
    g^4 \tilde{\mu}^2 x^+
    \left[
      L(0)
    -
    L(|\vec x_\perp-\vec y_\perp|)
    \right]
    \bigg] 
    \;,
  \end{split}
\end{align}
in which 
\begin{align}\label{eq:Lxy}
  \begin{split}
  L\left(r=|\vec x_\perp-\vec y_\perp|\right)
  =&
  \int_{\k} \frac{e^{-i \vec k_\perp \cdot (\vec x_\perp-\vec y_\perp)}}{\left(m_g^2+\vec k_\perp^2\right)^2}
  =\frac{m_g r K_1(m_g r)}{4\pi m_g^2}
  \;.
\end{split}
\end{align}
  The summation over the color state of the quark amounts to the effective $\bar q q$ state being in the color singlet state.

Then the momentum transfer in Eq.~\eqref{eq:psq_eik} can be evaluated by taking the order derivative of the Wilson line correlator at the zero separation limit,
\begin{align}\label{eq:psq_eik_res}
  \begin{split}
    \braket{p_\perp^2 (x^+)}_{Eik}
    =& \braket{p_\perp^2 (0)}
    -
    \nabla_r^2 S_F(0,x^+;\vec r_\perp)|_{\vec r_\perp=\vec 0_\perp}\\
    =& \braket{p_\perp^2 (0)}
   + \hat q_{Eik} x^+
    \;.
  \end{split}
\end{align}
An alternative way to perform the derivation is to keep the momentum integral $ \int_{\p}$ and carry out the coordinate integral  $\int_{\r}$ instead,  which we discuss in more detail in Appendix~\ref{app:qhat_alt}. 
The quenching parameter $\hat q$ as defined in Eq.~\eqref{eq:dpperp} follows as,
\begin{align}\label{eq:qhat_Eik_res}
  \begin{split}
  \hat q_{Eik}
    =
  4\pi \hat q_0
\mathcal{G}_2,
\qquad
\hat q_{0}  
\equiv
C_F
g^4 \tilde{\mu}^2 
\frac{1}{4\pi}\;.
\end{split}
\end{align}
The ``bare'' quenching parameter $\hat q_0 $ is the effective speed for a quark reaching the saturation scale with half the evolution time, i.e., $\hat q_0 =Q_s^2/2 L_\eta$.
The quantity $\mathcal{G}_2$ contains a logarithmic divergence,
\begin{align}\label{eq:G2_x}
  \begin{split}
  \mathcal G_2
  =&
  -\nabla_r^2
  L(r)\big|_{r=0}\\
  =
  &  \frac{1}{4\pi}
  \Biggl\{
    \log\left[1+\frac{1}{(m_g a_\perp/\pi)^2}\right]
    -
    \frac{1}{1+(m_g a_\perp/\pi)^2}
  \Biggr\} 
    \;.
  \end{split}
\end{align}
In analogy, one gets the $\hat q$ for a gluon state replacing $C_F=(N_c^2-1)/(2 N_c)$ by $C_A=N_c$ in Eq.~\eqref{eq:qhat_Eik_res}.

In the above derivation, we do not take into account the effect from the momentum space lattice edges, allowing transverse momentum square increase linearly over time boundlessly. 
However, on the finite lattice, there is an asymptotic value for $\braket{p^2_\perp}$. This happens when the particles are distributed uniformly in the whole momentum space, 
\begin{align}\label{eq:pperp_asy}
  \begin{split}
    \braket{p^2_\perp}_{\mathrm{asy}}=&
    \frac{1}{(2N_\perp)^2}\sum_{i=-N_\perp}^{N_\perp-1}\sum_{j=-N_\perp}^{N_\perp-1}(i^2+j^2)
    \left(\frac{\pi}{L_\perp}\right)^2
    \approx
    \frac{2}{3}\left(\frac{\pi}{a_\perp}\right)^2
    \;,
  \end{split}
\end{align}
where the approximated value is obtained by taking the continuum limit. \footnote{ In the continuum limit, this asymptotic value is evaluated by integration instead of summation,
\begin{align}\label{eq:lattice_saturation}
  \braket{p^2_\perp}_{\mathrm{asy}}
  = \frac{
    \int_{-\lambda_{UV}}^{\lambda_{UV}}\diff k^x 
    \int_{-\lambda_{UV}}^{\lambda_{UV}}\diff k^y 
    [(k^x)^2+(k^y)^2]
    }
    {
    \int_{-\lambda_{UV}}^{\lambda_{UV}}\diff k^x
    \int_{-\lambda_{UV}}^{\lambda_{UV}}\diff k^y
    }
  =\frac{2}{3}\lambda_{UV}^2
  \;,
\end{align}
then on the lattice $\lambda_{UV}=\pi/a_\perp$.
}
Thus the linear growth of the momentum broadening with $x^+$ will saturate when approaching this limit.

\subsection{The quark-gluon state}\label{sec:eikonal_qg}
For a quark-gluon state, the expectation value of its total momentum squared can be evaluated from its wavefunction, in analogy to that of a single particle state in Eq.~\eqref{eq:psq_eik}, 
\begin{align}\label{eq:psq_full_r_qg}
  \begin{split}
     &\braket{p_\perp^2 (x^+)}_{qg,c;Eik}
   = 
    \int_{\p_q}
    \int_{\p_g}
    (\vec p_{q,\perp}+\vec p_{g\perp})^2
    \\
    &\qquad
    \int_{\x_q}\int_{\y_q}
    \int_{\x_g}\int_{\y_g}
        \tilde \phi^* (\vec x_{q,\perp},\vec x_{g,\perp})
    \tilde \phi (\vec y_{q,\perp},\vec y_{g,\perp})\\
     &\qquad
    e^{-i \vec p_{q,\perp}\cdot (\vec x_{q,\perp}-\vec y_{q,\perp})}
    e^{-i \vec p_{g\perp}\cdot (\vec x_{g,\perp}-\vec y_{g,\perp})}\\
    &\qquad \mathcal{P}_{qg,c} (0,x^+;\vec x_{q,\perp},\vec x_{g,\perp},\vec y_{q,\perp},\vec y_{g,\perp})
    \;.
  \end{split}
\end{align}
Here the initial state is written in form of the wavefunction in the coordinate basis, $\ket{\psi;0}
      =\int_{\x_q}\int_{\x_g}
      \tilde \phi (\vec x_{q,\perp}, \vec x_{g,\perp})\ket{\vec x_{q,\perp}, \vec x_{g,\perp}}$.

In the discussion for the single quark (or gluon), $\braket{p_\perp^2 (x^+)}$ is evaluated in the entirety of the corresponding color space, which is irreducible by itself. 
For a quark-gluon state, we examine the momentum broadening in each of its invariant color subspaces as well as the full. The invariant color space is indicated by the subscript ``c'' in the expression. An extensive discussion on the quark-gluon color space can be found in Appendix~\ref{app:qg_color}. To calculate the probability, one must square the wavefunction, keeping the colors of the incoming particles the same in the amplitude and the conjugate amplitude, and summing over the outgoing color in the full color space. 
The probability function of a quark-gluon state in the color state $c$ is written as $\mathcal{P}_{qg,c}$, 
\begin{align}
  \begin{split}
    &\mathcal{P}_{qg,c} 
    \equiv 
    \sum_{i=1}^{N_c}\sum_{a=1}^{d_A}
     \braket{
     \psi_{\bar q\bar g q g,\{i,a,i,a\} }
     | S_{\bar q \bar g q g} |
     \psi_{\bar q \bar g q g, \bar c(\bar q\bar g) c(q g)}
     }\;,
    \end{split}
  \end{align}
in which 
  \begin{align} \label{eq:colordecpomposition}
    \begin{split}
    &\bm{S}_{\bar qgqg} (0,x^+;\vec x_{q,\perp},\vec x_{g,\perp},\vec y_{q,\perp},\vec y_{g,\perp}) =\\
    &\qquad \expconfig{
     U_F^\dagger(\vec  x_{q,\perp})
    \otimes
    U_A^\dagger(\vec x_{g,\perp})
    \otimes
    U_F(\vec y_{q,\perp})
    \otimes
    U_A(\vec y_{g,\perp})
    }\;.
    \end{split}
  \end{align}
  is the $\bar q \bar g qg$ four-point Wilson line correlator. 
  Similarly to the $q$ state, the summation over the color states of the $qg$ state amounts to the effective $\bar q\bar g qg$ state being color singlets.
  We leave out the time argument in the Wilson lines for simplicity in the above expression.
  The details of $\bm{S}_{\bar qgqg}$ and $\mathcal{P}_{qg,c} $ can be found in App.~\ref{sec:UqgUdqg}, see also Ref.~\cite{Lappi:2020srm} for a study and analysis of $\bm{S}_{\bar qgqg}$.
  Note that in spite of the bra-ket notation in Eq.~\eqref{eq:colordecpomposition}, for our purpose this color structure is already at the probability level, with the $\bar q \bar g$ corresponding to a quark (gluon) in the conjugate wavefunction.

The quark-gluon momentum squared in Eq.~\eqref{eq:psq_full_r_qg} can be split into three terms,
 \begin{align}\label{eq:p_qg}
  \begin{split}
    \braket{p_\perp^2 (x^+)}_{qg,c;Eik}
    =&\braket{\vec p_{q,\perp}^2 (x^+)}_{Eik}
    +\braket{\vec p_{g,\perp}^2 (x^+)}_{Eik}\\
    &+2\braket{\vec p_{q,\perp}(x^+)\cdot \vec p_{g,\perp} (x^+)}_{c;Eik}
    \;.
  \end{split}
\end{align}
The first two terms turn out to be the same as Eq.~\eqref{eq:psq_eik_res} with the corresponding Casimir. For the third term, it is more convenient to change to relative and center-of-mass coordinates 
\begin{align}\label{eq:uv_variables}
  \begin{split}
        &\vec v_{\perp}=\vec x_{q,\perp}-\vec x_{g,\perp}\;,
    \qquad
    \vec R_{\perp}=z\vec x_{q,\perp}+(1-z)\vec y_{g,\perp}
    \;,\\
    &
    \vec u_{q,\perp}=\vec x_{q,\perp}-\vec y_{q,\perp}\;,
    \qquad
    \vec u_{g,\perp}=\vec x_{g,\perp}-\vec y_{g,\perp}
    \;,
  \end{split}
\end{align}
which allows us to express the momentum correlation as
\begin{align}\label{eq:pperp_qg_P}
  \begin{split}
    & \braket{\vec p_{q,\perp}(x^+)\cdot \vec p_{g,\perp} (x^+)}_{c;Eik}
    =   
    \int_{\v}
    f_{Rel} (\vec v_{\perp})
    \vec \nabla_{u,q} 
    \cdot 
    \vec \nabla_{u,g}\\
    &~~\mathcal{P}_{qg,c}(
     0,x^+;
  \vec v_\perp+ \vec u_{q,\perp}, 
  \vec u_{q,\perp},
  \vec v_\perp,
  \vec u_{q,\perp}-\vec u_{g,\perp}
    )
    \bigg|_{\vec u_{q,\perp},\vec u_{g,\perp}=\vec 0_\perp}
    \;.
  \end{split}
\end{align}
Note that we take advantage of the Wilson line correlator being translationally invariant in the above expression; see Eq.~\eqref{eq:S_shift} and discussions for detail. 
The quantity $f_{Rel} (\vec v_{\perp})$ is the distribution function of the quark-gluon relative coordinate $\vec v_{\perp}$, and it can be obtained by integrating the wavefunction square over the center-of-mass coordinate $\vec R_\perp$, 
\begin{align}\label{eq:f_rel}
  \begin{split}   
    f_{Rel} (\vec v_\perp)
    \equiv
    \int_{\R} 
    \left|\tilde \phi (\vec x_{q,\perp},\vec x_{g,\perp})\right|^2 
    \;.
  \end{split}
\end{align}

We find that the cross term depends on the initial color configuration of the quark-gluon state and the separation between the quark and the gluon,
\begin{align}\label{eq:pperp_qg_red}
  \begin{split}
    \braket{\vec p_{q,\perp}(x^+)\cdot \vec p_{g,\perp} (x^+)}_{c;Eik}
    =&    
    \begin{cases}
      0, & c=3\otimes 8\\
      -\frac{N_c \sqrt{2}}{2}f_{12}, & c=3\\
      -\frac{\sqrt{2}}{2}f_{12}, & c=\bar 6\\
      \frac{\sqrt{2}}{2}f_{12}, & c=15
    \end{cases}
    \;,
  \end{split}
\end{align}
in which $f_{12} = -\int_{\v}f_{Rel} (\vec v_\perp) s_{12} (v_{\perp})$ and the explicit form of the function $s_{12}$ is given in Eq.~\eqref{eq:DDS_res} in the Appendix. 
We leave the details of the derivation to Appendix~\ref{app:Wilsonlines}.
From Eq.~\eqref{eq:pperp_qg_red}, one can see that if the quark and the gluon are in a color-uncorrelated state, as is effectively the case if one sums over the whole $3\otimes 8$-dimensional color space of the final state, then the cross term vanishes.~\footnote{Note that if one sums the coefficients of the representations 3, $\bar{6}$ and 15 in Eq.~\eqref{eq:pperp_qg_red} weighted by the dimension of the representation, one gets zero; see Eq.~\eqref{eq:P_qg_X_cf} and discussions around it.}
For a color-correlated quark-gluon state, the cross term is large if the separation between the two particles, $v$ is small, and becomes negligible when the separation set by the wavefunction $\tilde \phi$ gets large, $v>1/m_g$.
In particular, if the quark-gluon state is a single momentum state, it is maximally delocalized, and the relative coordinate distribution is uniform, $f_{Rel} (\vec v_\perp)=1/(2 L_\perp)^2$, leading to a very weak correlation between the quark and gluon.

\section{$\hat q$ in the non-eikonal regime, using numerical simulations}
\label{sec:numerical}
\subsection{The single quark state $\ket{q}$}
We first perform the simulations in the leading Fock sector of $\ket{q}$, and study the evolution of the quark's transverse momentum.
In the eikonal limit of $p^+=\infty$, the quark's transverse momentum square $\braket{p^2_\perp ( x^+)}$ is expected to grow linearly over the evolution time $x^+$, as we have obtained from Eq.~\eqref{eq:psq_eik_res} in Sec.~\ref{sec:eikonal_q}.
We verify our numerical calculations in the eikonal limit by comparing them to the analytical expectation and also go beyond this limit by letting the quark have finite $p^+$.
We then study the dependence on the medium IR regulator $m_g a_\perp/\pi$ and the saturation scale $Q_s a_\perp/\pi$. Note that though the quark mass enters the kinetic energy term, as in Eq.~\eqref{eq:S_KE}, it acts as an overall phase factor in calculating $\braket{p^2_\perp ( x^+)}$ and therefore does not contribute.

\subsubsection{Dependence on $p^+$}\label{sec:qhat_ppl}
The longitudinal momentum $p^+$ signifies how fast the quark jet travels through the medium, and its effect is related to the structure of the medium along $x^+$. The medium, according to the MV model, should be singularly uncorrelated as $\expconfig{\mathcal{A}(x^+)\mathcal{A}( y^+)}\propto \delta(x^+ - y^+)$ as in Eq.~\eqref{eq:chgcor}.
On the amplitude level, it means that $\mathcal{A}(x^+)$ is stochastic, which is realized numerically with a $x^+$-resolution of $\tau$; this is in analogy to having an $x_\perp$-resolution of $a_\perp$ in the transverse dimension, see also Eq.~\eqref{eq:chgcor_dis}. Consequently, to match the analytical formulation, one should take the continuous limit of $a_\perp\to 0$ ($N_\perp\to\infty$ while fixing $L_\perp$) and $\tau\to 0$ ($N_\eta\to\infty$ while fixing $L_\eta$). However, in reality, the medium, as a composition of quarks and gluons, is more likely to  have a finite correlation length. In the transverse dimension, the medium IR regulator $m_g$ plays such a role of screening. Considering the medium in its rest frame and assuming that it is isotropic, this would imply that the duration of each layer would be of the same magnitude, such that $\tau\sim 1/m_g$. When the medium is longitudinally boosted, e.g., in a frame with its $p^-\approx \infty$, both $\tau$ and $L_\eta$ get contracted, but $N_\eta$ will stay the same as in the medium rest frame. Thus, to study a finite energy jet it is interesting to also perform simulations with a finite $\tau$.

\begin{figure*}[htp]
  \centering
  \subfigure[$Q_s a_\perp/\pi=0.01$ \label{fig:Fig1_Neta_smallQs}]{
    \includegraphics[width=0.4\textwidth]{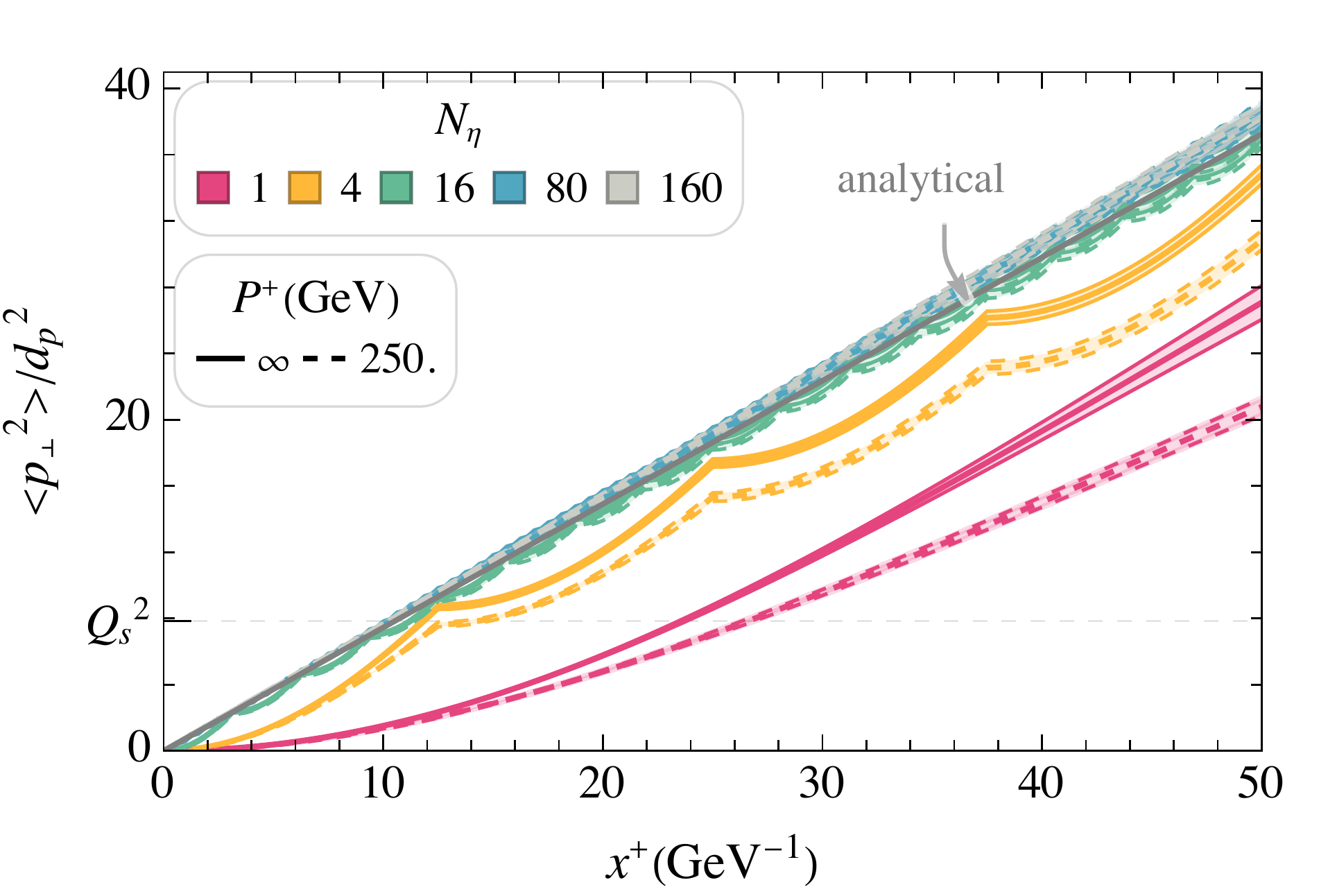} 
    \qquad
    \includegraphics[width=0.4\textwidth]{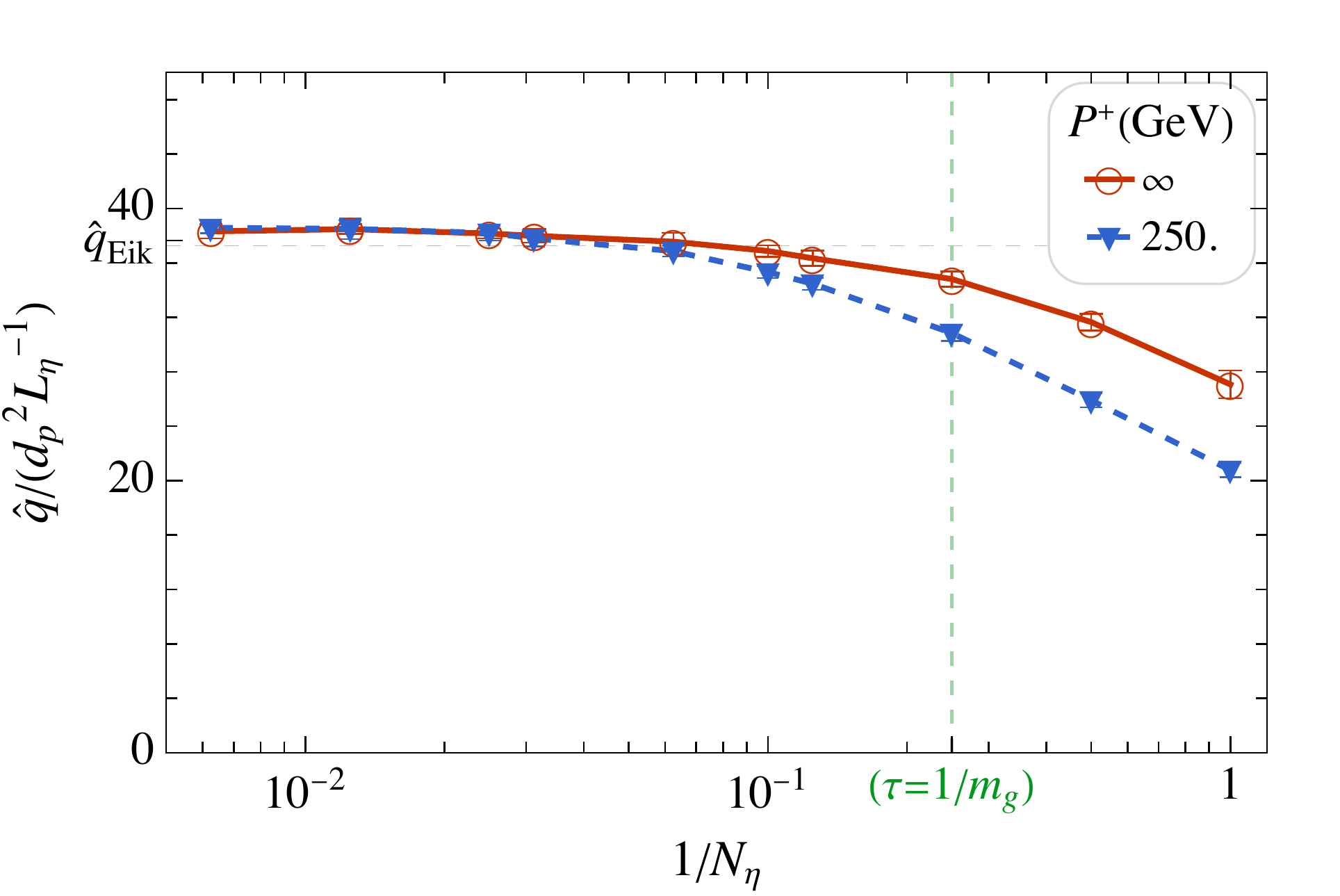} 
    }
  \subfigure[ $Q_s a_\perp/\pi=0.04$\label{fig:Fig1_Neta_largeQs}]{
    \includegraphics[width=0.4\textwidth]{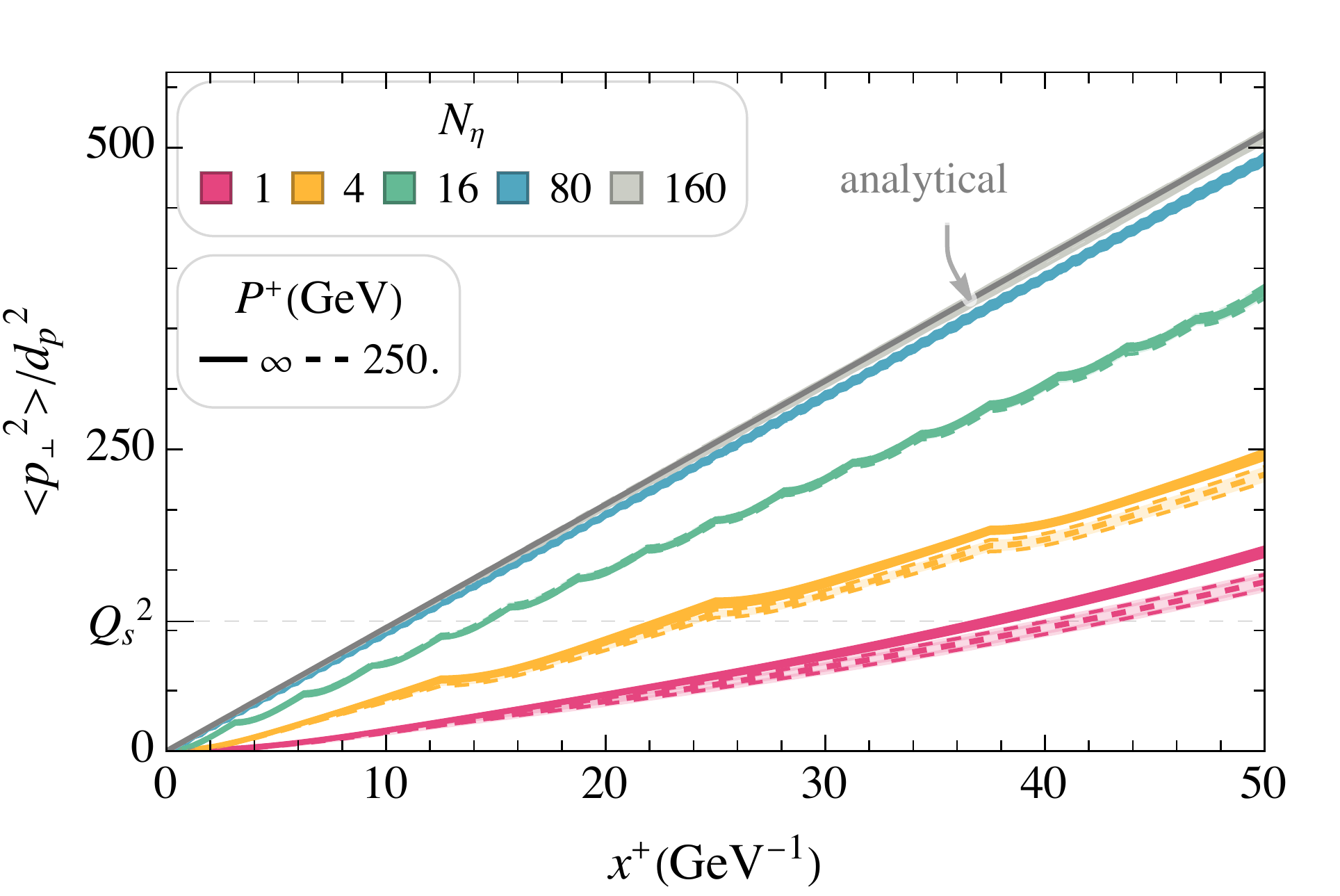} 
    \qquad
    \includegraphics[width=0.4\textwidth]{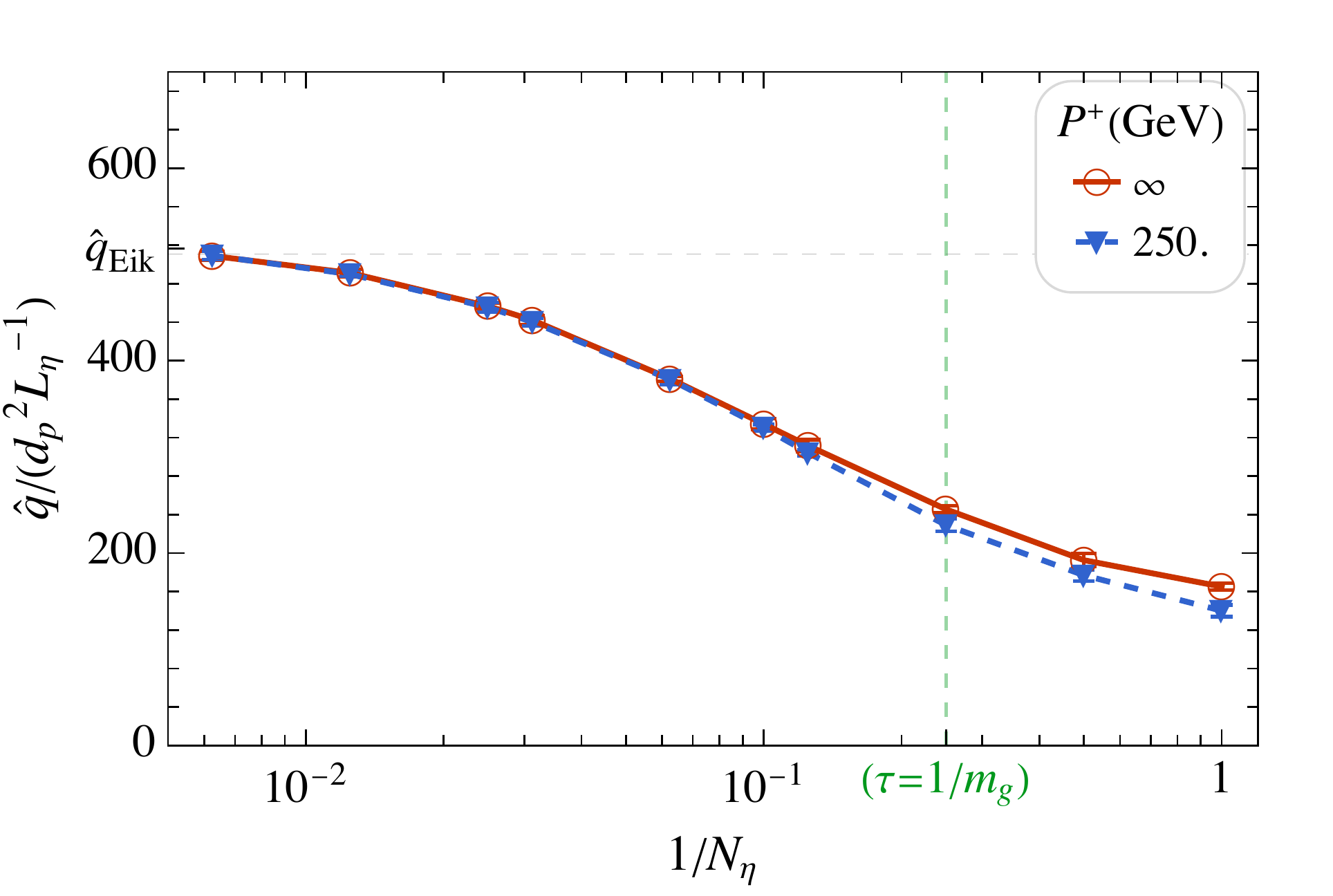} 
    }   
  \caption{\label{fig:Fig1_Neta} 
  The dependence of the momentum broadening on the number of medium layers $N_\eta$, at (a) a smaller and (b) a larger saturation scale $Q_s$.
  Left panels: the transverse momentum square $\braket{p_\perp^2 (x^+)}$ as a function of $x^+$ at various $N_\eta$ with fixed $m_g a_\perp/\pi=0.005$ and $Q_s$ (with the bare $Q_s^2/d_p^2$ given by Eq.~\eqref{eq:bareqs} indicated by the horizontal dashed line). 
  Results at infinite(finite) $p^+$ are shown in solid (dashed) lines, and the band indicates the fluctuation from 10 configurations. 
  Right panels: the quenching parameter $\hat q$ as a function of $1/N_\eta$. The green vertical dashed line indicates the value of $1/N_\eta$ at which $\tau=1/m_g$. 
  The eikonal analytical results according to Eq.~\eqref{eq:psq_eik_res} are in the solid black line indicated as ``analytical" in the left panels, and in the dashed grey line indicated as ``$\hat q_{Eik}$" in the right panels.
  }
\end{figure*}

To proceed, we first study the dependence of the momentum broadening on $N_\eta$.
In Fig.~\ref{fig:Fig1_Neta}, we present $\braket{p^2_\perp (x^+)}$ and the extracted values of the quenching parameter $\hat q$ [calculated according to Eq.~\eqref{eq:dpperp}, with $\Delta x^+=L_\eta$] at various $N_\eta$ while fixing $L_\eta$. 
The value at each data point is averaged over 10 configurations, and the band width is the standard deviation indicating the uncertainties from the configuration fluctuation.
Though in both infinite and finite $p^+$ cases, the simulation results converge to the analytical expectation as $N_\eta$ increases, the latter happens at a slower pace. 
We have three key observations. 
First, in the continuous limit of $N_\eta\to \infty$, the kinetic energy term does not affect the momentum broadening and the value of $\hat q$. Thus the analytical result derived in the eikonal limit in Sec.~\ref{sec:eikonal_q} also applies to finite $p^+$ cases; see also Ref.~\cite{Blaizot:2012fh}. Second, the convergence to the eikonal analytical result as $N_\eta\to \infty$  is slower at a larger saturation scale, as shown by the comparison between Fig.~\ref{fig:Fig1_Neta_smallQs} and Fig.~\ref{fig:Fig1_Neta_largeQs}.
Third, at a finite $N_\eta$, in particular when $\tau=1/m_g$, a finite $p^+$ leads to a decrease in $\hat q$. 

It is also interesting to observe the momentum evolution within a layer. The results at $N_\eta=4$ in Fig.~\ref{fig:Fig1_Neta} exemplify that $\braket{p^2_\perp ( x^+)}$ grows quadratically within a layer while linearly across layers. The analytical derivation in Sec.~\ref{sec:eikonal_q} inherits the treatment of $\expconfig{\rho(x^+)\rho( y^+)}\propto \delta(x^+ - y^+)$ from the MV model, therefore only accounts for the behavior across layers. To understand the evolution within a layer analytically, let us calculate the Wilson line in such a scenario, there the $x^+$ integral becomes
  \begin{align}
    \begin{split}
      \int_0^{\delta x^+} \diff x^+&
      \int_0^{\delta x^+} \diff y^+
      \expconfig{\rho_a(x^+, \vec x_\perp)\rho_b(y^+, \vec y_\perp)}\\
      =&\delta_{ab}\delta^2(\vec{x}_\perp-\vec{y}_\perp)  
      g^2 \tilde{\mu}^2 
      \begin{cases}
        \delta x^+, & \delta x^+ \ge \tau\\
        (\delta x^+)^2/\tau, & \delta x^+ <\tau
      \end{cases}
      \;.
    \end{split}
  \end{align}
The resulting Wilson line within a layer, i.e., $\delta x^+ <\tau$, is thus $S_F(0,\delta x^+ ;r) 
=e^{-C_F g^4 \tilde{\mu}^2  (\delta x^+)^2/\tau \left[L(0)-L(r)\right]} $, in comparison to Eq.~\eqref{eq:SF} for the Wilson line across multiple layers. 
The observation from the numerical simulation verifies this quadratic/linear behavior of the  squared momentum.

\begin{figure*}[t]
  \centering
    \includegraphics[width=0.8\textwidth]{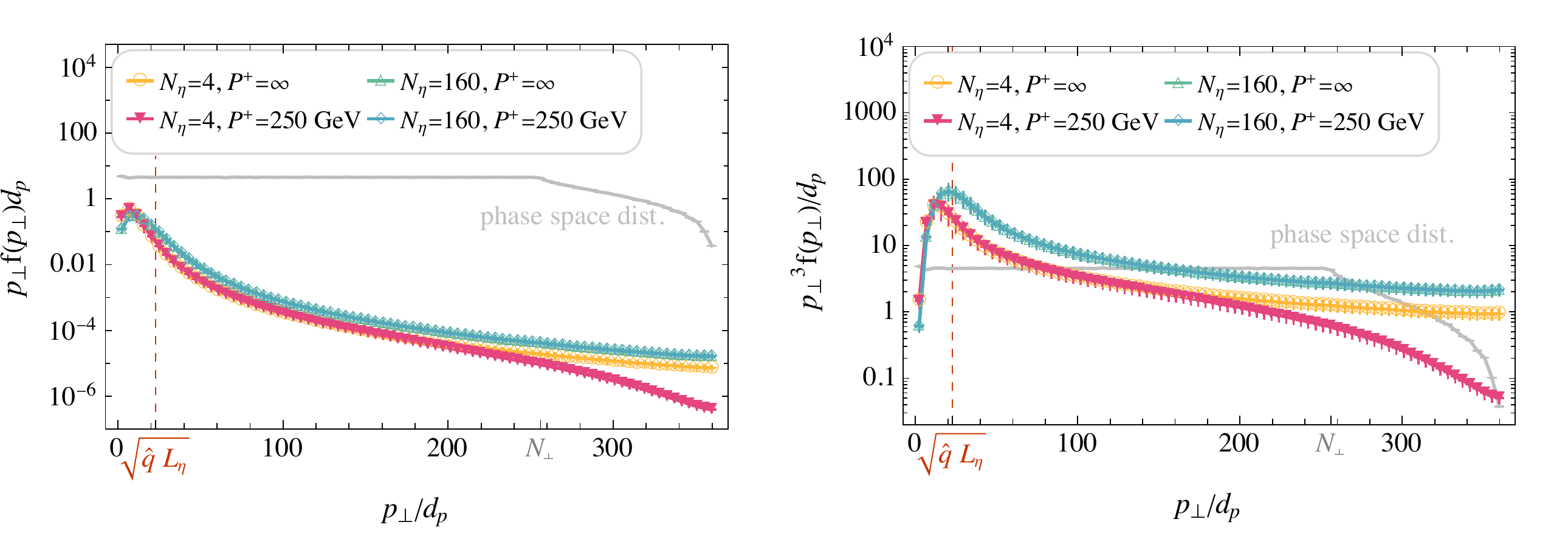}
  \caption{ 
  \label{fig:ffp_ppl}  
  The transverse momentum distribution $p_\perp f(p_\perp) d_p$ in the left panel and $p_\perp^3 f(p_\perp)/d_p $ in the right panel, both as a function of $p_\perp$ at infinite and finite $p^+$. 
  The light gray line indicates the relative size of the basis state phase space to the continuous case, according to $R(p_\perp)$ in Eq.~\eqref{eq:TMD_R}.}
\end{figure*}

For a closer examination, we present the transverse momentum distribution $f(p_\perp)$ at both infinite and finite $p^+$ in Fig.~\ref{fig:ffp_ppl}. Note that the phase space of states on the discrete square lattice is different from that in the continuous case, and their ratio is indicated by the dashed gray line in the left panel. More information can be found in Appendix~\ref{app:TMD}. We can see that the typical transverse momentum is of order $\sqrt{\hat{q}L_\eta}$, but there is a long power law tail up to higher $p_\perp$. The difference between different values of $N_\eta$ is smooth as a function of $p_\perp$, but the lower typical $p_\perp$ for finite $P^+$ at small $N_\eta$ is caused by a depletion of the highest $p_\perp$ modes.

\begin{figure}[tbp!]
  \centering
  \subfigure[$N_\eta=4$\label{fig:Fig2_ppl_Neta4}]{
      \includegraphics[width=0.4\textwidth]{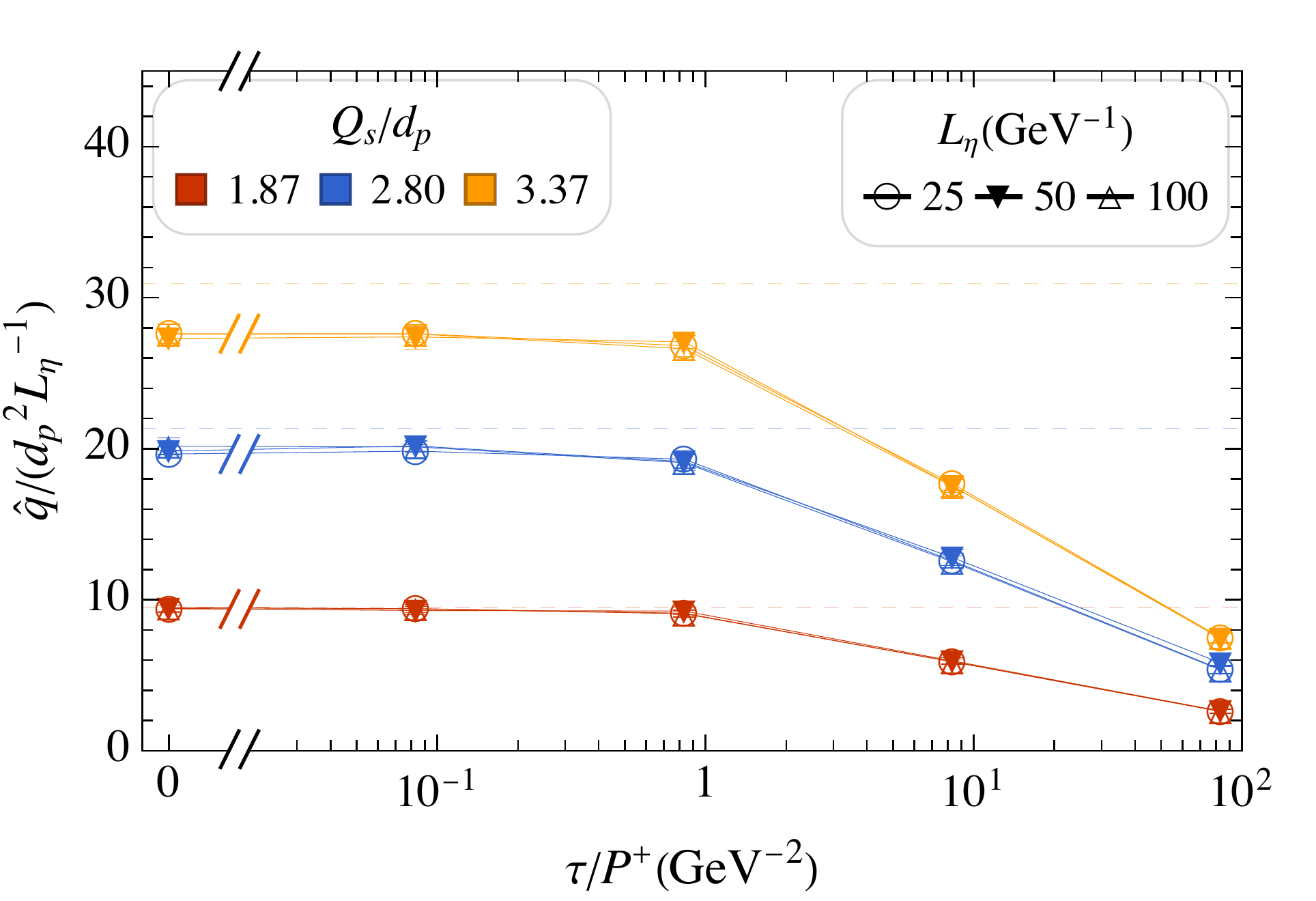} 
  }
  \subfigure[$N_\eta=8$\label{fig:Fig2_ppl_Neta8}]{
      \includegraphics[width=0.4\textwidth]{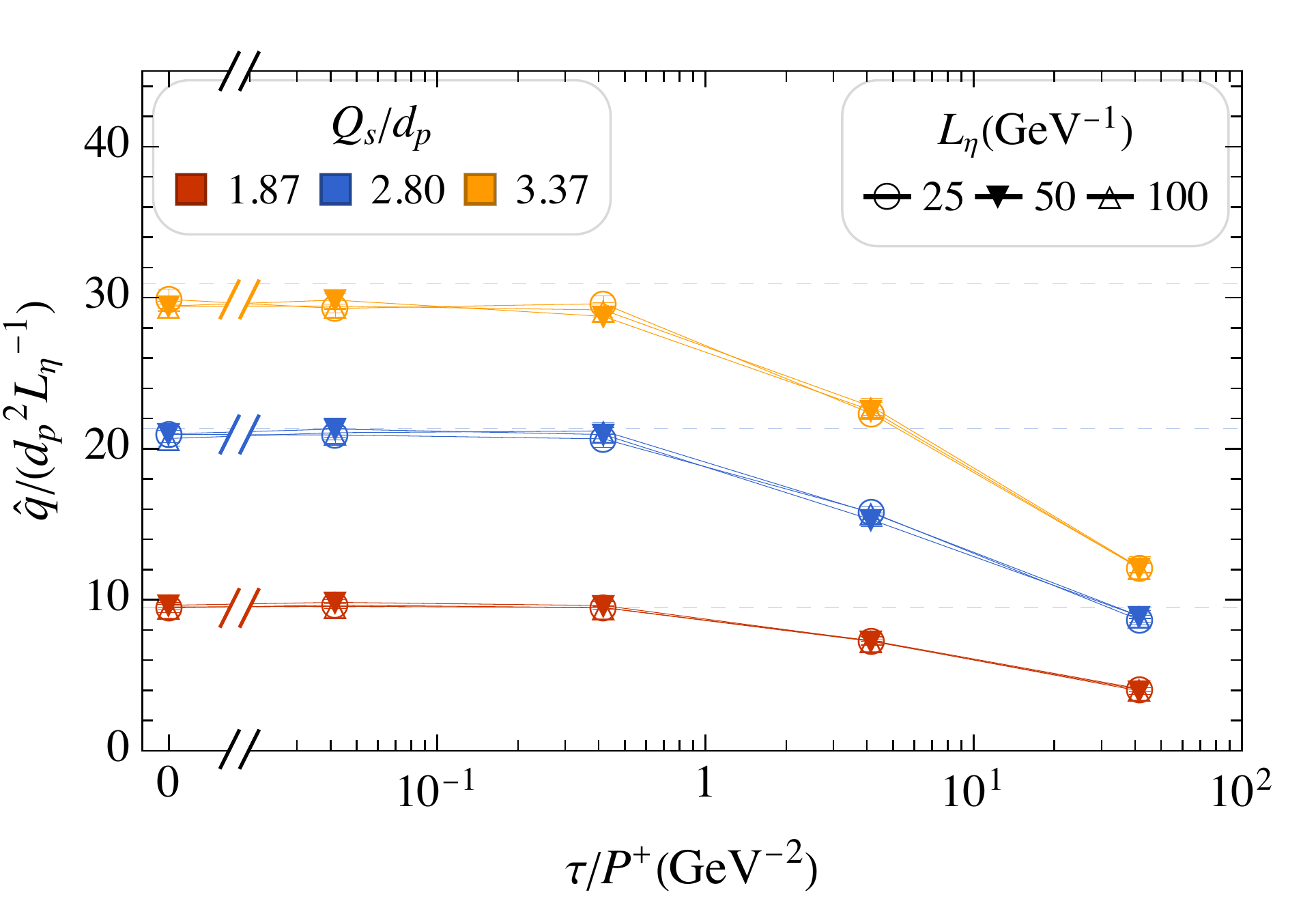} 
  }
  \caption{ 
    \label{fig:Fig2_ppl}
    The dependence of $\hat q$ on the particle longitudinal momentum $p^+$ at various saturation scale $Q_s$ and at a fixed $m_g a_\perp/\pi=0.04$, with (a) $N_\eta=4$ and (b) $N_\eta=8$.
    The results at different $L_\eta$ overlap with each other.
    The horizontal lines are the values of $\hat q_{Eik}$ at $\tau/p^+=0$ according to Eq.~\eqref{eq:psq_eik_res}.
  }
\end{figure}

Next, we take the estimation of $\tau= 1/m_g$ such that $N_\eta=4$ with $L_\eta=50 \GeV^{-1}$, and study the $p^+$ dependence. 
Note that both $N_\eta$ and $L_\eta/p^+$ are boost-invariant, and here we fix $L_\eta$ and $\tau$. One could alternatively fix $p^+$ and $N_\eta$ but vary $L_\eta$ (and thus $\tau$) to obtain the same results.  
The results are shown in Fig.~\ref{fig:Fig2_ppl_Neta4}, and the results with $N_\eta=8$ are in Fig.~\ref{fig:Fig2_ppl_Neta8} for comparison. 
We find that with a finite number of layers, a smaller $p^+$ leads to a smaller $\hat q$. 
Then towards the eikonal limit of $p^+=\infty$, the obtained $\hat q$ gets closer to the analytical result calculated with an infinite number of layers.
But even at $\tau/p^+=0$, i.e., in the eikonal case, the obtained $\hat q$ does not always agree with the analytical ones, especially at larger $Q_s$. This difference results from $N_\eta$ being finite in the setup. 
As we have seen earlier in Fig.~\ref{fig:Fig1_Neta}, a larger $Q_s$ requires a larger value of $N_\eta$ to restore the analytical results.
In addition, the results obtained at different $L_\eta$ overlap, showing that $\hat q$ in the units of $d_p^2 L_\eta^{-1}$ only depends on the boost-invariant quantity of $\tau/p^+$, but not $p^+$ or $L_\eta$ separately.

\subsubsection{Dependence on $m_g a_\perp/\pi$}
The quenching parameter $\hat q$ is expected to have a logarithmic dependence on $m_g a_\perp/\pi$ according to the analysis in Sec.~\ref{sec:eikonal_q}. 
In the numerical simulations, one can separately change one or multiple of the three quantities $L_\perp$, $N_\perp$, and $m_g$, but the physical results should remain the same if $m_g a_\perp/\pi$ are fixed. 
Yet there is a prerequisite, the medium IR regulator $m_g$ should be covered by the lattice resolution, i.e., $m_g d_p (= m_g L_\perp/\pi) >1$; otherwise, the lattice IR cutoff $d_p$ would effectively act as the IR regulator instead.

\begin{figure}[tbp!]
  \centering
    \includegraphics[width=0.4\textwidth]
    {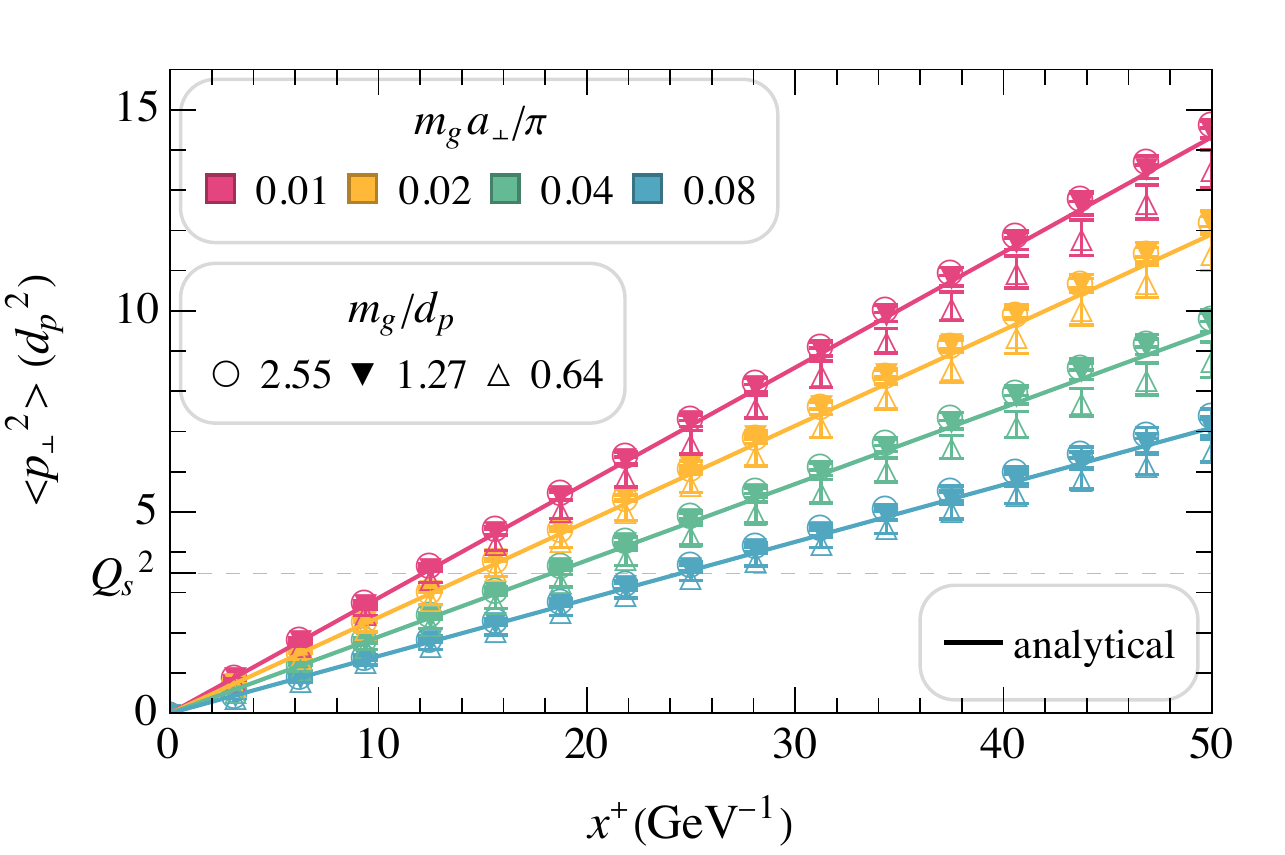} 
    \includegraphics[width=0.4\textwidth]{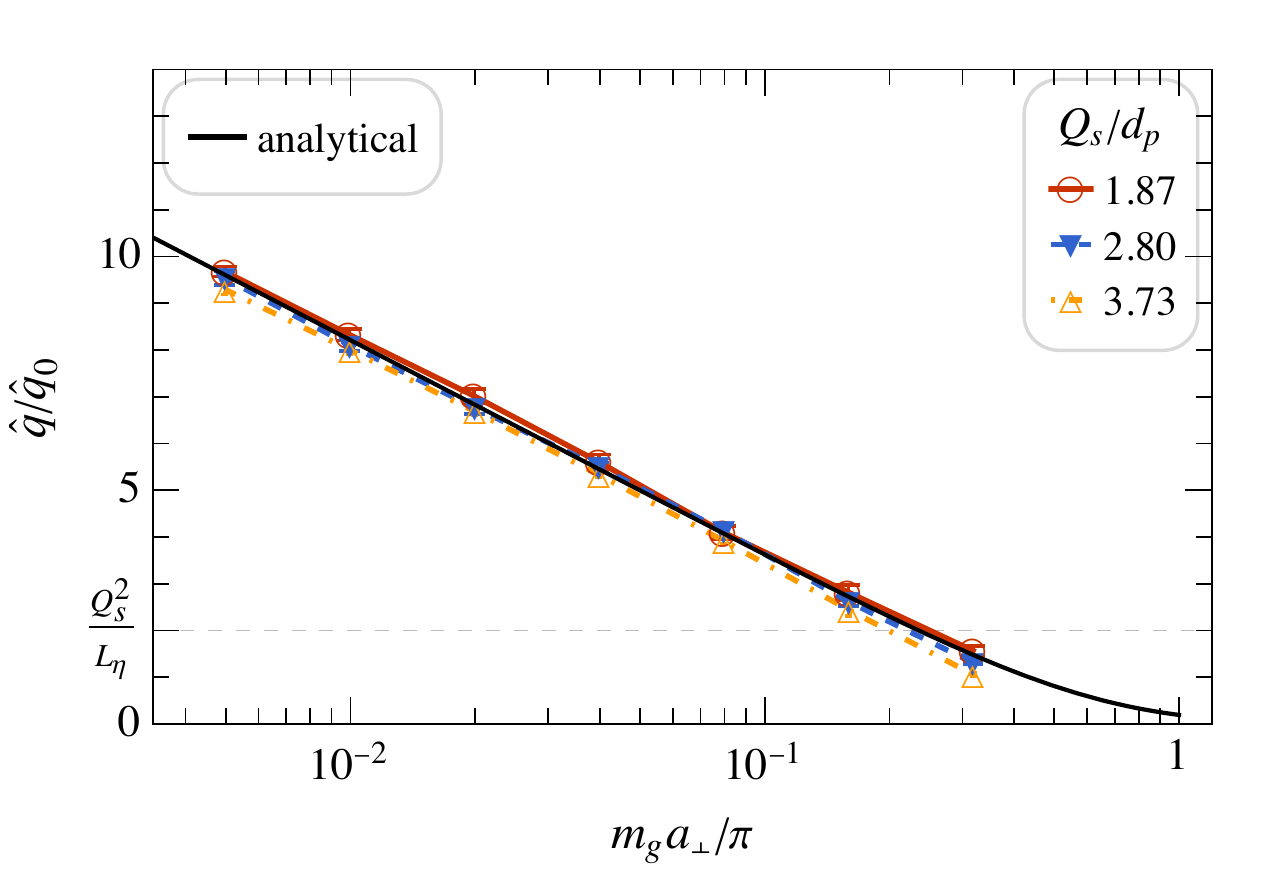} 
  \caption{ 
    \label{fig:Fig3_mg}
    The dependence on $m_g a_\perp/\pi$ of the momentum broadening.
    Top panel: The transverse momentum square $\braket{p_\perp^2 (x^+)}$ as a function of $x^+$ at various $m_g a_\perp/\pi$ and a fixed $Q_s$ (with the bare $Q_s^2/d_p^2$ given by Eq.~\eqref{eq:bareqs} indicated by the horizontal dashed line). 
    The data at the two larger $m_g d_p >1$, in open circles and solid triangles, overlap with each other, whereas that at the smaller $m_g d_p<1$, in the open triangle, deviates below the former. 
    Bottom panel: The quenching parameter $\hat q$ as a function of $m_g a_\perp/\pi$ at various $Q_s/d_p$ with $m_g d_p =1.27$. 
    The analytical results in the solid lines are given by Eq.~\eqref{eq:psq_eik_res}.
     }
\end{figure}

As shown in Fig.~\ref{fig:Fig3_mg}, the results obtained at $m_g d_p >1$ agree with the analytical results at each $m_g a_\perp/\pi$, whereas that at $m_g d_p < 1$ deviates below since the screening mass in the medium is no longer sufficiently represented by $m_g$. 
One could also see that the quenching parameter relative to its bare value, $\hat q/\hat q_0$, depends logarithmically on $m_g a_\perp/\pi$, but not on $Q_s/d_p$.

\subsubsection{Dependence on the saturation scale $Q_s$} 

\begin{figure}[tbp!]
  \centering
    \includegraphics[width=0.39\textwidth]
    {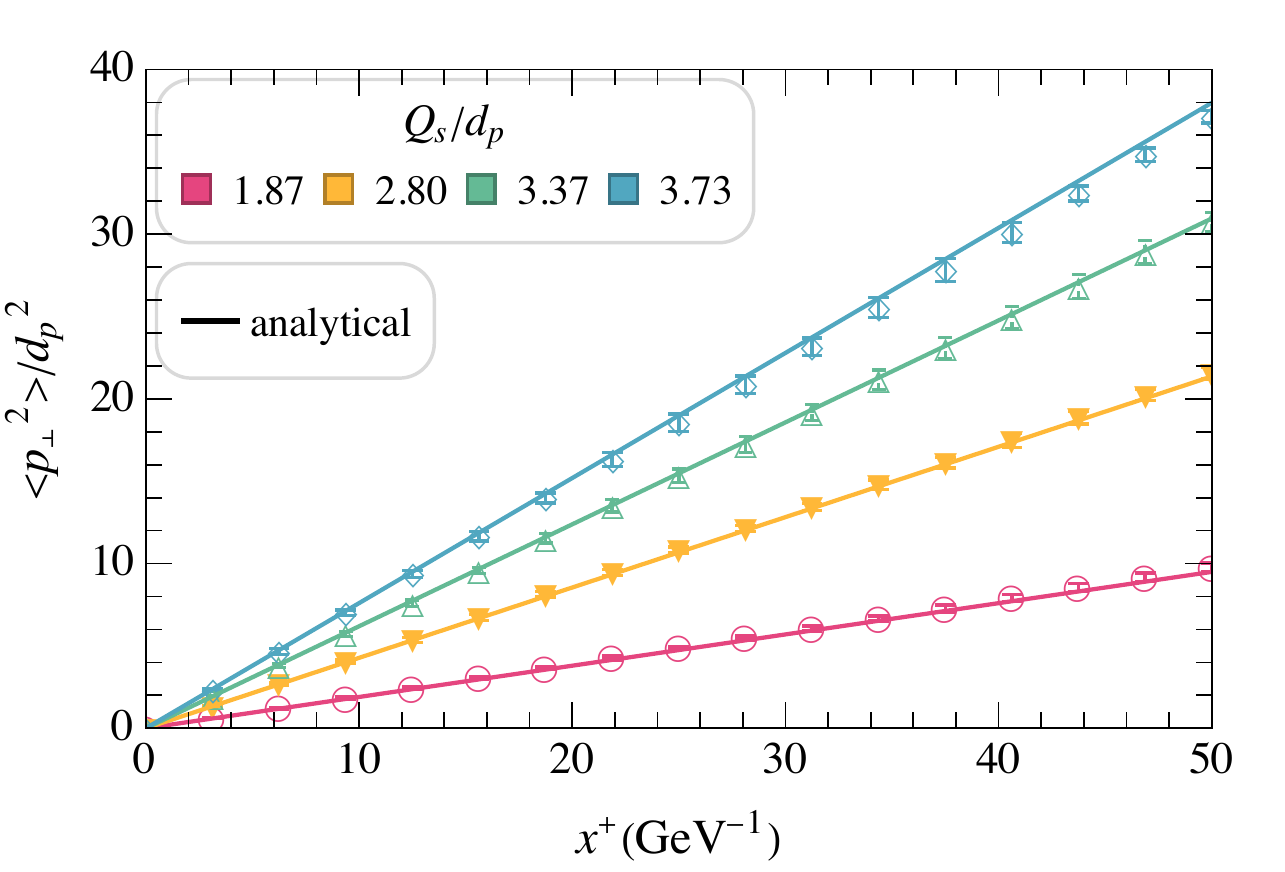} 
    \includegraphics[width=0.4\textwidth]{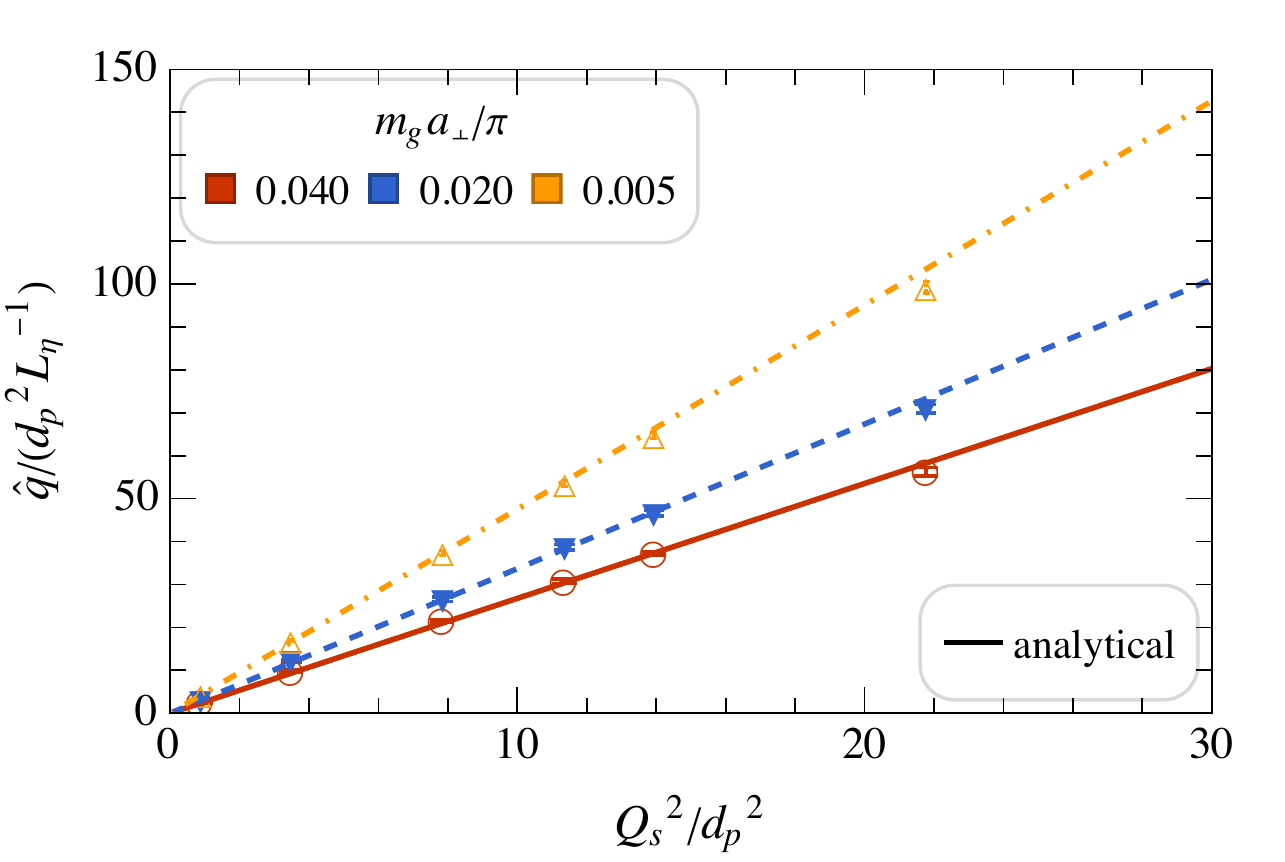} 
  \caption{ 
    \label{fig:Fig4_Qs}
    The dependence on $Q_s/d_p$ of the momentum broadening.
    Top panel: The transverse momentum square $\braket{p_\perp^2 (x^+)}$ as a function of $x^+$ at various $Q_s/d_p$ with a fixed $m_g a_\perp/\pi=.04$. 
    Bottom panel: The quenching parameter $\hat q$ as a function of $Q_s^2/d_p^2$ at various $m_g a_\perp/\pi$.
    The analytical results in the solid lines are given by Eq.~\eqref{eq:psq_eik_res}.
     }
\end{figure}

The quenching parameter $\hat q$ is expected to have a linear dependence on $Q_s$ defined by \eqref{eq:bareqs}, according to the analysis in Sec.~\ref{sec:eikonal_q}. 
We show in Fig.~\ref{fig:Fig4_Qs} that the transverse momentum squared $\braket{p_\perp^2 (x^+)}$ as a function of $x^+$ at various $Q_s/d_p$ agrees with the analytical expectation. 
Keeping $m_g a_\perp/\pi$ fixed, one could also see that $\hat q$ grows linearly with $Q_s^2/d_p^2$.

\subsection{The quark-gluon state $\ket{qg}$}\label{sec:qg}
Having studied the momentum broadening of a single particle state, we now proceed to the two-particle state $\ket{qg}$. When traversing through the medium, the quark and the gluon each exchange color and transverse momentum with the medium, in the manner of a single particle. Meanwhile, the correlation between them introduces additional complexity in examining their total and relative momenta. The quark-gluon correlation is characterized by their separation in the transverse coordinate space and their color configuration, as we have studied analytically in Sec.~\ref{sec:eikonal_qg}. The effect of such a correlation throughout the evolution is of our interest. To this purpose, we perform the simulations with quark-gluon states in different color-correlated configurations and separated by small and large distances.

Note that in a quantum formalism, as in this work, it is impossible to assign a particle state with a specific momentum and coordinate simultaneously, which one could do easily in a classical picture. Instead, the wavefunction in momentum space and that in coordinate space are related by the Fourier transform. Specifically, a single coordinate state is uniformly distributed in the momentum space. Consequently, a quark-gluon state with both particles in the same coordinate mode, which one would instinctively think of as the most correlated state, is not helpful in observing momentum broadening since all $p_\perp$ modes are already equally occupied before entering the medium. A more realistic and appropriate setup is to have the quark and the gluon as two Gaussian wavepackets, which we adopt in the following study.
The quark (gluon) is centered at $\vec s_{q,\perp}$ ($\vec s_{g,\perp}$) with a width of $ w_q$ ($w_g$), such that the wavefunction reads
\begin{align}\label{eq:qgWF}
  \begin{split}   
\tilde \phi (\vec r_{q,\perp},\vec r_{g,\perp})
=C e^{-\frac{1}{2}\left(\frac{\vec r_{q,\perp}-\vec s_{q,\perp}}{w_q}\right)^2}
\otimes 
e^{-\frac{1}{2}\left(\frac{\vec r_{g,\perp}-\vec s_{g,\perp}}{w_g}\right)^2}
    \;.
  \end{split}
\end{align}
Here, $C$ is a constant determined by the normalization condition.
In order to have the quark and the gluon each localized in transverse coordinate space, their Gaussian width $w_q$ and $w_g$ should be smaller than the inverse of their respective transferred momentum,
\begin{align}\label{eq:qgWF_condition}
    w_q, w_g< \frac{1}{\sqrt{\hat{q}_{Eik} L_\eta}}
    \;.
\end{align}

\begin{figure*}[tbp!]
  \centering
  \subfigure[ \label{paper_qg_fR2D_short_ti}]{   
    \includegraphics[width=0.8\textwidth]{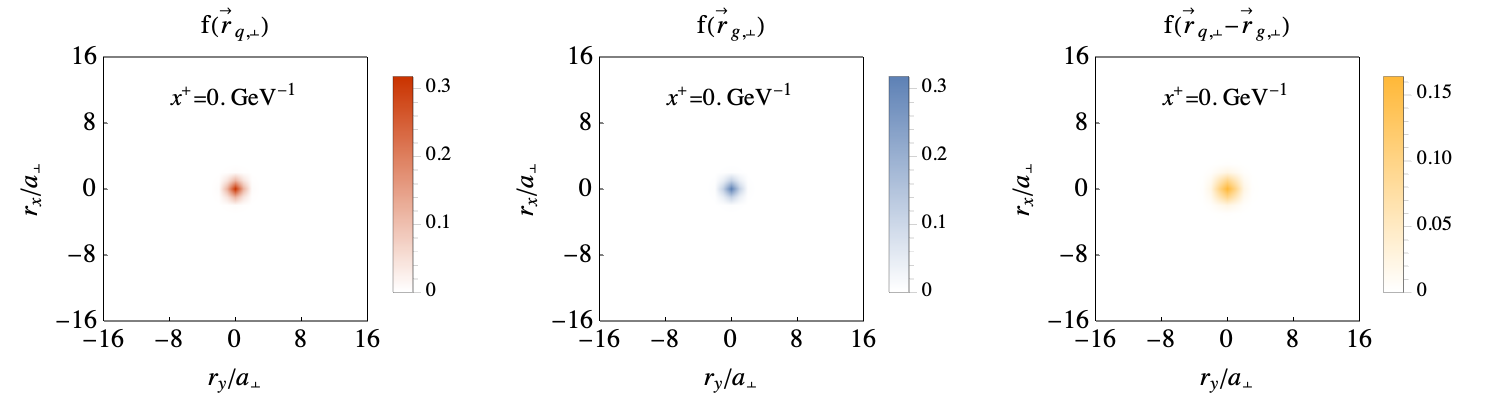}
  }
  \subfigure[\label{paper_qg_fR2D_long_ti}]{   
    \includegraphics[width=0.8\textwidth]{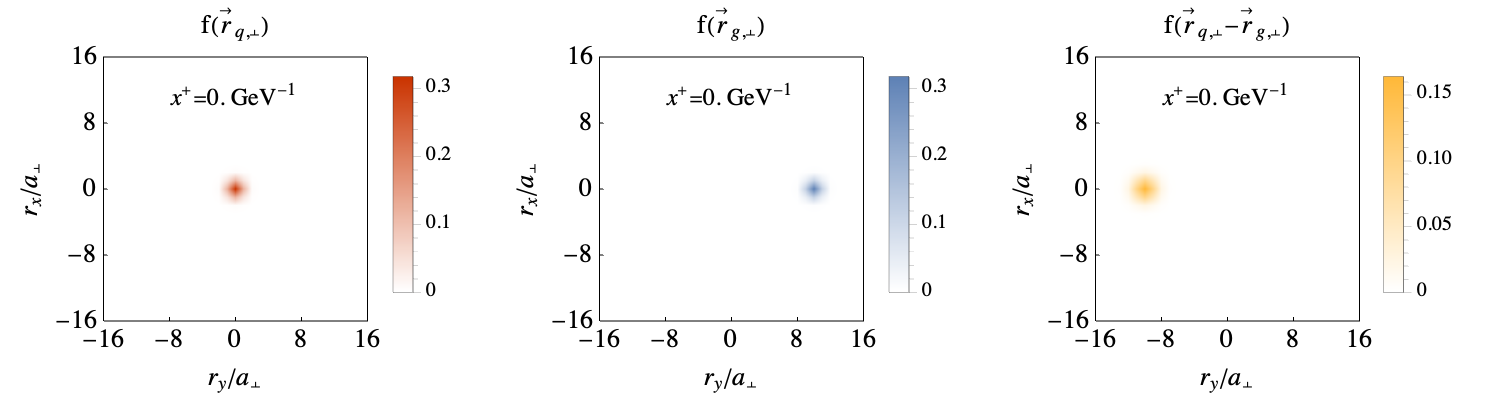}
    }
  \caption{ 
  \label{fig:qg_shortlong}  
   Initial distributions of the quark's, the gluon's, and their relative transverse coordinate for (a) a small-separation, and (b) a large-separation quark-gluon state, with $a_\perp\approx 0.6$ fm.
     }
\end{figure*}

We set the initial state as given in Eqs.~\eqref{eq:qgWF} and \eqref{eq:qgWF_condition} in two cases: (a) $\vec s_{q,\perp}/a_\perp=\vec s_{q,\perp}/a_\perp=\{0,0\}$, and (b) $\vec s_{q,\perp}/a_\perp=\{0,0\}$ and $\vec s_{q,\perp}/a_\perp=\{0,10\}$; for both cases, $w_q= w_g=a_\perp $. Figure~\ref{fig:qg_shortlong} shows the distributions of the quark and the gluon in the transverse coordinate space in both cases.
Note that the parameter in (b) is not special but just one choice that guarantees a large quark-gluon separation, where a separation of 10 lattice units is considerably large on a periodic 32 by 32 lattice.
The distribution function of the relative transverse coordinate is defined as in Eq.~\eqref{eq:f_rel}, by integrating over the center-of-mass part. The single particle distribution function could also be calculated from the wavefunction, by integrating over the dependence on the coordinate of the other particle,
\begin{align}\label{eq:f_single}
  \begin{split}   
    &f_q (\vec r_{q,\perp})
    \equiv
    \int_{\r_{g}} 
    \left|\tilde \phi (\vec r_{q,\perp},\vec r_{g,\perp})\right|^2 \;,\\
    &
    f_g (\vec r_{g,\perp})
    \equiv
    \int_{\r_{q}} 
    \left|\tilde \phi (\vec r_{q,\perp},\vec r_{g,\perp})\right|^2 
    \;.
  \end{split}
\end{align}

\subsubsection{Small-separation quark-gluon state}
First, we study a scenario where the two particles are extensively correlated in space. 
We set the initial state as given in Eqs.~\eqref{eq:qgWF} and \eqref{eq:qgWF_condition} with $\vec s_{q,\perp}/a_\perp=\vec s_{q,\perp}/a_\perp=\{0,0\}$ and $w_q= w_g=a_\perp $, i.e., case (a) in Fig. \ref{fig:qg_shortlong}. We study the correlation by examining the evolution of $\braket{ p_\perp^2}$ for the two-particle state. 

\begin{figure*}[tbp!]
  \centering
      \includegraphics[width=0.88\textwidth]{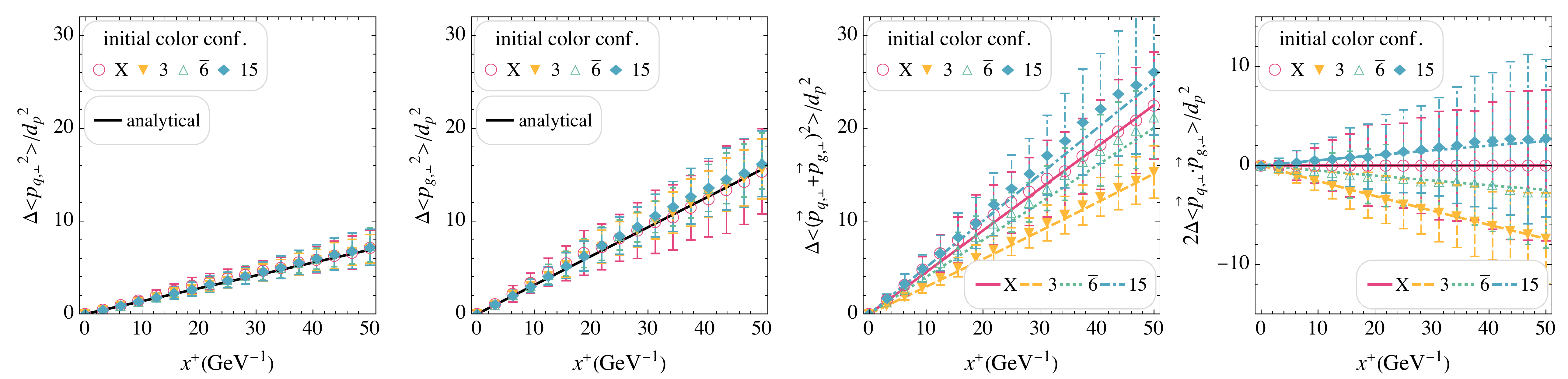}
  \caption{ 
  \label{fig:qg_short}  
  The transverse momentum evolution of a small-separation quark-gluon state [distribution as shown in Fig.~\ref{paper_qg_fR2D_short_ti}] in the eikonal limit of $p^+=\infty$, at $Q_s a_\perp/\pi=0.12$. Simulation parameters: $N_\perp=16$, $L_\perp=50 \GeV^{-1}$, and $\tau=10 \GeV^{-1}$.
  From left to right, it shows the $\braket{p^2_\perp}$ of the quark, the gluon, the total, and the cross terms $2\braket{(\vec p_{q,\perp}\cdot \vec p_{g,\perp})^2}$.
  The legends of open markers indicate the initial color configurations of the quark-gluon state, ``$X$" as uncorrelated,``$3$" as in the triplet, ``$\bar{6}$" as in the anti-sextet, and ``$15$" as in the deciquintuplet.
  The numerical values are averaged over 64 configurations, and the uncertainty bars indicate the standard deviation.
  The lines are analytical expectations at $p^+=\infty$ evaluated according to Eqs.~\eqref{eq:p_qg} and \eqref{eq:pperp_qg_red}.
  }
\end{figure*}

In the eikonal limit of $p^+=\infty$, the value of $\braket{ p_\perp^2(x^+)}$ can be calculated exactly according to Eq.~\eqref{eq:pperp_qg_red}. The results from numerical simulation agree with such expectations, as shown in Fig.~\ref{fig:qg_short}. 
To see the momentum that gets broadened, we present the change $\Delta\braket{ p_\perp^2}\equiv \braket{ p_\perp^2(x^+=L_\eta)}-\braket{ p_\perp^2(x^+=0)}$ in these plots.
The momentum broadening of the quark (gluon) individually is the same, regardless of the quark-gluon color configurations. This is because the individual particle's momentum broadening in the $\ket{qg}$ state is the same as that of the single particle. On the contrary, the total momentum differs for different color configurations. This can be understood in an intuitive way. Let us first think of the quark and the gluon coinciding in the transverse coordinate space. Then its momentum broadening would behave as a single particle state, and depends on its color configuration. Though in the simulated state, the quark and the gluon do not coincide exactly, their separation is small relative to the reaction scale of the medium; one could, therefore, still observe the difference.    

\begin{figure*}[tbp!]
  \centering
      \includegraphics[width=0.88\textwidth]{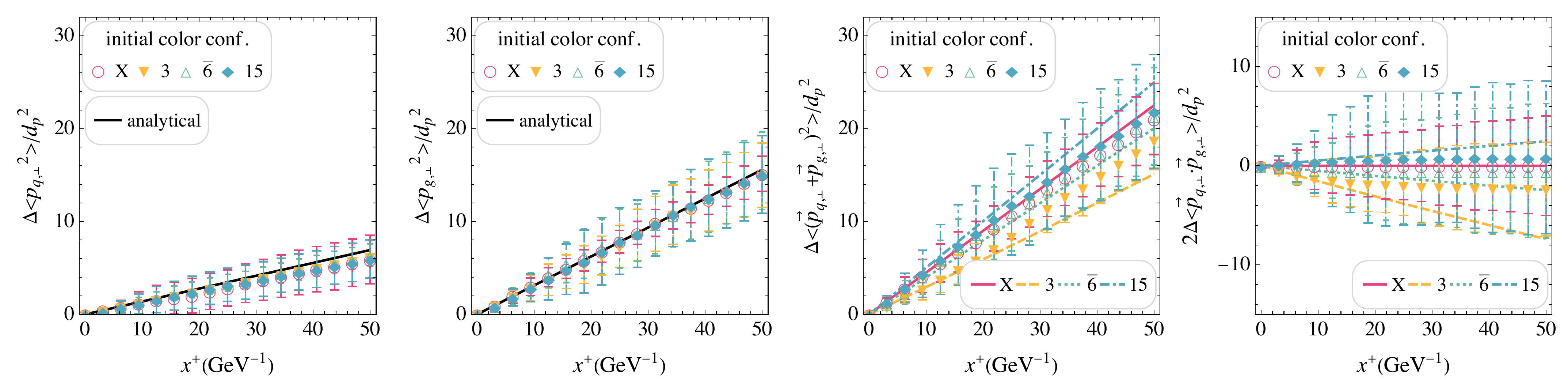}
  \caption{ 
  \label{fig:qg_short_ppl}  
  The transverse momentum evolution of a small-separation quark-gluon state [distribution as shown in Fig.~\ref{paper_qg_fR2D_short_ti}] at $p^+=1.5 \GeV$ ($p_q^+=0.5 \GeV$ and $p_g^+=1.0 \GeV$) with $m_q=0.15 \GeV$. 
  See simulation parameters and explanation of legends and data points in the caption of Fig.~\ref{fig:qg_short}.
  The lines are analytical expectations at $p^+=\infty$ evaluated according to Eqs.~\eqref{eq:p_qg} and \eqref{eq:pperp_qg_red}.
  }
\end{figure*}

\begin{figure*}[tbp!]
  \centering
  \subfigure[ \label{paper_qg_fR2D_short_tf}]{   
    \includegraphics[width=0.8\textwidth]{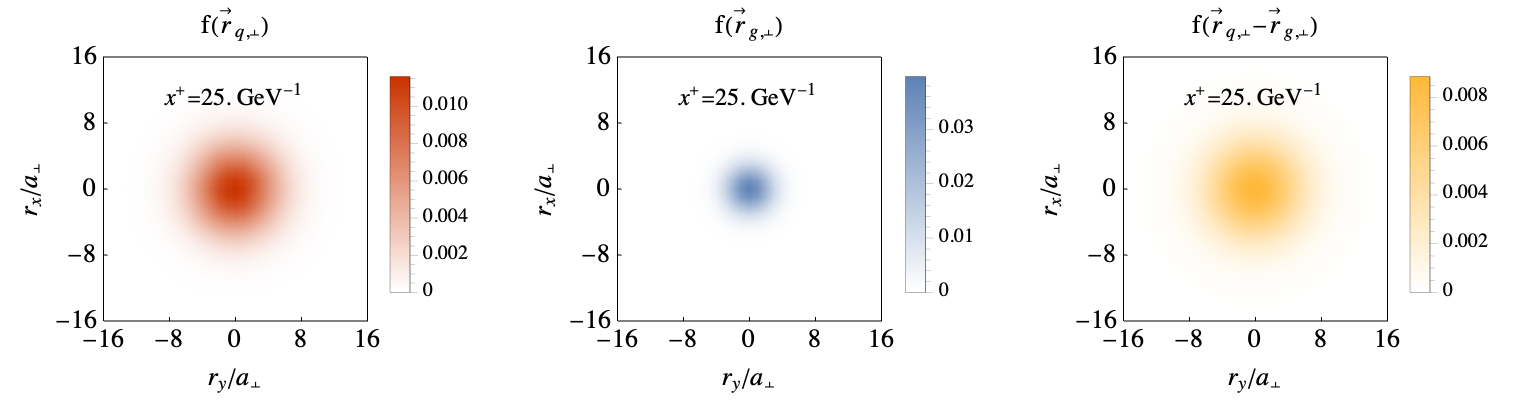}
  }
  \subfigure[\label{paper_qg_fR2D_long_tf}]{   
    \includegraphics[width=0.8\textwidth]{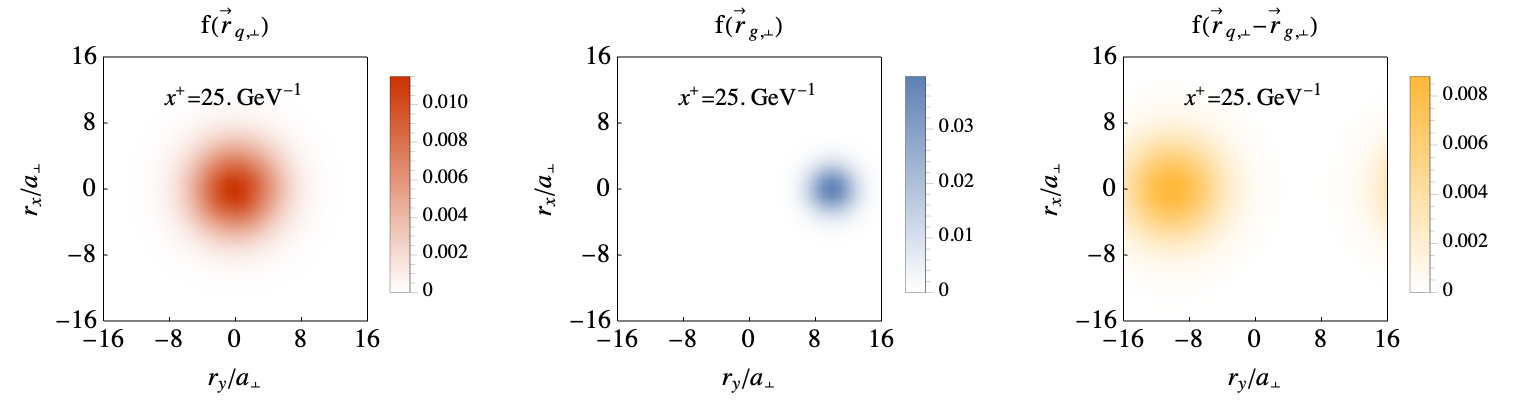}
    }
  \caption{ 
  \label{fig:qg_shortlong_tf}  
  Distributions of the quark's, the gluon's, and their relative transverse coordinate at a later time with $p^+=1.5~\GeV$. The initial state distribution of (a) is shown in Fig.~\ref{paper_qg_fR2D_short_ti}, and that of (b) in Fig.~\ref{paper_qg_fR2D_long_ti}.
     }
\end{figure*}

Things are more complicated at a finite $p^+$. First, we know from the previous section that the single particle momentum broadening would be different from the eikonal case given a finite $N_\eta$. Second, the transverse coordinate distribution would no longer stay the same as it is initially. The quark-gluon state would spread out, and so does its relative distribution $f_{Rel}(\vec v_\perp)$. This means that the spatial correlation between the quark and the gluon would decrease over time. 
The transverse coordinate distribution of the evolved state can be found in Fig.~\ref{paper_qg_fR2D_short_tf}.
As a result, the cross term, $\braket{\vec p_{q,\perp}(x^+)\cdot \vec p_{g,\perp} (x^+)}$ would get smaller than the eikonal expectation. Figure~\ref{fig:qg_short_ppl} shows the results, which agree with this expectation. The differences of the total momenta among different color configurations decrease over time, as compared to the eikonal case. 

\begin{figure*}[t]
  \centering
    \includegraphics[width=0.35\textwidth]{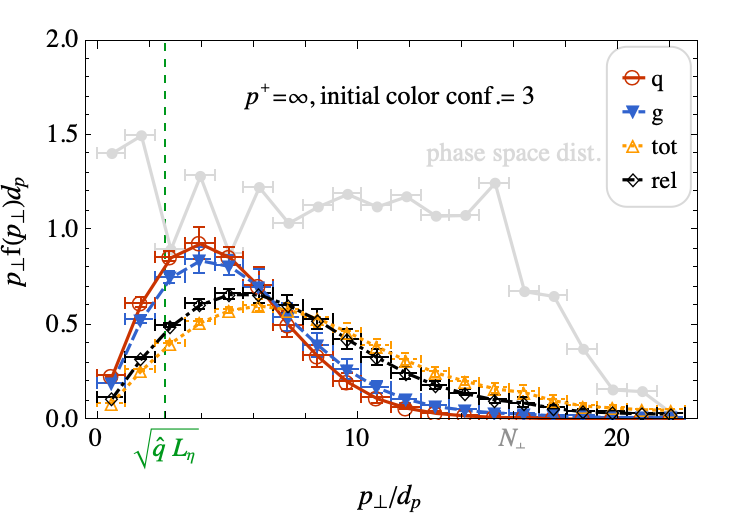}
    \includegraphics[width=0.35\textwidth]{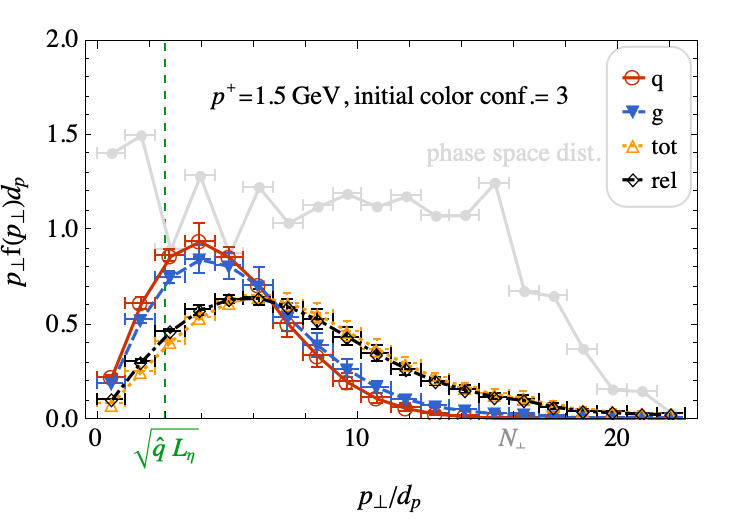}
  \caption{ 
  \label{fig:pfp_qg_3_small}  
  The transverse momentum distribution $p_\perp f(p_\perp) d_p$ of the small-separation color-triplet quark-gluon state at infinite and finite $p^+$, after the evolution through the medium, at $x^+=50~\GeV^{-1}$. 
  The legends ``q", ``g", ``tot", and ``rel", indicate the distribution function for the quark, the gluon, their total, and their relative transverse momentum, respectively.
  The light gray line indicates the relative size of the basis state phase space to the continuous case, according to $R(p_\perp)$ in Eq.~\eqref{eq:TMD_R}.
     }
\end{figure*}

Figure~\ref{fig:pfp_qg_3_small} presents the transverse momentum distribution $p_\perp f(p_\perp) d_p$ of the quark-gluon state in both the infinite and finite $p^+$ cases. 
The total~(relative) transverse momentum is defined as $\vec p_{q,\perp}\pm \vec p_{g,\perp} $. 
There is a difference between the total and the relative $p_\perp$ distributions, which, while small, implies a nonzero quark-gluon correlation. 

\subsubsection{Large-separation quark-gluon state}
We then move to a scenario where the two particles are not much correlated in space.
We set the initial state as given in Eqs.~\eqref{eq:qgWF} and \eqref{eq:qgWF_condition} with $\vec s_{q,\perp}/a_\perp=\{0,0\}$ and $\vec s_{q,\perp}/a_\perp=\{0,10\}$, $w_q= w_g=a_\perp $, i.e., case (b) in Fig. \ref{fig:qg_shortlong}.  We again study the correlation by examining the evolution of $\braket{ p_\perp^2}$ for the two-particle state. 

\begin{figure*}[tbp!]
  \centering
      \includegraphics[width=0.88\textwidth]{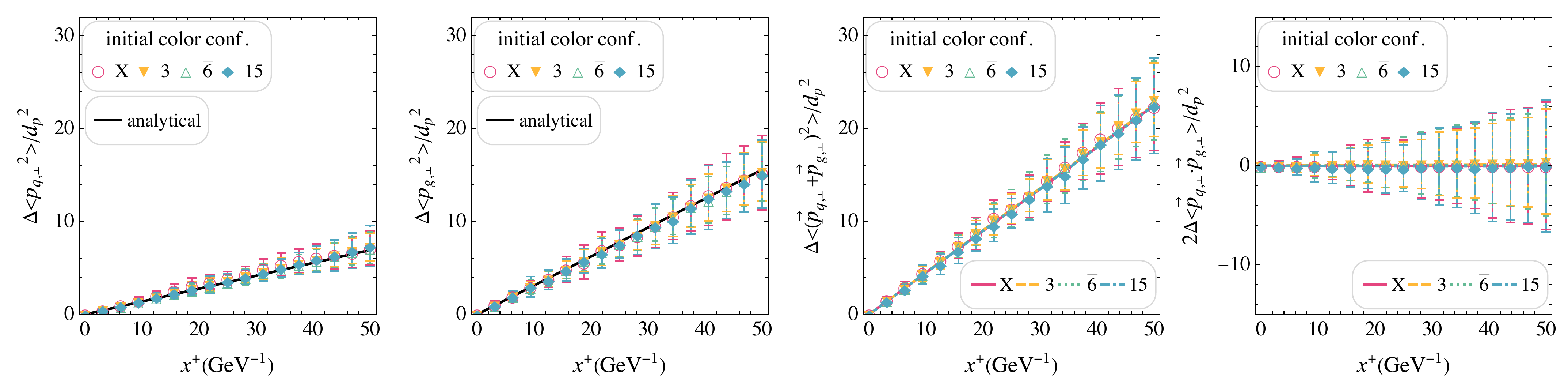}
  \caption{ 
  \label{fig:qg_long}  
  The transverse momentum evolution of a large-separated quark-gluon state [distribution as shown in Fig.~\ref{paper_qg_fR2D_long_ti}] in the eikonal limit of $p^+=\infty$. 
  See simulation parameters and explanation of legends and data points in the caption of Fig.~\ref{fig:qg_short}.
  The lines are analytical expectations at $p^+=\infty$ evaluated according to Eqs.~\eqref{eq:p_qg} and \eqref{eq:pperp_qg_red}.
  }
\end{figure*}

\begin{figure*}[tbp!]
  \centering
      \includegraphics[width=0.88\textwidth]{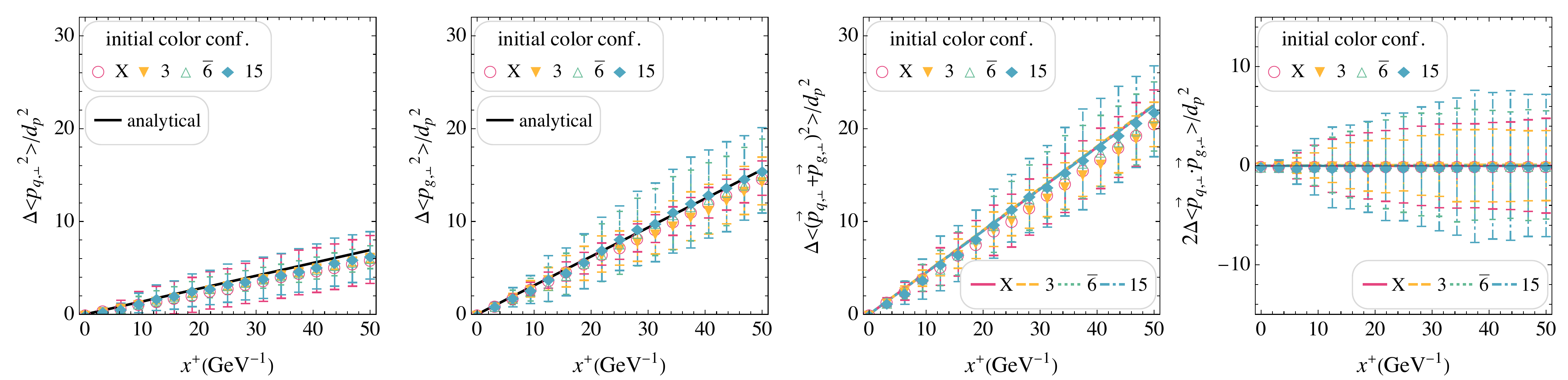}
  \caption{ 
  \label{fig:qg_long_ppl}  
  The transverse momentum evolution of a large-separated quark-gluon state [distribution as shown in Fig.~\ref{paper_qg_fR2D_long_ti}] at $p^+=1.5 \GeV$ ($p_q^+=0.5 \GeV$ and $p_g^+=1.0 \GeV$) with $m_q=0.15 \GeV$. 
  See simulation parameters and explanation of legends and data points in the caption of Fig.~\ref{fig:qg_short}.
  The lines are analytical expectations at $p^+=\infty$ evaluated according to Eqs.~\eqref{eq:p_qg} and \eqref{eq:pperp_qg_red}.
  }
\end{figure*}

In the eikonal limit of $p^+=\infty$, the value of $\braket{ p_\perp^2(x^+)}$ can be calculated exactly according to Eq.~\eqref{eq:pperp_qg_red}. The results from numerical simulation agree with such expectations, as shown in Fig.~\ref{fig:qg_long}. In contrast to the small-separation quark-gluon state, which is shown in Fig.~\ref{fig:qg_short}, the difference in the total momentum for different color configurations is negligible. This is because the quark and the gluon are too far away from each other, such that they are hardly correlated.
Then, as shown in Fig.~\ref{fig:qg_long_ppl}, at a finite $p^+$ the quark-gluon correlation is still negligible.

\begin{figure*}[t]
  \centering
    \includegraphics[width=0.35\textwidth]{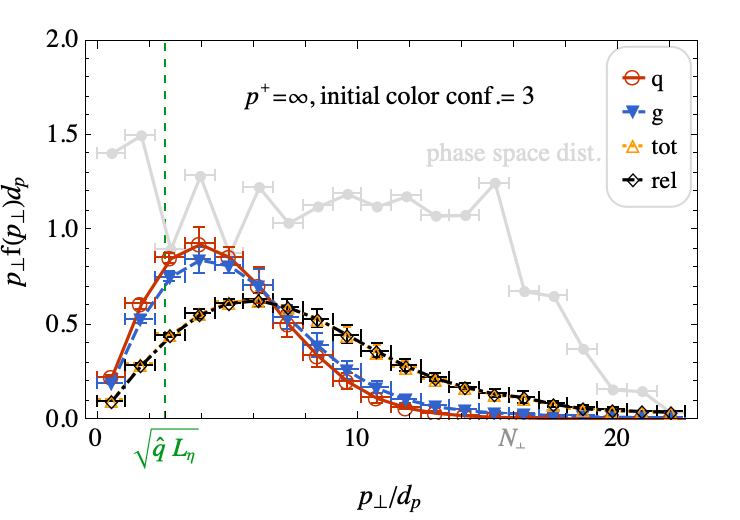}
    \includegraphics[width=0.35\textwidth]{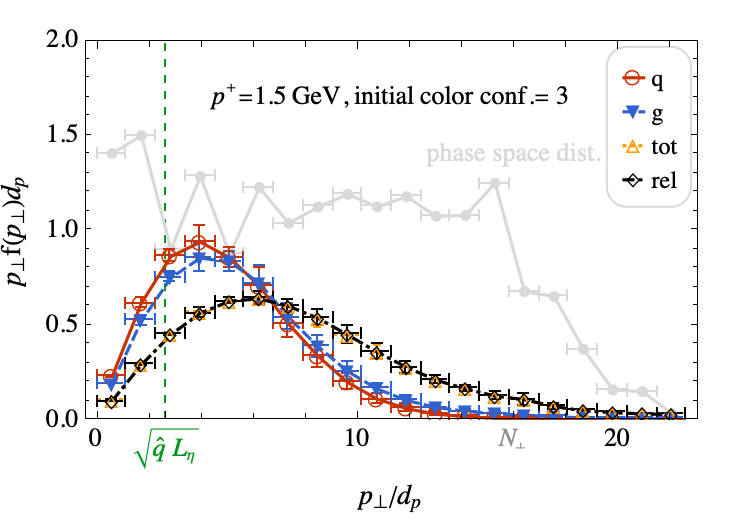}
  \caption{ 
  \label{fig:pfp_qg_3_long}  
  The transverse momentum distribution $p_\perp f(p_\perp) d_p$ of the long-separated color-triplet quark-gluon state at infinite and finite $p^+$, after the evolution through the medium, at $x^+=50~\GeV^{-1}$. 
  See simulation parameters and explanation of legends in the caption of Fig.~\ref{fig:pfp_qg_3_small}.
     }
\end{figure*}

Figure~\ref{fig:pfp_qg_3_long} presents the transverse momentum distribution $p_\perp f(p_\perp) d_p$ of the quark-gluon state in both the infinite and finite $p^+$ cases. From here, we can see that there is no sizable difference between the total and the relative momenta, implying a vanishing quark-gluon correlation.

\subsection{The dressed quark state $\ket{q}+\ket{qg}$}
We have studied the respective behaviors of a single quark state and a quark-gluon state in the preceding sections. In both cases, we find our numerics agree with the analytical expectation in the eikonal limit with a sufficiently large number of layers $N_\eta$.
We now proceed to a more realistic and interesting scenario by allowing the quark to emit and absorb a gluon throughout the quenching process. We initialize the simulation with a single quark state of $\vec p_\perp=\vec 0_\perp$ in the $\ket{q}+\ket{qg}$ space.

Specifically, by comparing the momentum broadening of the $\ket{q}$ sector and the total, we could study the contribution from gluon emission to $\hat{q}$. 
For a dressed quark state, the momenta of each sector can be evaluated as the following
\begin{align}\label{eq:psq_q_qg}
  \begin{split}
    &\braket{p_\perp^2}_{\ket{q}}= \braket{q|p_\perp^2|q} /P_q \;,\\
    &\braket{p_\perp^2}_{\ket{qg}}= \sum_{k_g^+=1}^{K-1/2} \braket{qg;k_g^+|p_\perp^2|qg;k_g^+} /P_{qg;k_g^+} \;,
  \end{split}
\end{align}
in which $P_q$ is the probability of the state to be in the $\ket{q}$, and $P_{qg;k_g^+}$ to be in the $\ket{qg}$ sector, with the gluon longitudinal momentum fraction $z=k_g^+/K$.
The total momentum is, therefore
\begin{align}
  \braket{p_\perp^2}_{total}=P_q \braket{p_\perp^2}_{\ket{q}}
  +P_{\ket{qg}} \braket{p_\perp^2}_{\ket{qg}}\;,
\end{align}
with
\begin{align}
  P_{\ket{qg}}\equiv \sum_{k_g^+=1}^{K-1/2} P_{qg;k_g^+},\qquad
  P_q +P_{\ket{qg}} =1\;.
\end{align}

In order to understand the interplay of the medium scattering and the gluon emission, we run the simulation at various medium intensities (quantified by $Q_s a_\perp/\pi$) and jet energies (quantified by $p^+$).  

Recall that the longitudinal ($x^+$) structure of the medium, quantified by the boost invariance $\tau/p^+$, also has an effect on the jet momentum broadening, as we have discussed extensively in Sec.~\ref{sec:qhat_ppl}.
Here we explore the $\tau/p^+<1$ regime, or in other words, the large-$N_\eta$ regime
. 
This corresponds to the physics that the medium has infinite uncorrelated color charges along its longitudinal direction, allowing us to have analytically tractable limits to compare with.

\begin{figure*}[tbp!]
  \centering
  \subfigure[In vacuum\label{fig:g2mu0_ppl17}]{
    \includegraphics[width=0.8\textwidth]{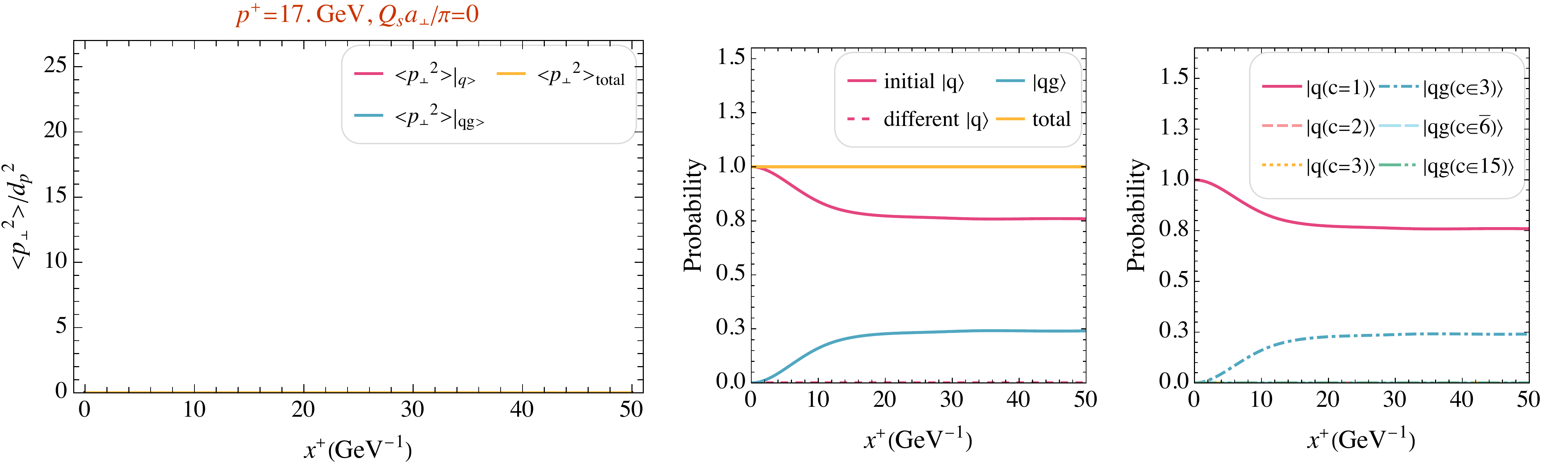}
  }  
  \subfigure[In a weak medium\label{fig:g2mu2_ppl17}]{
    \includegraphics[width=0.8\textwidth]{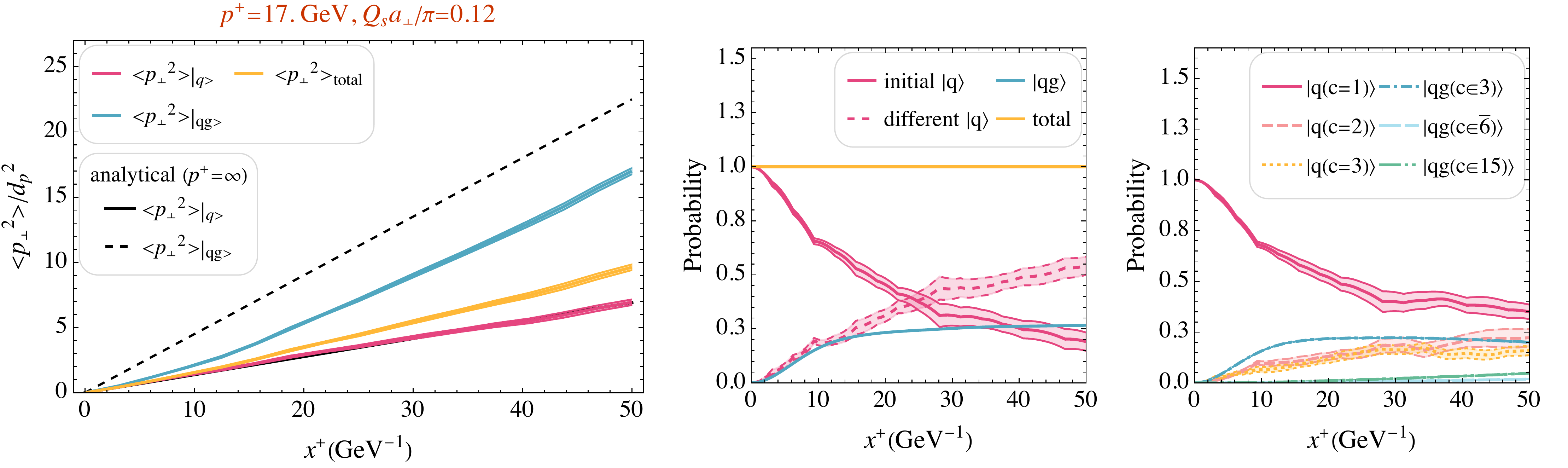}
  }
  \subfigure[In a strong medium\label{fig:g2mu5_ppl17}]{
    \includegraphics[width=0.8\textwidth]{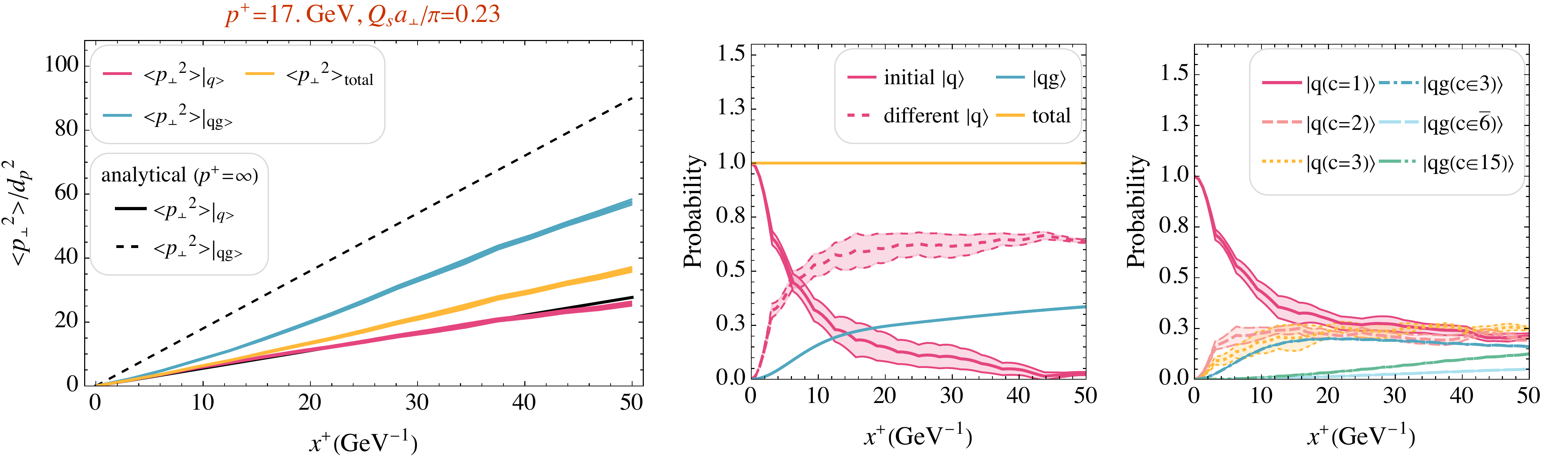}
  }
  \subfigure[In a weak medium, with a smaller $p^+$\label{fig:g2mu2_ppl8p5}]{
    \includegraphics[width=0.8\textwidth]{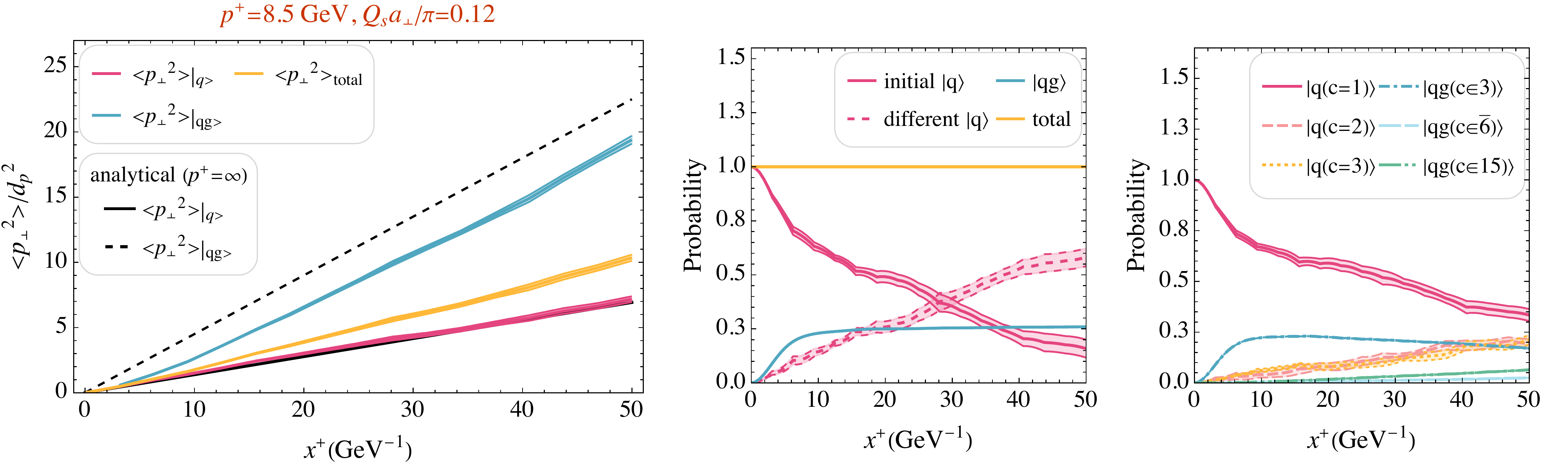}
  }
  \caption{ 
  \label{fig:qg_qdress}  
  The transverse momentum evolution of a quark jet in the $\ket{q}+\ket{qg}$ space in vacuum and at different media and $p^+$ (values indicated in each subfigure).
  The middle and the rightmost panels are the probability functions in the Fock and color phase space, respectively.
  The initial state is a single quark with $\vec p_\perp=\vec 0_\perp$ and $p^+_Q=p^+$. Simulation parameters: $N_\perp=16$, $L_\perp=50~\GeV^{-1}$, $\tau=3.125~\GeV^{-1}$ and $K=8.5$. The analytical results at $p^+=\infty$ in the solid and dashed black lines are calculated according to Eq.~\eqref{eq:psq_eik_res}.
  }
\end{figure*}
Figure~\ref{fig:qg_qdress} shows the evolution of the a quark jet in the $\ket{q}+\ket{qg}$ space in four different scenarios, Fig.~\ref{fig:g2mu0_ppl17} in vacuum with $p^+=17.0~\GeV$, Fig.~\ref{fig:g2mu2_ppl17} at $Q_s a_\perp/\pi=0.117$ and $p^+=17.0~\GeV$, Fig.~\ref{fig:g2mu5_ppl17} at $Q_s a_\perp/\pi=0.234$ and $p^+=17.0~\GeV$, and Fig.~\ref{fig:g2mu2_ppl8p5} at $Q_s a_\perp/\pi=0.117$ and $p^+=8.5~\GeV$. We take $\tau=3.125~\GeV^{-1}$, such that there are a number of $N_\eta=16$ layers of independent sources for the medium at $ x^+=[0, L_\eta=50~\GeV^{-1}]$.
The initial state is a single quark with $\vec p_\perp=\vec 0_\perp$ and $p^+_Q=p^+$. For all configurations we plot the total $p_\perp^2$ of the jet, and the probability to be in the original momentum mode in the $\ket{q}$ sector, another $\ket{q}$ mode or the $\ket{qg}$ sector, and the probabilities for the different color states.
We have the following key observations.

First, the inclusion of the $\ket{qg}$ sector substantially enhances the jet momentum broadening. In all the three cases with medium shown in Fig.~\ref{fig:qg_qdress}, the transverse momentum of the jet, $\braket{p_\perp^2}_{total}$, increases at a larger rate than that of the single quark, $\braket{p_\perp^2}_{\ket{q}}$. 
The former, by its definition, is a weighted sum of $\braket{p_\perp^2}_{\ket{q}}$ and $ \braket{p_\perp^2}_{\ket{qg}}$.
We see that $\braket{p_\perp^2}_{\ket{q}}$ agrees with the analytical result calculated in the eikonal case of $p^+=\infty$, Eq.~\eqref{eq:psq_eik_res}, just as one would expect in the $\tau/p^+<1$ regime. 
The momentum broadening of the $\ket{qg}$ component resembles that of the $\ket{q}$ initially and gradually migrates to a much larger rate. The initial quark-like broadening is due to the fact that the $\ket{qg}$ components appear by transiting from the $\ket{q}$ sector, a process that conserves momentum. Later, when the occupation of the $\ket{qg}$ components stabilizes, its momentum value reveals the effect of the direct medium interaction. Since the quark-gluon state has a larger color phase space, its momentum broadens at a larger rate than that of the single quark, due to the Casimir effect.
The dashed black line shows the momentum broadening of a single $\ket{qg}$ state, with a zero initial total transverse momentum, according to Eq.~\eqref{eq:p_qg}. Note that in this case, the quark and the gluon are spatially decorrelated; thus, the cross term is negligible, as we have seen in Sec.~\ref{sec:qg}.
Overall, one can think that the jet momentum broadening in the $\ket{q}+\ket{qg}$ Fock space is larger compared to that in $\ket{q}$ due to the larger phase space.

Second, the medium interaction interferes with the gluon emission process. The medium is absent, relatively weak and strong in Fig.~\ref{fig:g2mu0_ppl17}, Fig.~\ref{fig:g2mu2_ppl17} and Fig.~\ref{fig:g2mu5_ppl17}, respectively. 
In the vacuum case, the $\ket{qg}$ probability stabilizes at about $x^+=20~\GeV^{-1}$, and similarly in the case with a weaker medium.
 Consequently, in a weak medium,  $ \braket{p_\perp^2}_{\ket{qg}}$ increases almost at a constant rate, which is seen as the line almost in parallel with the dashed analytical line. 
Differently, in the stronger-medium case, the $\ket{qg}$ probability is still increasing till the end of the evolution. 
As a result, $ \braket{p_\perp^2}_{\ket{qg}}$ increases at a slower rate than the dashed analytical line because at each intermediate time instant, the newly generated $\ket{qg}$ component broadens as a quark. 
The rightmost panels show the evolution in color space. The color configuration is dominated by the triplet states initially in all cases. The medium interaction results in the color transitions, as can be seen by comparing Fig.~\ref{fig:g2mu2_ppl17} and Fig.~\ref{fig:g2mu5_ppl17} to Fig.~\ref{fig:g2mu0_ppl17}.  
The color transition is faster in a stronger medium by comparing Fig.~\ref{fig:g2mu5_ppl17} to Fig.~\ref{fig:g2mu2_ppl17}. 
We summarize this effect as the following: the medium interaction enhances gluon emission by promoting the $\ket{qg}$ occupation, which in return in fact slows down jet momentum broadening as compared to a pure $\ket{qg}$ state.

Third, jet longitudinal momentum drags down the process of gluon emission. From Fig.~\ref{fig:g2mu2_ppl17} to Fig.~\ref{fig:g2mu2_ppl8p5}, the jet $p^+$ is halved, and the $\ket{qg}$ probability stabilizes twice faster, at about $x^+=10~\GeV^{-1}$. 
Accordingly, $ \braket{p_\perp^2}_{\ket{qg}}$ increases at the rate of a pure quark-gluon state much earlier, which is seen as its line in Fig.~\ref{fig:g2mu2_ppl8p5} is closer to the dashed analytical line than that in Fig.~\ref{fig:g2mu2_ppl17}. Therefore, a smaller $p^+$ leads to a larger jet momentum broadening.

\begin{figure}[tbp!]
  \centering
    \includegraphics[width=0.4\textwidth]{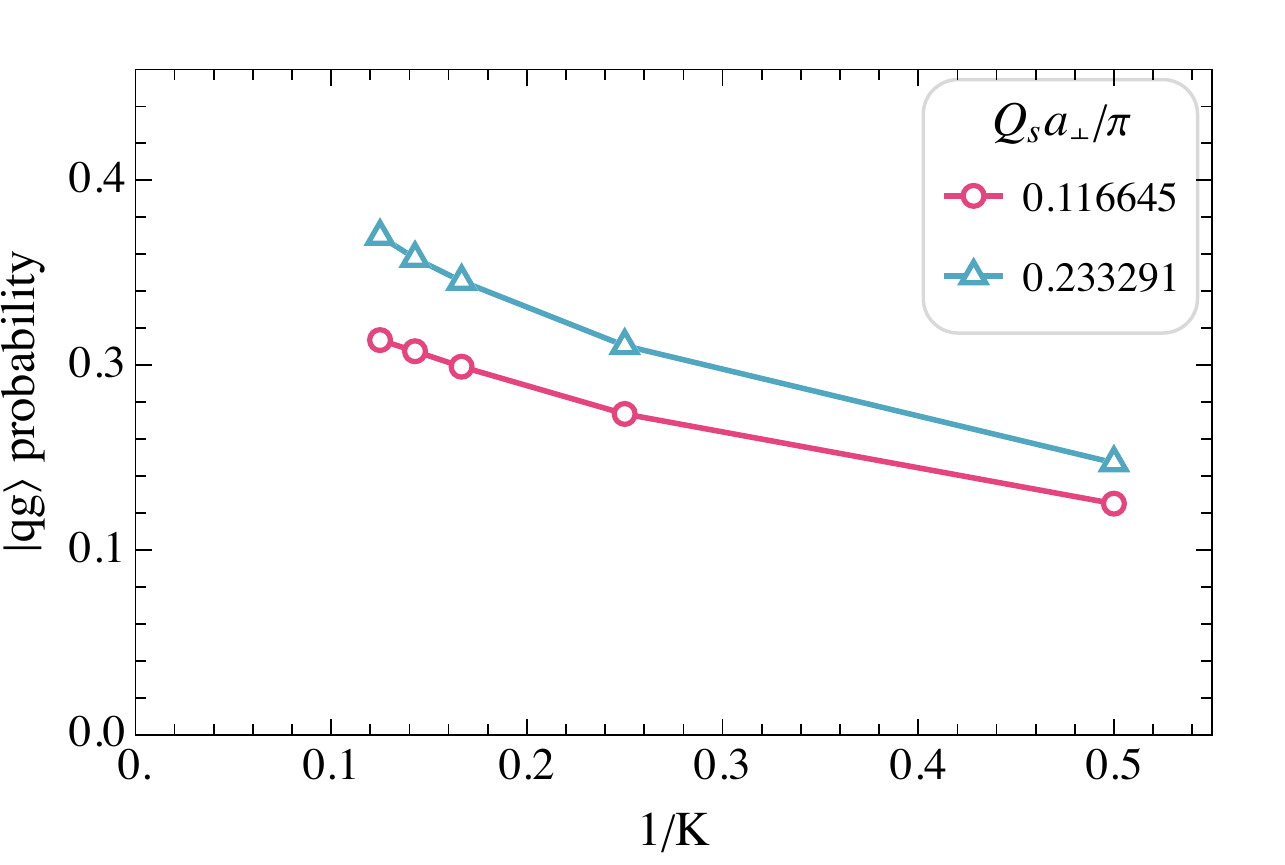}
  \caption{ 
  \label{fig:pqg_K}  
  The probabilities of the $\ket{qg}$ sector after the evolution at various $K$ and two different saturation scales, at $x^+=50~\GeV^{-1}$.
  }
\end{figure}

To further examine the effect of gluon emission in the medium, we plot the probabilities of the $\ket{qg}$ sector after evolution in the relatively weak and strong media in Fig.~\ref{fig:pqg_K}. Here, we also show the dependence on $1/K$, with $K=2.5,4.5,6.5,7.5,8.5$. The leftmost data points are the same as those in Fig.~\ref{fig:g2mu2_ppl17} and Fig.~\ref{fig:g2mu5_ppl17}. 
As we have observed in Fig.~\ref{fig:qg_qdress}, now at each $K$, the $\ket{qg}$ probability is higher in the stronger medium.
On the other hand, as $K$ increases, the $\ket{qg}$ probability also increases. This can be understood in the sense that as $K$ increases, the resolution on longitudinal momentum fractions increases, and the available phase space for the $\ket{qg}$ sector gets larger. 
In addition, the smallest value of $z_g=1/K$ also gets smaller as well, and the $\ket{q}\to\ket{qg}$ process is largest around $z_g=0$. 
\footnote{At each small time step, it is true that the $\ket{q}\to\ket{qg}$ process also favors the large $z_g (\approx 1)$ modes, e.g., $V_{qg}\propto 1/z_g^{3/2}/(1-z_g)$ in the spin-non-flip case.  But over a longer time span, the transition to those large $z_g$ modes gets suppressed by the large fluctuation of the energy $\Delta P^-_{KE}\propto 1/z_g/(1-z_g)$ whereas the small $z_g$ modes survive. We refer to our preceding work~\cite{Li:2021zaw} and its Fig. 17 for illustration. }
However, the total momentum square does not increase substantially when $K$ increases from 2.5 to 8.5. The reason is that the different $p^+$ segments of the $\ket{qg}$ sector have approximately the same broadening effect caused by the medium, a process independent of $p^+$. Thus $\braket{p^2_\perp}_{qg}$ is not sensitive to $K$. 
The contribution of $K$ into $\braket{p^2_\perp}$ mainly comes from the probability $P_{\ket{qg}}$ through Eq.~\eqref{eq:psq_q_qg}.
Our calculation is different from the study of keeping track of only one daughter particle (in our setup, this means counting the quark's instead of the total momentum of the $\ket{qg}$ state), in which the recoil effects from other daughter particles can be large compared to the eikonal term~\cite{Liou:2013qya, Blaizot:2013vha}.
Though in each time step, the splitting $\ket{q}\to\ket{qg}$ favors the quark-gluon state with large relative transverse momentum (more discussions can be found in our previous work Ref.~\cite{Li:2021zaw}), the total transverse momentum is conserved, and the recoil effect is not counted into $\braket{p^2_\perp}_{qg}$.

\begin{figure*}[tbp!]
  \centering
\includegraphics[width=.8\textwidth]{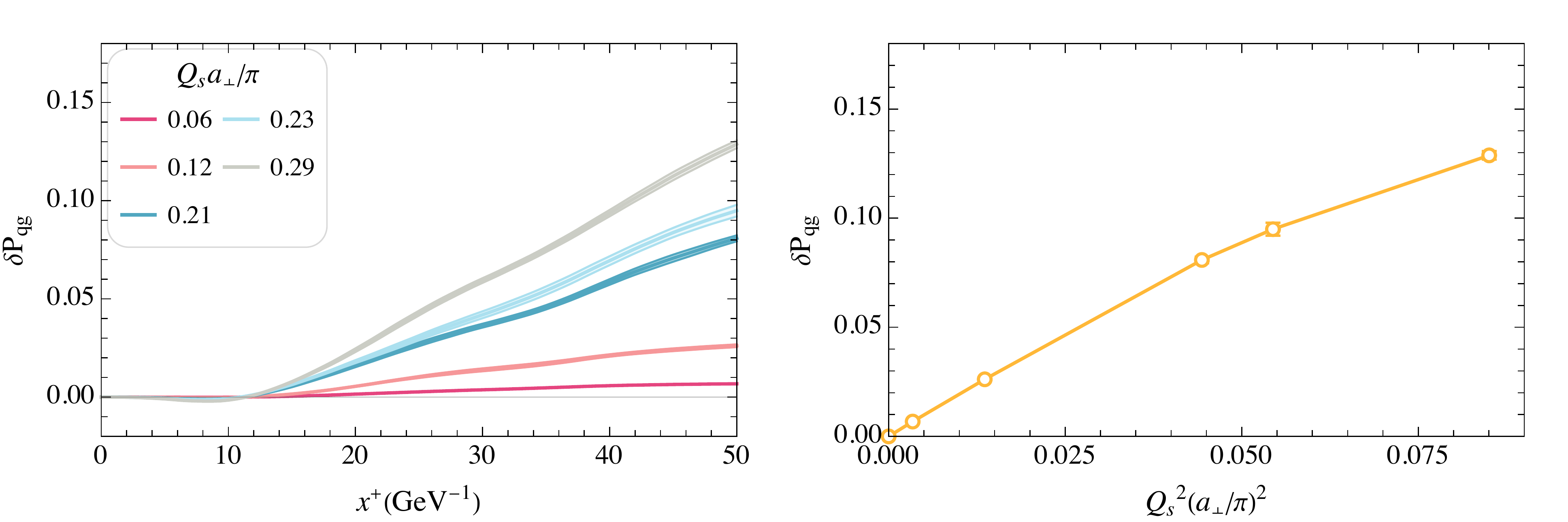}
  \caption{
  The difference of the probability of the quark jet in the $\ket{qg}$ sector in the medium and that in the vacuum, as defined in Eq.~\eqref{eq:delta_Pqg}, as a function of evolution time $x^+$ and saturation scale $Q_s$. 
 }
  \label{fig:delta_Pqg}
\end{figure*}

To quantify the medium-induced gluon emission, we define $\delta P_{\ket{qg}}$ as the difference of the probability of the quark jet in the $\ket{qg}$ sector in the medium and that in the vacuum,
\begin{align}\label{eq:delta_Pqg}
  \delta P_{\ket{qg}}(Q_s, x^+)\equiv P_{\ket{qg}}(Q_s, x^+)-P_{\ket{qg}}(Q_s=0, x^+)\;.
\end{align}
We present one set of results in Fig.~\ref{fig:delta_Pqg}, from simulations with the same parameters taken in Fig.~\ref{fig:qg_qdress}, at $p^+=17~\GeV$. 
On the left panel, we see that the $\delta P_{\ket{qg}}$s at various saturation scale $Q_s$ have a similar behavior: each curve forms a very small dip in the early time region, then after around the point $x^+=12~\GeV^{-1}$, grows linearly in time.
Comparing the curves at different $Q_s$, one can see that the denser the medium, the more the $\ket{qg}$ component develops.
This is also shown on the right panel, where the $\delta P_{\ket{qg}}$ of the final state ($x^+=L_\eta=50~\GeV^{-1}$) is approximately proportional to $Q_s^2$.
A similar observation has been made in Ref.~\cite{Zhang:2021tcc}, using the high-twist approach and agreeing with the Gyulassy-Levai-Vitev result in the first-order opacity expansion. There, the induced gluon radiation spectrum, counterpart to $\delta P_{\ket{qg}}$ here, is proportional to the transverse-momentum gluon distribution density. Then the integration of such distribution's $\braket{\vec p_\perp^2}$ over the evolution time is the saturation scale $Q_s^2$.

\begin{figure*}[tbp!]
  \centering
\includegraphics[width=.8\textwidth]{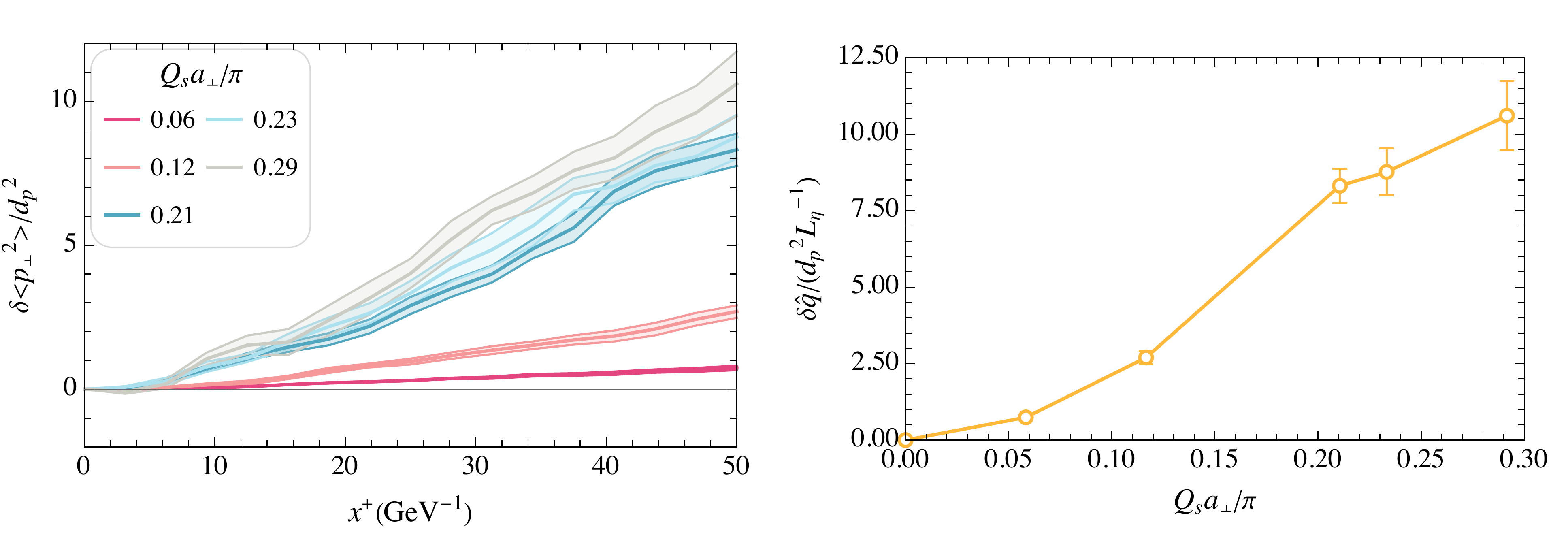}
  \caption{
  The non-eikonal correction to $\hat q$ as defined in Eq.~\eqref{eq:delta_qhat}, as a function of evolution time $x^+$ and saturation scale $Q_s$. 
 }
  \label{fig:delta_qhat}
\end{figure*}

It is also interesting to analyze the non-eikonal and radiative correction to the quenching parameter $\hat{q}$. We define $\delta \hat q$ as the difference of the $\hat q$ that is calculated from the total momentum of the quark jet in the $\ket{q}+\ket{qg}$ space, and the eikonal $\hat q$ of a bare quark (as in Eq.~\eqref{eq:qhat_Eik_res}),
\begin{align}\label{eq:delta_qhat}
  \delta \hat q \equiv \hat q-\hat q_{Eik}\;.
\end{align}
We present the results in Fig.~\ref{fig:delta_qhat}, from the same set of simulations presented in Fig.~\ref{fig:delta_Pqg}. 
The left panel shows that $\delta \braket{p_\perp^2}(\equiv \braket{p_\perp^2}-\braket{p_\perp^2}_{Eik})$ increases over the evolution time at various $Q_s$. In the right panel, the $\delta \hat q$ extracted from the final state $\delta \braket{p_\perp^2}$ is plotted, and it increases non-trivially when $Q_s$ increases.

In summary, the momentum broadening of the quark jet in the $\ket{q}+\ket{qg}$ Fock space is larger compared to that of the bare quark state $\ket{q}$, due to gluon emission. 
The difference between the two leads to a correction to $\hat q$.

\section{Conclusions}\label{sec:conclusions}
We present in this paper an extensive study on the momentum broadening of in-medium jet evolution using the tBLFQ approach.
We perform the numerical simulation of the real-time jet evolution in the Fock space of $\ket{q}$, $\ket{qg}$, and $\ket{q}+\ket{qg}$, at various jet energies $p^+$ and medium densities.
  We obtain jet light-front wavefunction and we extract the jet transverse momentum distribution, the quenching parameter, and the gluon emission rate. 
  We analyze the dependence of momentum broadening on $p^+$, medium density, color configuration, spatial correlation, and medium-induced gluon emission. 
  For comparison, we also derive analytically in the eikonal limit the expectation value of the transverse momentum of a quark-gluon state for any color configuration and in an arbitrary spatial distribution.

  This work provides an enhanced understanding of jet quenching beyond the eikonal limit. 
 We have set up the initial quark projectile as a bare quark on its mass shell for the simplicity of the study.
 This setup is close to the scenario of a quark produced inside the medium from a hard scattering. This means that even in the vacuum, it will radiate gluons. In the language of our formalism, the bare quark is not an eigenstate of the full interacting Hamiltonian $P^-_{KE}+V_{qg}$. It only has a partial overlap with a dressed quark state, with the difference corresponding to vacuum radiation. This is possible because $p^-$ need not be conserved at an interaction vertex, a convention known in the ``old-fashioned perturbation theory''~\cite{Brodsky:1997de,Schwartz:2014sze}. 
 In a covariant formalism our initial quark should be thought of as an off-shell one, as required in order to have final state radiation and opposed to a quark on its mass shell that cannot emit gluons in the vacuum~\cite{Casalderrey-Solana:2007knd}. In this case, the four-momentum, including $p^-$, must be conserved at any interaction vertex, and the tradeoff is the initial quark being off-shell.

 A different and physically equally interesting scenario is the quark coming from outside the medium, described by the fully developed wave function that contains a gluon cloud~\cite{Kovner:2003zj}.
 We leave this task for our next work, where we will treat the initial quark as the eigenstate of the light-front QCD Hamiltonian in the $\ket{q}+\ket{qg}$ space.

 \section*{Acknowledgments}
We are very grateful to Guillaume Beuf, Fabio Dominguez, Miguel A. Escobedo, Sigtryggur Hauksson, Cyrille Marquet, Wenyang Qian, Andrecia Ramnath, Andrey Sadofyev, Xin-Nian Wang, and Bin Wu for helpful and valuable discussions.

XB is supported by new faculty startup funding by the Institute of Modern Physics, Chinese Academy of Sciences, by Key Research Program of Frontier Sciences, Chinese Academy of Sciences, Grant No. ZDBS-LY-7020, by the Natural Science Foundation of Gansu Province, China, Grant No. 20JR10RA067, by the Foundation for Key Talents of Gansu Province, by the Central Funds Guiding the Local Science and Technology Development of Gansu Province, Grant No. 22ZY1QA006 and by the Strategic Priority Research Program of the Chinese Academy of Sciences, Grant No. XDB34000000.
ML and CS are supported by Xunta de Galicia (Centro singular de Investigacion de Galicia accreditation 2019-2022), European Union ERDF, the “Maria de Maeztu” Units of Excellence program under project CEX2020-001035-M, the Spanish Research State Agency under project PID2020-119632GB-I00, and European Research Council under project ERC-2018-ADG-835105 YoctoLHC.
TL is supported by the Academy of Finland, the Centre of Excellence in Quark Matter (project 346324)
and project 321840.
This work was also supported under the European Union’s
Horizon 2020 research and innovation  by the STRONG-2020 project (grant agreement No. 824093). The content of this article does not reflect the official opinion of the European Union and responsibility for the information and views expressed therein lies entirely with the authors.

\appendix

\section{Discretization}

The conventions and notations in this paper follow Ref.~\cite{Li:2021zaw}.
We adopt the following shorthand notation for transverse integrals in position and momentum space, respectively,
\begin{align}
\int_\r \equiv \int \diff^2 r_\perp \, , \qquad     \int_\p \equiv \int \frac{\diff^2 p_\perp}{(2\pi)^2} \; , 
\end{align}
in which $r_\perp=|\vec r_\perp|$ and $p_\perp=|\vec p_\perp|$. 
On the discrete basis space, the integrations become summations over the corresponding transverse quantum numbers ($\vec r_\perp=\{n_1, n_2\}a_\perp$ and $\vec p_\perp=\{k_1, k_2\}d_p$), 
\begin{align}
\sum_\r \equiv a_\perp^2 \sum_{n_1, n_2=-N_\perp}^{N_\perp-1}
\, , \qquad   
\sum_\p \equiv \frac{1}{(2L_\perp)^2}\sum_{k_1, k_2=-N_\perp}^{N_\perp-1}  \;.
\end{align}

\subsection{The transverse distribution function}\label{app:TMD}
In studying the angularly-integrated transverse distribution function $f(p_\perp)$ (recall that $p_\perp=|\vec p_\perp|=|\p|$) and similarly its Fourier transform $\tilde f(r_\perp)$ ($r_\perp=|\vec r_\perp|=|\r|$), one needs to be aware that the phase space of states on the discrete square lattice is different from that in the continuous case. 
As $p_\perp$ increases, the number of momentum modes within the range $p_\perp\sim p_\perp+\delta_p$ (with $\delta_p$ a small positive value) first increases up till the edge of the lattice $\Lambda_{UV}$ and then drop down, whereas the continuous phase space would keep increasing as $2\pi p_\perp \delta_p $.

The 2D transverse distribution function $f(\vec p_\perp)$ can be obtained directly by squaring the wavefunction $|\psi(\vec p_\perp)|^2$. 
The normalization is therefore $\sum_\p f(\vec p_\perp) =1$ and $\int_\p f(\vec p_\perp) =1$ in the discrete and continuous case, respectively. 
By integrating/summing over the azimuthal angle, $f(\vec p_\perp)$ becomes the 1D distribution function $f(p_\perp)$, the average value of $f(\vec p_\perp)$ over momenta with the same $p_\perp$.
In the continuous space, the normalization reads $\int_\p f(\vec p_\perp) =\int p_\perp \diff p_\perp f(p_\perp)/(2\pi)=1$. 
To obtain $f(p_\perp)$ from a distribution on the square lattice, we first group $f(\vec p_\perp)$ into adjacent bins with equal span on $p_\perp$, then take the average value of the data in each bin as $f(p_\perp)$. 

To be specific, choosing $N_b$ as the total number of bins in the radial direction, then the dimensionless width of each bin is $\omega_b=\sqrt{2}N_\perp /N_b$. The dimensional bin width is $\omega_b d_p$ and $\omega_b a_\perp$ for the momentum and coordinate space, respectively. The $i_b$-th bin would contain all the states with $p_\perp/d_p ( r_\perp/a_\perp)\in [i_b-1, i_b]\omega_b$; note that states on the bin boundary should be counted only into one bin. Next, count the number of states in each bin, $h(i_b)$, and they should sum up as
\begin{align}
  \sum_{i_b=1}^{N_b} h(i_b)=(2 N_\perp)^2  \;.
\end{align}
The normalization in terms of the summation reads
\begin{align}
\sum_{\p} f(\vec p_\perp)=\frac{1}{(2 L_\perp)^2}\sum_{i_b=1}^{N_b} h(i_b) f(p_\perp)=1
  \;.
\end{align}
Here, $p_\perp(i_b)=(i-1/2)\omega_b d_p$ inside the summation over bins, the center point of the $i_b$-th bin.
Similarly in the conjugate $\vec r_\perp$ space, with $h(i_b)$ the number of states in the $i_b$-th bin along $r_\perp$,  
\begin{align}
  \sum_{\r} \tilde f(\vec r_\perp)=a_\perp^2\sum_{i_b=1}^{N_b} h(i_b) \tilde f(r_\perp)=1
    \;,
\end{align}
where $r_\perp(i_b)=(i-1/2)\omega_b a_\perp$.

\begin{figure}[tbp!]
\centering 
\includegraphics[width=0.4\textwidth]{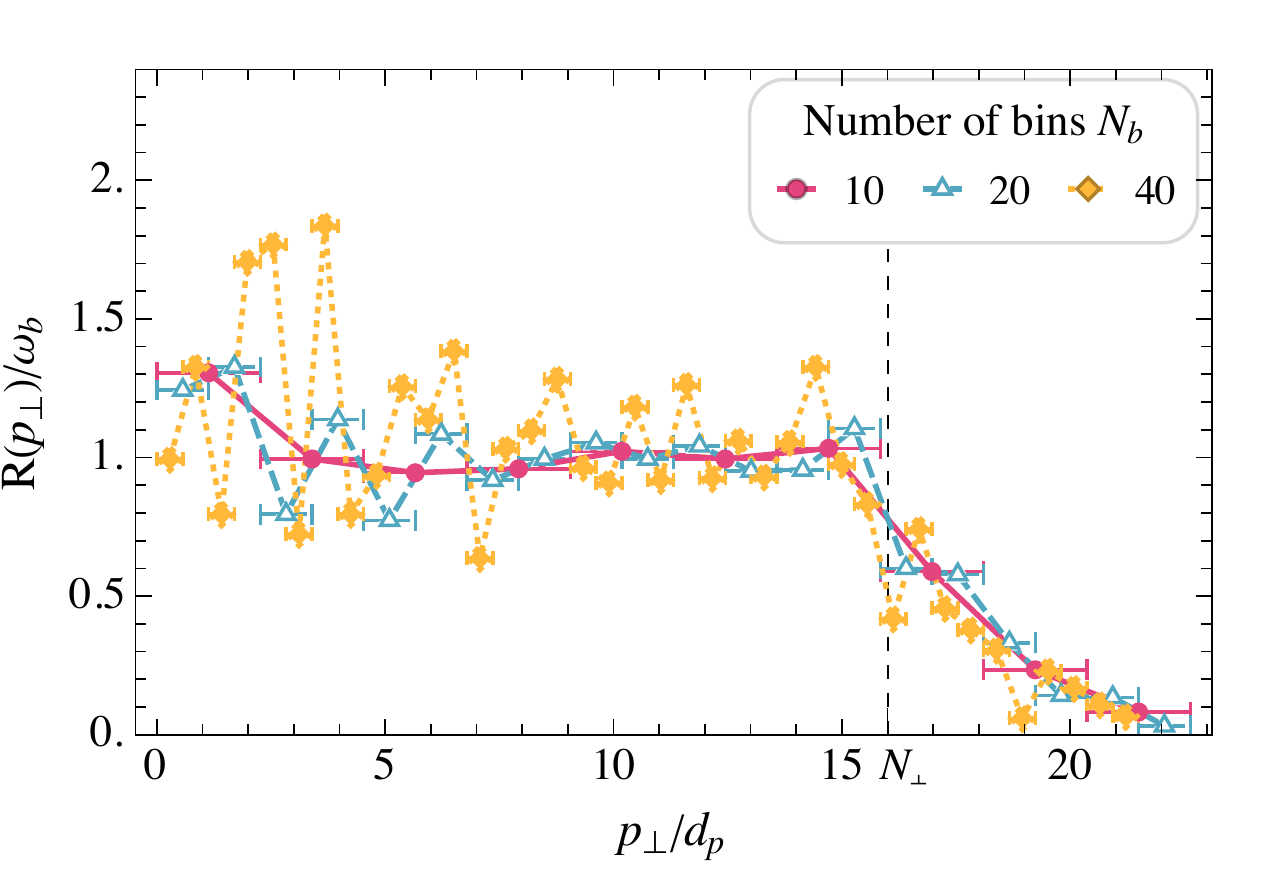}
\caption{ 
\label{fig:TMD_R}  
The ratio function $R(p_\perp)$ [as in Eq.~\eqref{eq:TMD_R}] divided by bin width $\omega_b$, at $N_\perp=16$ and at various $N_b$. 
 }
\end{figure}

With $h(i_b)$, we can quantify the ratio between the discrete and the continuous space size at each bin, that is, $h(i_b)d_p^2/(2\pi p_\perp d_p)$ [$h(i_b)a_\perp^2/(2\pi r_\perp a_\perp)$ ] in the momentum (coordinate) space. With a simplification, it reads,
\begin{align}\label{eq:TMD_R}
  R(i_b)\equiv \frac{h(i_b) }{2\pi (i-1/2)\omega_b }
  \;.
\end{align}
This ratio function depends on the lattice size $N_\perp$ and the number of bins $N_b$. Since $i_b$ corresponds to a specific range of $p_\perp$ or $r_\perp$, one can also write $R$ in terms of $p_\perp$ or $r_\perp$.
Figure \ref{fig:TMD_R} exemplifies the behavior of $R(p_\perp)/\omega_b $ at $N_\perp=16$ and at various $N_b$. The horizontal error bar indicates the range covered by each bin. The ratio function is overall flat when the momentum mode is below $N_\perp$, and decreases when going above. This is expected from the above discussion on $h(i_b)$. In addition, as $N_b$ increases, the ratio function admits more ``zig-zag" patterns. 
With $R$, the normalization of the probability function is written as
\begin{align}
  \begin{split}
  &\frac{1}{2\pi}\sum_{p_\perp} R(p_\perp) p_\perp  f(p_\perp) d_p=1\;,\\
 &2\pi\sum_{r_\perp} R(r_\perp) r_\perp  \tilde f(r_\perp) a_\perp =1
  \;.
  \end{split}
\end{align}

\subsection{Numerical simulation of the medium}\label{app:medium}
In the discrete basis space, the correlation relation of the color charge in Eq.~\eqref{eq:chgcor} takes the form,
\begin{multline}\label{eq:chgcor_dis}
  \expconfig{\rho_a(n^x,n^y,n_\tau)\rho_b({n'}^x,{n'}^y,n_\tau')}\\
 =g^2\tilde{\mu}^2\delta_{ab}\frac{\delta_{n^x,{n'}^x}\delta_{n^y,{n'}^y}}{a_\perp^2}\frac{\delta_{n_\tau,n_\tau'}}{\tau}\;,
\end{multline}
as we have implemented in Refs.~\cite{Li:2020uhl, Li:2021zaw}.
The sources generating the medium are stochastic random variables with a Gaussian distribution on each site, with the transverse indices $n^x,n^y=-N_\perp,-N_\perp+1,\ldots, N_\perp-1$, and the layer indices $n_\tau=1,2,\ldots, N_\eta$.

 Then $\mathcal A$, the field of the medium is solved from Eq.~\eqref{eq:poisson}, a Poisson's equation in the two dimensions for each $x^+$ layer. Numerically, it is efficient to solve in the momentum space, accompanied by a Fourier transform,
\begin{align}\label{eq:CGCA_Green_full_discrete}
  \begin{split}
\mathcal{A}^-_a (n^x,n^y, x^+)
=&
\frac{1}{{(2N_\perp)}^2}
\sum_{\bar n_x, \bar n_y=-N_\perp}^{N_\perp-1} 
\rho_a(\bar n^x,\bar n^y, n_\tau(x^+))\\
&
\sum_{ k_x, k_y=-N_\perp}^{N_\perp-1} 
\frac{ e^{-i[(n_x-\bar n_x) k_x + (n_y-\bar n_y)  k_y]\pi/N_\perp}  }{m_g^2 a_\perp^2/\pi^2/N_\perp^2+k_x^2+k_y^2} 
\;.
\end{split}
\end{align}
We write $n_\tau(x^+)$ to indicate that the larger indices $n_\tau$ can be determined by the position of $x^+$ in the entire duration of $[0, L_\eta]$.
For each layer, $\rho$ is sampled independently, so the resulting $\mathcal{A}^-$ is also independent.

\section{An alternative derivation of $\hat q$}\label{app:qhat_alt}
Here we demonstrate an alternative derivation of the single particle quenching parameter in Eq.~\eqref{eq:qhat_Eik_res}. Starting from Eqs.~\eqref{eq:psq_eik} and \eqref{eq:SF}, we have
\begin{align}\label{eq:psq_eik_app}
  \begin{split}
    \braket{p_\perp^2 (x^+)}_{Eik}
    =& 
    \int_{\p}
    \vec p_\perp^2
    \int_{\y,\r}  \tilde \phi^* (\vec r_\perp+\vec y_\perp)
    \tilde \phi (\vec y_\perp)
    e^{-i \vec p_\perp\cdot \vec r_\perp}\\
    &\times  S_F(0,x^+;r)
    \;,
  \end{split}
\end{align}
with a change of variables $\vec r_\perp=\vec x_\perp-\vec y_\perp$ and $r=|\vec r_\perp|$.
We first expand the Wilson line correlator,
\begin{align}
  \begin{split}
     S_F(0,x^+;r)
    &=
    \sum_{n=0}^\infty
    \frac{1}{n!}
    \left(-C_F
    g^4 \tilde{\mu}^2 L(0) x^+\right)^n
    \left[
      1
    -
    \frac{L(r)}{L(0)}
    \right]^n\\
    =&
    \sum_{n=0}^\infty
    \frac{1}{n!}
    \left(- 
    \frac{C_F
    g^4 \tilde{\mu}^2}{4\pi m_g^2}
    x^+\right)^n
    \sum_{k=0}^n
    \binom{n}{k}
    \left[
    -
    \frac{L(r)}{L(0)}
    \right]^k
    \;.
  \end{split}
\end{align}
Next, we evaluate the integral for the power terms of $L(r)$,
\begin{widetext}
  \begin{align}\label{eq:Lr_power}
    \begin{split}
      \int_{\p}
      \vec p_\perp^2
      \int_{\r,\y} & \tilde \phi^* (\vec r_\perp+\vec y_\perp)
      \tilde \phi (\vec y_\perp)
      e^{-i \vec p_\perp\cdot \vec r_\perp} L(r)^m\\
      =& 
      \int_{\p}
      \vec p_\perp^2
      \int_{\l_1}\int_{\l_2}\cdots\int_{\l_m}
      \int_{\r,\y}  \tilde \phi^* (\vec r_\perp+\vec y_\perp)
      \tilde \phi (\vec y_\perp)
      e^{-i (\vec p_\perp+\sum_{i=1}^m \vec l_{\perp,i} )\cdot \vec r_\perp}
      \left[
        \prod_{j=1}^m
        \frac{1}{(m_g^2+ \vec l_{\perp,j}^2)^2}
        \right]\\
        =& \int_{\p}
      \vec p_\perp^2
      \int_{\l_1}\int_{\l_2}\cdots\int_{\l_m}
       \phi^* (\vec p_\perp+\sum_{i=1}^m \vec l_{\perp,i})
      \phi (\vec p_\perp+\sum_{i=1}^m \vec l_{\perp,i})
      \left[
        \prod_{j=1}^m
        \frac{1}{(m_g^2+ \vec l_{\perp,j}^2)^2}
        \right]\\
        =& \int_{\q}
        \int_{\l_1}\int_{\l_2}\cdots\int_{\l_m}
        \left(\vec q_\perp-\sum_{i=1}^m \vec l_{\perp,i}\right)^2
         \phi^* (\vec q_\perp)
        \phi (\vec q_\perp)
        \left[
          \prod_{j=1}^m
          \frac{1}{(m_g^2+ \vec l_{\perp,j}^2)^2}
          \right]
    \end{split}
  \end{align}
\end{widetext}
In the third line, the integral over $\r$ and $\y$ apply a Fourier transform to the wavefunctions, bringing them to the momentum space. In the last line, we made a change of variable $\vec q_\perp\equiv\vec p_\perp+\sum_{i=1}^m \vec l_{\perp,i} $.
There are three terms coming from the square $(\ldots)^2$. The first term containing $\vec q_\perp^2$, gives the transverse momentum squared of the initial state, 
\begin{align}
  \begin{split}
    I_0(m) & \equiv \int_{\q}
    \vec q_\perp^2
    \phi^* (\vec q_\perp)
    \phi (\vec q_\perp)
    \left[
      \prod_{i=1}^m
      \int_{\l_i}
      \frac{1}{(m_g^2+ \vec l_{\perp,i}^2)^2}
      \right]\\
      =&\braket{p_\perp^2(0)}
      L(0)^m\;,
\end{split}
\end{align}
in which the $l_{\perp,i}$-integral gives $L(0)=1/(4\pi m_g^2)$, according to Eq.~\eqref{eq:Lxy}.

The second term, containing $\vec q_\perp\cdot \sum_i \vec l_{\perp,i}$, vanishes after integrating over the angle of $\vec l_{\perp,i}$, 
\begin{align}\label{eq:l_angle_0}
      \int_0^{2\pi}\diff \theta_i
      \frac{l_i \cos(\theta_i-\theta_q)}{(m_g^2+ \vec l_{\perp,i}^2)^2}
      =0\;,
\end{align}
where $\theta_i$ ($\theta_q$) is the angle of the vector $\vec l_{\perp,i} $ ($\vec q_\perp$).

The last term, containing $( \sum_i \vec l_{\perp,i})^2$, reads
\begin{align}
  \begin{split}
    I_1(m)\equiv 
    & 
    \int_{\q}
    \phi^* (\vec q_\perp)
    \phi (\vec q_\perp)
    \int_{\l_1}\int_{\l_2}\cdots\int_{\l_m}
    \left(\sum_{i=1}^m \vec l_{\perp,i}\right)^2\\
    &\times
    \left[
      \prod_{j=1}^m
      \frac{1}{(m_g^2+ \vec l_{\perp,j}^2)^2}
      \right]\\
      =& 
      \int_{\l_1}\int_{\l_2}\cdots\int_{\l_m}
        \sum_{i=1}^m \vec l_{\perp,i}^2
      \left[
        \prod_{j=1}^m
        \frac{1}{(m_g^2+ \vec l_{\perp,j}^2)^2}
        \right]\\
      =&
      \sum_{i=1}^m 
      \int_{\l_i}
      \frac{\vec l_{\perp,i}^2}{(m_g^2+ \vec l_{\perp,i}^2)^2}
      \left[
      \prod_{j\neq i}^m
      \int_{\l_j}
        \frac{1}{(m_g^2+ \vec l_{\perp,j}^2)^2}
      \right]\\
      =&
      m \mathcal G_2
    L(0)^{m-1}
    \;.
  \end{split}
\end{align}
The $\q$-integral is the normalization of the initial state wavefunction, thus giving unity. 
The cross term $\vec l_{\perp,i}\cdot \vec l_{\perp,k}$ vanishes after the angular integral, as in Eq.~\eqref{eq:l_angle_0}.
The $l_{\perp,i}$-integral is logarithmically divergent, introducing a pair of IR and UV cutoffs, 
\begin{align}
  \begin{split}
  \mathcal G_2\equiv &
  \int_{\p}
  \frac{\vec p_\perp^2 }{(m_g^2+\vec p_\perp^2)^2}
  =  \int_{\lambda_{IR}}^{\lambda_{UV}} \frac{\diff p}{2\pi} 
  \frac{p^3 }{(m_g^2+p^2)^2}\\
  =&
  \frac{1}{4\pi}
  \biggl[
    \log(\frac{\lambda_{UV}^2+m_g^2}{\lambda_{IR}^2+m_g^2})
    -
    m_g^2
    \bigg(
    \frac{1}{\lambda_{IR}^2+m_g^2}
    -
    \frac{1}{\lambda_{UV}^2+m_g^2}
    \bigg)
  \biggr]    
    \;.
  \end{split}
\end{align}
In the numerical calculation, one should always let $\lambda_{IR}\ll m_g$, such that the result does not depend on the numerical cutoff $\lambda_{IR}$, and $m_g$ plays the role of the IR regulator,
\begin{align}
  \begin{split}
  \mathcal G_2
  \big|_{\lambda_{IR}=0}
  =
  &   \frac{1}{4\pi}
  \biggl[
    \log(\frac{\lambda_{UV}^2+m_g^2}{m_g^2})
    -\frac{\lambda_{UV}^2}{\lambda_{UV}^2+m_g^2}
  \biggr] 
    \;.
  \end{split}
\end{align}
We thereby obtain Eq.~\eqref{eq:G2_x}. 

Back to Eq,~\eqref{eq:psq_eik_app}, we have
\begin{align}
  \begin{split}
    \braket{p_\perp^2 (x^+)}_{Eik}
    =& 
    \sum_{n=0}^\infty
    \frac{1}{n!}
    (- 
    \frac{C_F
    g^4 \tilde{\mu}^2}{4\pi m_g^2}
    x^+)^n
    \sum_{k=0}^n
    \binom{n}{k}
    (-1)^k\\
    &\times
    \left[
    \braket{p^2_\perp(0) }+k \mathcal G_2\frac{1}{L(0)}
    \right]\\
    =& 
    \braket{p^2_\perp(0) }+\mathcal G_2
    C_F g^4 \tilde{\mu}^2
   x^+
    \;.
  \end{split}
\end{align}
We have used the following relations in the above equation,
\begin{align}
  \sum_{k=0}^n
    \binom{n}{k}
    (-1)^k
    =\delta_{n,0}\;,
    \qquad
    \sum_{k=0}^n
    \binom{n}{k}
    (-1)^k
    k 
    =-\delta_{n,1}\;.
\end{align}
From here, we see that only the linear term survives in the acquired momentum, and we arrive at the result of $\hat q$ in Eq.~\eqref{eq:qhat_Eik_res}.

\section{The color dimension of the quark-gluon state}\label{app:qg_color}
In this appendix, we first present two sets of the basis for the quark-gluon color space and the transformation between the two. Then, we write out the antiquark-gluon-quark-gluon color singlet states. This transformation is helpful in deriving the four-point Wilson line correlators.

\subsection{The quark-gluon color states}
The color space of the quark-gluon state is the tensor product of the color spaces of a single quark and a single gluon. 
This 24-dimensional space, built up as a product of a triplet and an octet, reduces into a direct sum of three irreducible representations,
\begin{align}\label{eq:qg_color_decomp}
  3\otimes 8= 3\oplus \bar 6 \oplus 15\;.
\end{align}
This is a Clebsch–Gordan (CG) series of SU(3), and can be obtained from the Young tableaux method, e.g., see Chapter 12 of Ref.~\cite{Georgi:2000vve}. 
For the generalization to SU$(N_C)$, we still take 3, 8, $\bar 6$, and 15, as the names of the corresponding representations, for convenience.  
We use $\dim(R)$ to denote the dimension of the $R$ representation, such that $\dim(3)=N_c$, $\dim(8)=N_c^2-1$, $\dim(\bar 6)=N_c(N_c+1)(N_c-2)/2$, and $\dim(15)=N_c(N_c-1)(N_c+2)/2$.

The two representations, given on the left- and right-hand sides of Eq.~\eqref{eq:qg_color_decomp}, provide us with two bases for the quark-gluon color state.
The uncoupled color basis is indexed by iterating the quark and gluon color, in the tuple form as $[c_q, c_g]$ ($c_q=1,2,3, c_g=1,2,\ldots,8$), or in the number form as $c_{qg}=(c_q-1)\times 8+c_g$ in which $c_g$ is iterated over first. This uncoupled basis is convenient for simulating the interaction between an individual particle and the background field, as we have adopted in formulating the numerical calculations \cite{Li:2021zaw}.
The coupled color basis is indexed by enumerating the representations in the right-hand side $\{h_{qg}\}$.

The basis expansion of a $\ket{qg}$ color state on the two bases are, respectively,
\begin{align}\label{eq:color_ch}
  \begin{split}\ket{\psi}=&\sum_{c_q=1}^{3}\sum_{c_g=1}^{8}\braket{c_q,c_g|\psi}\ket{c_q,c_g}
    =  \sum_{i=1}^{24}\alpha_i\ket{c_{qg}=i}\;,\\
  \ket{\psi}=&\sum_{h_{qg}=1}^{24}\braket{h_{qg}|\psi}\ket{h_{qg}}
  =  \sum_{j=1}^{24}\beta_j\ket{h_{qg}=j}
  \;.
  \end{split}
\end{align}
In the above equation, $\alpha_i$ is the coefficient of the wavefunction in the uncoupled basis, and $\beta_j$ in the coupled basis. Their column vector forms are $\bm\alpha$ and $\bm\beta$, respectively.
The transformations between the two bases are specified by the SU(3) CG coefficients, $ \mathcal{C}(h_{qg}; c_q, c_g)$,
\begin{align}\label{eq:CG_3p8}
  \begin{split}
    &\ket{h_{qg}}=\sum_{c_q=1}^{3}\sum_{c_g=1}^{8}
    \mathcal{C}(h_{qg}; c_q, c_g) \ket{c_q,c_g}\;,\\
    &\ket{c_q,c_g}=\sum_{h_{qg}=1}^{24}\mathcal{C}^{-1}(h_{qg}; c_q, c_g)\ket{h_{qg}}\;.  
  \end{split}
\end{align}
Writing the CG coefficients $\mathcal C(h_{qg}; c_q,c_g)$ in the matrix form $\mathcal C$ with row index $h_{qg}$ and column index $c_{qg}(c_q,c_g)$, we see that $\bm \alpha=\mathcal C\bm\beta$. 
The transformation matrix is unitary, $\mathcal C^{-1}=\mathcal C^\dagger$. 
For convenience, we partition $\bm  C$ into blocks $\bm  C=\{    \bm  C_{Q},\bm  C_{h6},\bm  C_{h15}\}^\intercal$ such that the three operators extract the $3$, the $\bar 6$, and the $15$ components of a state respectively.

\begin{figure*}[tbp!]
  \centering 
  \includegraphics[width=.8\textwidth]{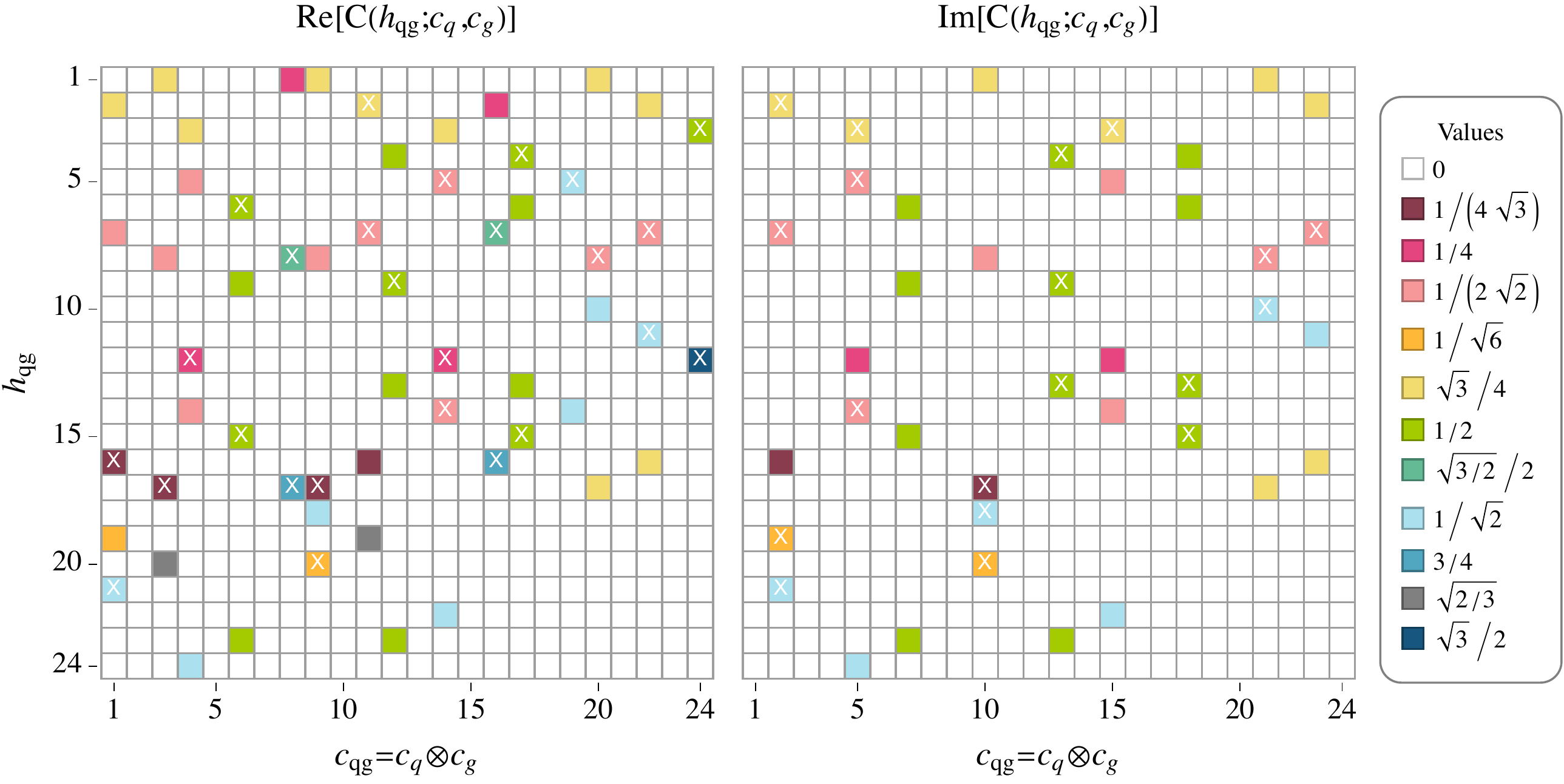}
  \caption{ The CG coefficients $\mathcal C(h_{qg}; c_q,c_g)$ as defined in Eq.~\eqref{eq:CG_3p8}. 
  The white cross indicates negative values in the corresponding grids.
  }
  \label{fig:CG_3p8_table_matrix_rep} 
\end{figure*}

There are multiple ways of computing $\mathcal C$, and here we proceed with the tensor method as illustrated in Ref.~\cite{Georgi:2000vve}. 
 The generic quark-gluon state is described by a tensor, 
\begin{align}
\begin{split}
        u^i v^j_k=&\frac{1}{8}(3\delta^i_k u^l v^j_l-\delta^j_k u^l v^j_l)\\
        &+\frac{1}{4}\epsilon^{ijk}(\epsilon_{lmn} u^m v^n_k +\epsilon_{kmn}u^m v^n_l)\\
         &+\frac{1}{2}(u^i v^j_k +u^j v^i_k-\frac{1}{4}\delta^i_k u^l v^j_l-\frac{1}{4}\delta^j_k u^l v^j_l)\;.
\end{split}
\end{align}
which acts in the tensor product space $\ket{_i} \ket{^k_j} $. It relates to the uncoupled quark-gluon basis by the Gell-Mann matrices $\ket{c_q=i, c_g=a}=T^{a}_{kj}\ket{_i} \ket{^k_j}$. 
The right-hand side is the CG decomposition, and the three terms correspond to the quark-gluon in the $3$, the $\bar 6$, and the $15$ state, respectively. 
We thereby find the transformation between the uncoupled and the coupled states. However, the transformation matrix is expressed in the tensor representation. 
The transformations to the states in the triplet, $h_{qg}=c_Q=1,2,3$, are given by the Gell-Mann matrices,
\begin{align}\label{eq:CG_3t8_3}
    \mathcal{C}_{Q} (h_{qg}=c_Q; c_g, c_q) =\frac{\sqrt{3}}{2}T^{c_g}_{c_q, c_Q} \;.
\end{align}
The coefficient is determined by state normalization.
The transformations to the states in the anti-sextet are specified by two symmetric indices $m,n=1,2,3$, 
  \begin{align}\label{eq:CG_3t8_6}
    \tilde{C}_{h6}(m,n; i, a) 
    =
    \frac{1}{2}
    \sum_{j}^3
    ( T^a_{nj}
    \epsilon_{ijm}
    + T^a_{mj}
    \epsilon_{ijn})\;.
  \end{align}
  The $\{m,n\}$ states must be related to the coupled states $h_{qg}=4,\ldots,9$ by some linear combination, thus $\mathcal{C}_{h6}(h_{qg};  c_q,c_g) = \sum_{m,n=1}^3 h_{qg}(m, n)\tilde{C}_{h6}(m,n; c_q, c_g) $.
  The transformations to the 15-plet states are specified by three indices $\bar i,\bar j,\bar k=1,2,3$, 
    \begin{align}\label{eq:CG_3t8_15}
    \begin{split}
            &\tilde{C}_{h15}(\bar i,\bar j,\bar k; i, a) \\
    &\qquad
    =  \frac{\sqrt{2}}{2}(
      T^a_{\bar k\bar j}
      \delta^{\bar i}_i  
      +  T^a_{\bar k\bar i}
      \delta^{\bar j}_i  
      -\frac{1}{4}
      T^a_{i \bar j}
      \delta^{\bar i}_{\bar k}
      -  \frac{1}{4}
      T^a_{i \bar i}
      \delta^{\bar j}_{\bar k}
      )\;.
    \end{split}
  \end{align}
  The $\{\bar i,\bar j,\bar k\}$ states must be related to the coupled states $h_{qg}=10,\ldots,24$ by some linear combination, thus $\mathcal{C}_{h15}(h_{qg}; c_q,c_g) = \sum_{\bar i,\bar j,\bar k=1}^3 h_{qg}(\bar i,\bar j,\bar k)\tilde{C}_{h15}(\bar i,\bar j,\bar k; c_q, c_g) $.
The specific form of $h_{qg}(m,n)$ and $h_{qg}(\bar i,\bar j,\bar k) $ depends on the choice of the coupled basis states. 
We provide here one set of $\mathcal{C}$ in Fig.~\ref{fig:CG_3p8_table_matrix_rep}.
We construct it by an online CG coefficients generator~\cite{SU3_CG_calculator} based on the numerical algorithm proposed in Ref.~\cite{doi:10.1063/1.3521562}. 
\footnote{
Note, however, the CG coefficients obtained directly from Ref.~\cite{SU3_CG_calculator} are in the Gelfand-Tsetlin (GT) basis (cf. the Cartan-Weyl basis, see Ref.~\cite{doi:10.1063/1.3521562} for more details), not on the Gell-Mann (GM) basis that one usually adopts in high-energy physics, this work as well. 
An easy way to find the transformation from the GT to the GM basis is by comparing the CG coefficients for the known decomposition $3\otimes \bar 3=1\oplus 8$. }

It is also useful to define the projectors, using the transformation matrix $\mathcal{C}$, or equivalently $\tilde{C}$ in the tensor product basis,
\begin{align}\label{eq:projectors}
  \begin{split}
    &\bm P_{Q}=\bm C_{Q}^{-1}\bm C_{Q}\\
    &\bm P_{h6}=\bm C_{h6}^{-1}\bm C_{h6}=\tilde C_{h6}^{-1}\tilde  C_{h6}\\
   & \bm P_{h15}=\bm C_{h15}^{-1}\bm C_{h15}=\tilde  C_{h15}^{-1}\tilde  C_{h15}
    \;. 
  \end{split}
\end{align}
The expectation value of each projector operator $\braket{\psi|\bm P|\psi}$ gives the probability of the state in the corresponding color subspace.
In the component form, $\bm P|_{j,b;i,a}=\sum_{\{s\}} \mathcal C^{-1}(j,b; \{s\})\mathcal C( \{s\};I,a)$, in which the summation is over the full basis space indexed by $ \{s\}$.
One can find the projectors using Eqs.~\eqref{eq:CG_3t8_3}, \eqref{eq:CG_3t8_6} and \eqref{eq:CG_3t8_15}.
Here, we write out the expressions for SU($N_c$)~\cite{cougoulic:tel-02011152, Florian_rapport},
\begin{subequations}
  \begin{align}
    &\bm P_Q|_{j,b;i,a} =\frac{2 N_c}{N_c^2-1}(T^a T^b)_{ij}\;.\\
    & \bm P_{h6}|_{j,b;i,a} =\frac{1}{2}\delta_{ab}\delta_{ij}-(T^b T^a)_{ij}-\frac{1}{N_c-1}(T^a T^b)_{ij}
    \;,\\
    & \bm P_{h15}|_{j,b;i,a} =\frac{1}{2}\delta_{ab}\delta_{ij}+(T^b T^a)_{ij}-\frac{1}{N_c+1}(T^a T^b)_{ij}
    \;.
    \end{align}
\end{subequations} 
The summation of those projectors is identity, as should be, $\bm I_{N_F\times N_A}=\bm P_{Q}+\bm P_{h6}+\bm P_{h15}$.
\subsection{The antiquark-gluon-quark-gluon color singlet states}\label{app:Wilsonlines}
\begin{figure}[tbp!]
  \centering  
  \subfigure[s-basis \label{fig:qdgqg_Ss}]{   
    \includegraphics[width=0.23\textwidth]{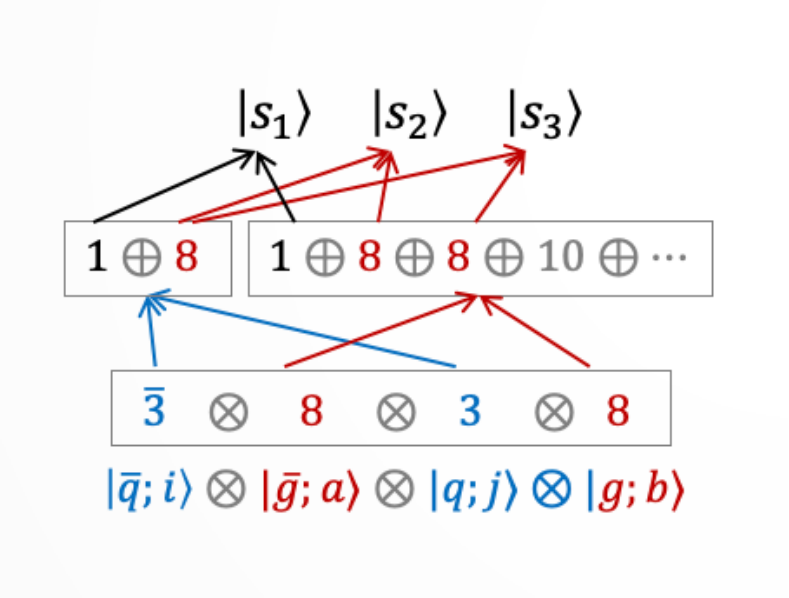}}
  \subfigure[v-basis \label{fig:qdgqg_Sv}]{ 
   \includegraphics[width=0.23\textwidth]{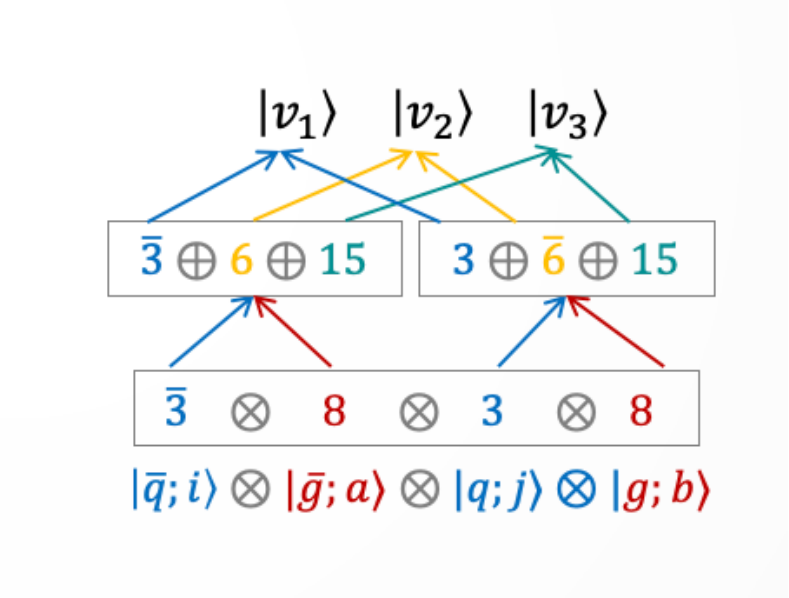}}
  \caption{An illustration of constructing the color singlet subspace of the $\bar q\otimes \bar g\otimes q\otimes g $ state into (a) the s-basis as in Eq.~\eqref{eq:def_sbais} and (b) the v-basis as in Eq.~\eqref{eq:def_vbais}. }
  \label{fig:qdgqg_Ssv}
  \end{figure}

In the full color space of $\bar q\otimes \bar g\otimes q\otimes g $, there are 3 singlets, $\bar 3\otimes 8\otimes 3\otimes 8 = 1\oplus 1\oplus 1\oplus 8\oplus\ldots$.
We consider two sets of basis in this singlet subspace, namely, the s-basis, and the v-basis, as illustrated in Figure.~\ref{fig:qdgqg_Ssv}.
The s-basis as an expansion in the uncoupled basis space reads,
\begin{align}\label{eq:def_sbais}
  \begin{split}
    &\ket{s_1}=
    -\sqrt{\frac{1}{N_c d_A}}\sum_{i,j=1}^{N_c}\sum_{a,b=1}^{d_A} \delta_{i,j}\delta_{a,b}\ket{i,a,j,b}\;,\\
    &\ket{s_2}=
   \frac{1}{i} \sqrt{ \frac{2}{N_c d_A}}\sum_{i,j=1}^{N_c}\sum_{a,b=1}^{d_A} f_{abc}T^c_{ji}\ket{i,a,j,b}\;,\\
   &\ket{s_3}=
  \sqrt{ \frac{2}{C_d d_A}}\sum_{i,j=1}^{N_c}\sum_{a,b=1}^{d_A} d_{abc}T^c_{ji}\ket{i,a,j,b}\;,
\end{split}
\end{align}
in which $d_A=N_c^2-1=8$, $C_d=(N_c^2-4)/N_c=5/3$, 
  $d_{abc}=2\Tr[\{T_a,T_b\}T_c]$, and
  $f_{abc}=-2 i\Tr[[T_a,T_b]T_c]$.
The convention in defining the s-basis is the same as those in Ref.~\cite{Lappi:2020srm}, by replacing two $\bar q q$-dipoles by two gluons, viz., $\ket{c_g}=\sqrt{2}\sum_{\bar c_q, c_q}T^{c_g}_{ c_q, \bar c_q}\ket{\bar c_q, c_q}$. 

The v-basis as an expansion in the uncoupled basis space can be written in terms of the qg-projectors defined in Eq.~\eqref{eq:projectors},
\begin{align}\label{eq:def_vbais}
  \begin{split}
   \ket{v_1}= &
 \frac{1}{\sqrt{\dim(3)}}
   \sum_{i,j=1}^{N_c}\sum_{a,b=1}^{d_A} 
   \bm P_Q|_{i,a;j,b}
   \ket{i,a,j,b} \;,\\
    \ket{v_2}=&
    \frac{1}{\sqrt{\dim(\bar 6)}}\sum_{i,j=1}^{N_c}\sum_{a,b=1}^{d_A} 
    \bm P_{h6}|_{i,a;j,b}
    \ket{i,a,j,b}\;,\\
   \ket{v_3}=&
   \frac{1}{\sqrt{\dim(15)}}\sum_{l=1}^{15}\sum_{a,b=1}^{d_A} 
    \bm P_{h15}|_{i,a;j,b}
    \ket{i,a,j,b}\;.
\end{split}
\end{align}

The transformation between the two bases is
\begin{align}
  \ket{v_I}=\sum_{J=1}^3 \ket{s_J}\braket{s_J|v_I}\,,
  \quad
  \ket{s_I}=\sum_{J=1}^3 \ket{v_J}\braket{v_J|s_I}\;,
\end{align}
for which we write out the transformation matrix $V$ with element $V_{JI}=\braket{s_J|v_I}=\braket{v_I|s_J}$, 
\begin{align}
  \begin{split}
  V=  &
  \begin{pmatrix}
    -\frac{1}{\sqrt{N_c^2-1} }  
    &  -\frac{1}{\sqrt{2}}\sqrt{\frac{N_c-2}{N_c-1}}  
    &-\frac{1}{\sqrt{2}}\sqrt{\frac{N_c+2}{N_c+1}}   \\
    \frac{1}{\sqrt{2}}\frac{N_c}{\sqrt{N_c^2-1}} 
    &\frac{1}{2}\sqrt{\frac{N_c-2}{N_c-1}}
    &  -\frac{1}{2}\sqrt{\frac{N_c+2}{N_c+1}}\\
    \frac{1}{\sqrt{2}}\sqrt{\frac{N_c^2-4}{N_c^2-1}}  
    & -\frac{1}{2}\sqrt{\frac{N_c+2}{N_c-1}}  &\frac{1}{2}\sqrt{\frac{N_c-2}{N_c+1}}  \\
    \end{pmatrix}\\
    =&
   \begin{pmatrix}
    -\frac{1}{2\sqrt{2} }  &  -\frac{1}{2}  &-\frac{\sqrt{5}}{2\sqrt{2}}  \\
    \frac{3}{4} &\frac{1}{2\sqrt{2}}&  -\frac{\sqrt{5}}{4}\\
     \frac{\sqrt{5}}{4}  & -\frac{\sqrt{5}}{2\sqrt{2}}  &\frac{1}{4} 
    \end{pmatrix}
    \;.
  \end{split}
\end{align}

\section{The four-point Wilson line correlator}\label{sec:UqgUdqg}
The quark-gluon Wilson line is built as the tensor product of a quark and a gluon Wilson line,
  \begin{align}\label{eq:Uqg_def}
    U_{qg}(0,L_\eta; \vec x_\perp, \vec y_\perp) \equiv U_F(0,L_\eta; \vec x_\perp)\otimes U_A(0,L_\eta; \vec y_\perp)
    \;.
  \end{align}
The dimension of the $qg$ Wilson line is $N_c d_A=24$. In the component form,
\begin{multline}\label{eq:Uqg_def_component}
  U_{qg}(0,L_\eta; \vec x_\perp, \vec y_\perp)_{\bar c_{qg}, c_{qg}}\\
  =
  U_F(0,L_\eta; \vec x_\perp)_{\bar c_q, c_q} 
  U_A(0,L_\eta; \vec y_\perp)_{\bar c_g, c_g} 
  \;.
\end{multline}

The single $qg$ Wilson line gives the amplitude of a quark-gluon state propagating through the color medium and can be related to its total cross section through the optical theorem, for which, see Ref.~\cite{Li:2021zaw}. Here, we would like to understand the probability distribution of a quark-gluon state, so what we are interested in is the correlation function of two $qg$ Wilson lines, i.e., the four-point function, 
  \begin{align}\label{eq:S_qgqg_def}
    \begin{split}
      & S_{\bar{q}\bar{g}qg}(0,x^+;\vec x_{q,\perp},\vec x_{g,\perp},\vec y_{q,\perp},\vec y_{g,\perp})_{\beta_1\beta_2\beta_3\beta_4; \alpha_1\alpha_2\alpha_3\alpha_4}\\
      = & \Bigl \langle U_F^\dagger(0,x^+;\vec x_{q,\perp})_{\alpha_1,\beta_1} 
      U_A^\dagger(0,x^+;\vec x_{g,\perp})_{\alpha_2,\beta_2}\\
      & \qquad U_F(0,x^+;\vec y_{q,\perp})_{\beta_3,\alpha_3}
      U_A(0,x^+;\vec y_{g,\perp})_{\beta_4,\alpha_4}
      \Bigr \rangle_\rho \\
      =&
      e^{ - (C_F+C_A)\xi L(0)}
      e^{
        \xi \mathcal M_{\bar{q}\bar{g}qg}
       } \bigg|_{\beta_1\beta_2\beta_3\beta_4; \alpha_1\alpha_2\alpha_3\alpha_4} \;,
    \end{split}
  \end{align}
 in which $ \xi\equiv g^4 \tilde{\mu}^2 x^+ $ and
    \begin{align}\label{eq:Mqgqg}
      \begin{split}
    -\mathcal M_{\bar{q}\bar{g}qg}\equiv&
     (-T^{a*}) \otimes I_A \otimes T^a\otimes I_A  
    L(|\vec x_{q,\perp}-\vec y_{q,\perp}|)
     \\
    &
    + I_F \otimes (-t^{a*}) \otimes I_F\otimes t^a
    L(|\vec x_{g,\perp}-\vec y_{g,\perp}|)
    \\
    &
  + 
  (-T^{a*})\otimes  (-t^{a*}) \otimes I_F\otimes I_A 
    L(|\vec x_{q,\perp}-\vec x_{g,\perp}|)
    \\
      &
    +I_F\otimes I_A\otimes T^a \otimes t^a 
    L(|\vec y_{q,\perp}-\vec y_{g,\perp}|) \\
    &
    + 
    (-T^{a*}) \otimes I_A  \otimes I_F\otimes t^a
    L(|\vec x_{q,\perp}-\vec y_{g,\perp}|) \\
    &
    + I_F \otimes  (-t^{a*})\otimes T^a\otimes I_A
    L(|\vec x_{g,\perp}-\vec y_{q,\perp}|)
      \;.
    \end{split}
  \end{align}
The above expression gives the amplitude of a $\bar q g q g $-quadruple going from color configuration $\{\bar c_{q}= \alpha_1, \bar c_g=\alpha_2, c_q= \alpha_3, c_g=\alpha_4\}\ $ to $\{\bar c_{q}= \beta_1, \bar c_g=\beta_2, c_q= \beta_3, c_g=\beta_4\}\ $.
This function $S_{\bar{q}\bar{g}qg} $ is $576\times 576$ in the full color space of the $\bar q\otimes \bar g\otimes q\otimes g $ state. 
But since we are interested in the probability of the $q g $ state transferring from some color state $\{ c_{q}= \alpha_1, c_g=\alpha_2 \}\ $ to $\{  c_{q}= \beta_1,  c_g=\beta_2\}\ $, the quantity we should look at is in the form of 
$ S_{\bar{q}\bar{g}qg}\big|_{\beta_1\beta_2 \beta_1\beta_2; \alpha_1 \alpha_2 \alpha_1\alpha_2}
$.
Therefore, we only need to study the $\bar q g q g $ singlet states.
A closely related study in the color structure can be found in Ref.~\cite{Lappi:2020srm}, which is carried out in terms of six-point fundamental Wilson line correlators. 

In the s-basis as given by Eq.~\eqref{eq:def_sbais}, we have [equivalent to Eq. (49) in Ref.~\cite{Lappi:2020srm}],
\begin{align}\label{eq:M_qgqg_s}
  \begin{split}
  \mathcal M_{\bar{q}\bar{g}qg}^s=&
   \begin{pmatrix}
     \Gamma_0 &  -\frac{1}{\sqrt{2}}\Gamma_2 & 0\\
    -\frac{1}{\sqrt{2}}\Gamma_2 &\frac{N_C}{4} \Gamma_1& \frac{\sqrt{N_c C_d}}{4}\Gamma_2\\
     0  & \frac{\sqrt{N_c C_d}}{4}\Gamma_2 & \frac{N_C}{4} \Gamma_1 \\
   \end{pmatrix}
    \;,
  \end{split}
\end{align}
where
\begin{align}\label{eq:Gammas}
  \begin{split}
    \Gamma_0=&  C_F L^{13} + N_c  L_{24}\\
    \Gamma_1=& (L^1_{2} + L^1_{4} +L^3_2  +  L_4^3)
    -\frac{2}{N_c^2}L^{13}+2 L_{24}\\
    \Gamma_2=&  L^1_{2}- L^1_{4} -L^3_2 +  L_4^3 \;.
  \end{split}
\end{align}
 For the convenience of reading, let us define a short-hand notation for the scalar function $L(r)$, by putting the particle index that labels the contracted quark (gluon) into the superscript (subscript), $L^{[\text{quarks}]}_{[\text{gluons}]}$. 
The four particles are labeled by numbers 1 to 4 from left to right, the same order used in the color indices $\alpha_i$ as in Eq.~\eqref{eq:S_qgqg_def}. For example,  
\begin{align*}
  L^{13}= 
  L(|\vec x_{q,\perp}-\vec y_{q,\perp}|)\;,\qquad
  L_{24}= L(|\vec x_{g,\perp}-\vec y_{g,\perp}|)\;.
\end{align*}
The four-point Wilson line correlator in the s-basis reads,
\begin{align}\label{eq:S_qgqg_s_res_red}
  \begin{split}
     S_{\bar{q}\bar{g}qg}^s
     =&     e^{ - (C_F+C_A)\xi L(0)}
      \sum_{i=1}^3
     \frac{e^{-z_i \xi/4}}{d(z_i)}\\
     &
     \begin{pmatrix}
       m_{11} (z_i) & m_{12} (z_i)& m_{13} (z_i)\\
       m_{21} (z_i) & m_{22} (z_i)& m_{23} (z_i)\\
       m_{31} (z_i) & m_{32} (z_i)& m_{33} (z_i)
     \end{pmatrix}
    \;,
  \end{split}
\end{align}
in which 
\begin{align}
  z_i=&-\frac{1}{3}
     \left[b+\zeta^i C+ \frac{\Delta_0}{\zeta^i C}\right],\qquad \zeta=\frac{-1+i\sqrt{3}}{2} \;,
\end{align}
the roots of characteristic polynomial $\det (-z/4 I-M_{\bar{q}\bar{g}qg}^s)=0$,
\begin{align}
  \begin{split}
     a=&1,\\
     b= &2(2\Gamma_0+N_c\Gamma_1),\\
     c=&
     N_c\Gamma_1(N_c\Gamma_1+8\Gamma_0)-(N_c^2+4)\Gamma_2^2,\\
     d= &4
    \left\{
      N_c^2\Gamma_0\Gamma_1^2-[(N_c^2-4)\Gamma_0+2N_c\Gamma_1]\Gamma_2^2
     \right\}
    \;,
  \end{split}
\end{align}
the coefficients,
and
\begin{align}
  \begin{split}
     \Delta_0=&b^2-3ac,\\
     \Delta_1=&2b^3-9abc+27a^2 d,\\
     C=&\sqrt[3]{\frac{\Delta_1+\sqrt{\Delta_1^2-4 \Delta_0^3}}{2}}
    \;.
  \end{split}
\end{align}
The matrix elements read
\begin{subequations}
  \begin{align}
    & d(z_i)\equiv \prod_{\substack{j=1,\\ j\neq i}}^3(z_i-z_j)= 3 z_i^2+ 4 c_1 z_i+c\;,\\
    & m_{11} (z_i) = (z_i+N_c\Gamma_1 )^2+(4-N_c^2)\Gamma_2^2\;,\\
    & m_{12} (z_i)=m_{21} (z_i) = 2\sqrt{2}\Gamma_2 (z_i+N_c \Gamma_1)\;,\\
    & m_{13} (z_i)=m_{31} (z_i) = -2\sqrt{2C_d N_c}\Gamma_2^2\;,\\
    & m_{22} (z_i) = z_i^2+ (4\Gamma_0+N_c\Gamma_1) z_i+4 N_c\Gamma_1\Gamma_0\;,\\
    & m_{23} (z_i) = m_{32} (z_i)  = -\sqrt{C_d N_c}\Gamma_2(4\Gamma_0+ z_i)\;,\\
    & m_{33} (z_i) = m_{22} (z_i)-8\Gamma_2^2\;.
  \end{align}
\end{subequations}
We, therefore, arrive at Eq. (55) in Ref.~\cite{Lappi:2020srm}. The result is translationally invariant, which can also be seen from the definition in Eq.~\eqref{eq:Mqgqg}, in the sense that it only depends on the relative positions among the four particles but not the center of mass of the system. 
Explicitly, we can write
\begin{align}\label{eq:S_shift}
  \begin{split}
  S_{\bar{q}\bar{g}qg} & (0,x^+;\vec x_{q,\perp},\vec x_{g,\perp},\vec y_{q,\perp},\vec y_{g,\perp})\\
  = &S_{\bar{q}\bar{g}qg} ( 0,x^+; \vec x_{q,\perp}, \vec x_{g,\perp}, \vec x_{q,\perp}-\vec u_{q,\perp}, \vec x_{g,\perp}-\vec u_{g,\perp} )\\
 = &
  S_{\bar{q}\bar{g}qg}(0,x^+;
  \vec v_\perp+ \vec u_{q,\perp}, 
  \vec u_{q,\perp},
  \vec v_\perp,
  \vec u_{q,\perp}-\vec u_{g,\perp}
  )\;.
  \end{split}
\end{align}
In the second line, we make the change of variables according to Eq.~\eqref{eq:uv_variables}; in the third line, we shift all the four position arguments by $-\vec x_{g,\perp}+\vec u_{q,\perp}$. 
This rewriting is for the convenience of identifying the relevant physical quantities, without making any actual change to the content.
Let us also write out $\Gamma$s defined in Eq.~\eqref{eq:Gammas} in terms of the three independent vectors $\vec v_{\perp}, \vec u_{q,\perp}$, and $\vec u_{q,\perp}$,
\begin{align}\label{eq:Gammas_v2}
  \begin{split}
    \Gamma_0=&  C_F L(u_{q,\perp}) + N_c  L(u_{g,\perp})\\
    \Gamma_1=& (L(v_\perp) + L(|\vec v_\perp+ \vec u_{g,\perp} |) +L(|\vec v_\perp- \vec u_{q,\perp} |) \\
    & +  L(|\vec v_\perp- \vec u_{q,\perp}+ \vec u_{g,\perp} |))
    -\frac{2}{N_c^2}L(u_{q,\perp})+2 L(u_{g,\perp})\\
    \Gamma_2=&  L(v_\perp)-  L(|\vec v_\perp+ \vec u_{g,\perp} |)-L(|\vec v_\perp- \vec u_{q,\perp} |) \\
    &+ L(|\vec v_\perp- \vec u_{q,\perp}+ \vec u_{g,\perp} |) \;.
  \end{split}
\end{align}
We denote the modulus of a vector using the same variable without the arrow, $u_l=|\vec u_l|$.

Knowing the explicit form of $S_{\bar{q}\bar{g}qg}^s$, we can now evaluate the probability function of a quark-gluon state.
In doing so, we interpret the $\bar{q}\bar{g}qg$-state as a $qg$ state and its conjugate. 
Consider a quark-gluon state in a gauge-invariant color space $c$ with dimension $d_c$, then the corresponding $\bar{q}\bar{g}qg$ state is in the color configuration $\bar c c$.
The color space $c$ could be the $3$, the $\bar 6$, the $15$, and the full $N_c d_A$ space denoted by ``$X$''. 
The first three cases correspond to the three v-basis states, and the last one the $\ket{s_1}$ state.
The initial state is averaged over the corresponding qg color space,
\begin{align}
  \begin{split}
    \ket{\psi_{\bar q\bar g 
    qg, \bar c_i(\bar q\bar g) c_i(qg)}}
    = &\frac{1}{d_c}\ket{\bar \psi_{qg,c_i}}\otimes \ket{\psi_{qg,c_i}}\\
    =&\begin{cases}
      \frac{1}{\sqrt{\dim(R_i)}}\ket{v_i}, &c_i=3, \bar 6, 15\\
      -\frac{1}{\sqrt{N_c d_A}}\ket{s_1}, & c_i=X
    \end{cases}
    \;.
  \end{split}
\end{align}
The final state is summed over the corresponding qg color space,
\begin{align}
  \begin{split}
    \ket{\psi_{\bar q\bar g 
    qg, \bar c_f(\bar q\bar g) c_f(qg)}}
    =& d_c\ket{\bar \psi_{qg,c_f}}\otimes \ket{\psi_{qg,c_f}}\\
    =& \begin{cases}
      \sqrt{\dim(R_f)}\ket{v_f}, &c_f=3, \bar 6, 15\\
      -\sqrt{N_c d_A}\ket{s_1}, & c_f=X
    \end{cases}
    \;.
  \end{split}
\end{align}

The probability function of a $qg$ state is therefore, from one color subspace to the other,
\begin{align}\label{eq:P_qg_ci_cf}
  \begin{split}
    \mathcal{P}_{qg,c_i\to c_f}
    \equiv &
     \braket{
     \psi_{\bar q \bar g q g, \bar c_f(\bar q\bar g) c_f(qg)}
     | S_{\bar q \bar g q g} |
     \psi_{\bar q \bar g q g, \bar c_i(\bar q\bar g) c_i(qg)}
     }\\
     =& \frac{\sqrt{\dim(R_f)}}{\sqrt{\dim(R_i)}} \braket{
      v_f
      | S_{\bar q \bar g q g} |
     v_i
      }
    \;,
  \end{split}
\end{align}
and from one color subspace to the full space,
\begin{align}\label{eq:P_qg_c}
  \begin{split}
    \mathcal{P}_{qg,c}
    \equiv &
    \sum_{i=1}^{N_c}\sum_{a=1}^{d_A}
     \braket{
     \psi_{\bar q\bar g qg,\{i,a,i,a\} }
     | S_{\bar q \bar g q g} |
     \psi_{\bar q \bar g q g, \bar c(\bar q\bar g) c(qg)}
     }\\
     = &
     -\sqrt{N_c d_A
     }
     \braket{
     s_1
     | S_{\bar q \bar g q g} |
     \psi_{\bar q \bar g q g, \bar c(\bar q\bar g) c(qg)}
     }
    \;.
  \end{split}
\end{align}
The relation between the color-differential and the color-inclusive cross sections is,
\begin{align}
  \begin{split}
    \mathcal{P}_{qg,c_i} =\mathcal{P}_{qg,c_i\to X}
   =\sum_{c_f=3,\bar 6,15}\mathcal{P}_{qg,c_i\to c_f}
    \;.
  \end{split}
\end{align}
In analogy, the relation between the cross sections of the color-differential and the color-inclusive incoming states is,
\begin{align}\label{eq:P_qg_X_cf}
  \begin{split}
    N_c d_A \mathcal{P}_{qg,X \to c_f}
   =\sum_{c_f=3,\bar 6,15}\dim (R_i)\mathcal{P}_{qg,c_i\to c_f}
    \;.
  \end{split}
\end{align}
The probability function in the full-color space is
\begin{align}\label{eq:P_qg_X}
  \begin{split}
    \mathcal{P}_{qg,X}
     = &
     \braket{
     s_1
     | S_{\bar q \bar g q g} |
     s_1
     }
     =S^s_{11}
    \;.
  \end{split}
\end{align}
The probability function of the triplet is,
\begin{align}\label{eq:P_qg_3}
  \begin{split}
    \mathcal{P}_{qg,3}
     =&
      S^s_{11}
      -\frac{N_c \sqrt{2}}{2}S^s_{12}
      -\frac{\sqrt{2(N_c^2-4)} }{2} S^s_{13}
    \;.
  \end{split}
\end{align}
The color differential probability is
\begin{align}\label{eq:P_qg_3diff}
  \begin{split}
    \mathcal{P}_{qg,3\to 3}
     & =
     \frac{1}{2(N_c^2-1)}\bigg[
       2 S^s_{11}
       -2\sqrt{2}N_c S^s_{12}\\
       &
       -2\sqrt{2(N_c^2-4)} S^s_{13}
       +N_c^2 S^s_{22}\\
       &
       +2N_c\sqrt{N_c^2-4} S^s_{23}
       +(N_c^2-4)
       S^s_{33}
       \bigg]\;,\\
       \mathcal{P}_{qg,3\to \bar 6}
      &=
       \frac{1}{4(N_c-1)}\bigg[
        2(N_c-2) S^s_{11}\\
        &
        -(N_c+1)(N_c-2)\sqrt{2} S^s_{12}\\
        &
        -(N_c-3)\sqrt{2(N_c^2-4)}S^s_{13}\\
        &
        +N_c (N_c-2) S^s_{22}\\
        &
        -2\sqrt{N_c^2-4} S^s_{23}
        -(N_c+2)(N_c-2) S^s_{33}
        \bigg]\;,\\
       \mathcal{P}_{qg,3\to 15}
       =&
       \frac{1}{4(N_c+1)}\bigg[
        2(N_c+2) S^s_{11}\\
        &
        -(N_c-1)(N_c+2)\sqrt{2} S^s_{12}\\
        &
        -(N_c+3)\sqrt{2(N_c^2-4)}S^s_{13}\\
        &
        -N_c (N_c+2) S^s_{22}\\
        &
        +2\sqrt{N_c^2-4} S^s_{23}
        +(N_c+2)(N_c-2) S^s_{33}
        \bigg]
    \;.
  \end{split}
\end{align}
The probability function of the $\bar 6$ is,
\begin{align}\label{eq:P_qg_b6}
  \begin{split}
    \mathcal{P}_{qg,\bar 6}
      =&
      S^s_{11}
      -\frac{\sqrt{2}}{2}S^s_{12}
      +\sqrt{\frac{N_c+2}{2(N_c-2)}}S^s_{13}
    \;,
  \end{split}
\end{align}
and color-differentially,
\begin{align}\label{eq:P_qg_6diff}
  \begin{split}
    \mathcal{P}_{qg,\bar 6\to 3}
    =&\frac{\dim(3)}{\dim(\bar 6)}\mathcal{P}_{qg,3\to\bar 6 }\;,\\
      \mathcal{P}_{qg,\bar 6\to \bar 6}
      =&
      \frac{1}{4(N_c-1)}\bigg[
       2(N_c-2) S^s_{11}\\
       &
       -2\sqrt{2}(N_c-2)  S^s_{12}\\
       &
       +2\sqrt{2(N_c^2-4)} S^s_{13}
       +(N_c-2)  S^s_{22}\\
       &
       -2\sqrt{N_c^2-4} S^s_{23}
      +(N_c+2)  S^s_{33}
       \bigg]\;,\\
       \mathcal{P}_{qg,\bar 6\to 15}
       =&
       \frac{1}{4(N_c+1)}\sqrt{\frac{N_c+2}{N_c-2}}
       \bigg[
        2\sqrt{N_c^2-4} S^s_{11}\\
        &
        +4\sqrt{2} S^s_{13}
      -\sqrt{N_c^2-4} S^s_{22}
      +2 N_c S^s_{23}\\
      &
      -\sqrt{N_c^2-4} S^s_{33}
      \bigg]
    \;.
  \end{split}
\end{align}

The probability function of the $15$-plet is,
\begin{align}\label{eq:P_qg_15}
  \begin{split}
    \mathcal{P}_{qg,15}
     =
      S^s_{11}
      +\frac{\sqrt{2}}{2}S^s_{12}
      -\sqrt{\frac{N_c-2}{2(N_c+2)}}
      S^s_{13}
    \;,
  \end{split}
\end{align}
and
\begin{align}\label{eq:P_qg_15diff}
  \begin{split}
    \mathcal{P}_{qg,15\to 3}
     =&\frac{\dim(3)}{\dim(15)}\mathcal{P}_{qg,3\to 15 }\;,\\
      \mathcal{P}_{qg,15\to \bar 6}
    =&\frac{\dim(\bar 6)}{\dim(15)}\mathcal{P}_{qg,\bar 6\to 15 }\;,\\
       \mathcal{P}_{qg,15\to 15}
             =&
       \frac{1}{4(N_c+1)}\bigg[
        2(N_c+2) S^s_{11}\\
        &
        +2\sqrt{2}(N_c+2) S^s_{12}\\
        &
        -2\sqrt{2(N_c^2-4)} S^s_{13}
      +(N_c+2) S^s_{22}\\
      &
      -2\sqrt{N_c^2-4} S^s_{23}
      + (N_c-2)S^s_{33}
      \bigg]
    \;.
  \end{split}
\end{align}
We have written above the color-differential and inclusive probability functions $\mathcal P_{qg}$ in terms of the components of $S^s_{\bar q\bar g q g}$. 
In the dilute limit, one can replace $S^s_{\bar q\bar g qg}$ by its exponent $\mathcal M^s_{\bar q\bar g qg} $ as given in Eq.~\eqref{eq:M_qgqg_s}.
Then, the probability function can be written in terms of the dipole cross sections $L^{13}, L_{24},\ldots$; see discussion in Ref.~\cite{Nikolaev:2005dd} in the context of quark-gluon dijet production off nuclei.
The curious $N_c\to -N_c$ symmetry between the $\bar 6$ and $15$ states is observed here as 
\begin{subequations}
\begin{align}
  &\mathcal{P}_{qg, 3\to \bar 6}(N_c)=-\mathcal{P}_{qg,3\to 15}(-N_c)\;,\\
  &\mathcal{P}_{qg, \bar 6\to \bar 6}(N_c)=-\mathcal{P}_{qg, 15\to 15}(-N_c)\;,
\end{align}
\end{subequations}
by noting that $ \mathcal M^s_{ii}(N_c)=-\mathcal M^s_{ii}(-N_c), i=1,2,3$ according to Eq.~\eqref{eq:M_qgqg_s}.

Then, the derivative is
\begin{align}
  \begin{split}
    \lim_{\vec u_{g,\perp},\vec u_{q,\perp}\to \vec 0_\perp}
    \vec \nabla_{u,q}\cdot \vec \nabla_{u,g} S_{\bar{q}\bar{g}qg}^s
    =
    \begin{pmatrix}
      0& s_{12}(v)& 0\\
      s_{12}(v) &s_{dd}(v)& s_{23}(v)\\
      0 & s_{23}(v)& s_{dd}(v)
    \end{pmatrix}
    \;,
  \end{split}
\end{align}
in which
\begin{align}\label{eq:DDS_res}
  \begin{split}
  & \tilde \alpha_1 =4 N_c [L(0)-L(v)]\;,\\
    &s_{dd}(v)=-\nabla^2_v L(v)\frac{N_c \xi}{4}e^{ - \xi \tilde\alpha_1/4}  - [L'(v)]^2 \frac{N_c^2 \xi^2 }{4}e^{ - \xi \tilde\alpha_1/4} 
    \;,\\
    &s_{12}(v)=
    -\nabla^2_v L(v)\frac{2\sqrt{2}}{\tilde\alpha_1}[e^{ - \xi \tilde\alpha_1/4}-1] 
    \;,\\
    &s_{23}(v)=
    -\nabla^2_v L(v)\frac{\sqrt{C_d N_c}}{4}e^{ - \xi \tilde\alpha_1/4} \xi \;.
  \end{split}
\end{align}
In the quark-gluon coincidence limit,
\begin{align}\label{eq:DDs_v0lim}
  \begin{split}
    &\lim_{v\to 0}s_{dd}(v)=\frac{N_c}{4}\xi \mathcal G_2\;,\\
    &\lim_{v\to 0}s_{12}(v)=-\frac{\sqrt{2}}{2}\xi \mathcal G_2\;,\\
    &\lim_{v\to 0}s_{23}(v)=\frac{\sqrt{C_d N_c}}{4}\xi \mathcal G_2\;.
  \end{split}
\end{align}
It is straightforward to verify that in the $v\to 0$ limit, the total momentum of the quark-gluon system behaves as a single particle in the corresponding color representation c,
\begin{align}
  \begin{split}
    \lim_{v\to 0}(\vec p_{q,\perp}+\vec p_{g,\perp})^2\vert_{qg,c}
   =C_c \mathcal G_2 \xi
    \;,
  \end{split}
\end{align}
in which $C_c$ is the corresponding Casimir, and specially $C_X=C_3+C_8$ for color-uncorrelated state for which the cross term $\vec p_{q,\perp}\cdot\vec p_{g,\perp}$ vanishes. 
The total momentum can be calculated as
\begin{multline}
  (\vec p_{q,\perp}+\vec p_{g,\perp})^2\vert_{qg,c} \\
 =
  \lim_{\vec u_{g,\perp},\vec u_{q,\perp}\to \vec 0_\perp}
  -\left(
  \vec \nabla_{u,q}^2+
  \vec \nabla_{u,g}^2
  +2\vec \nabla_{u,q}\cdot \vec \nabla_{u,g}
  \right) 
  \mathcal{P}_{qg,c}
  \;.
\end{multline}

\begin{figure*}[tbp!]
  \centering
    \subfigure[\ $s_{dd}$]
  { \includegraphics[scale=0.48]{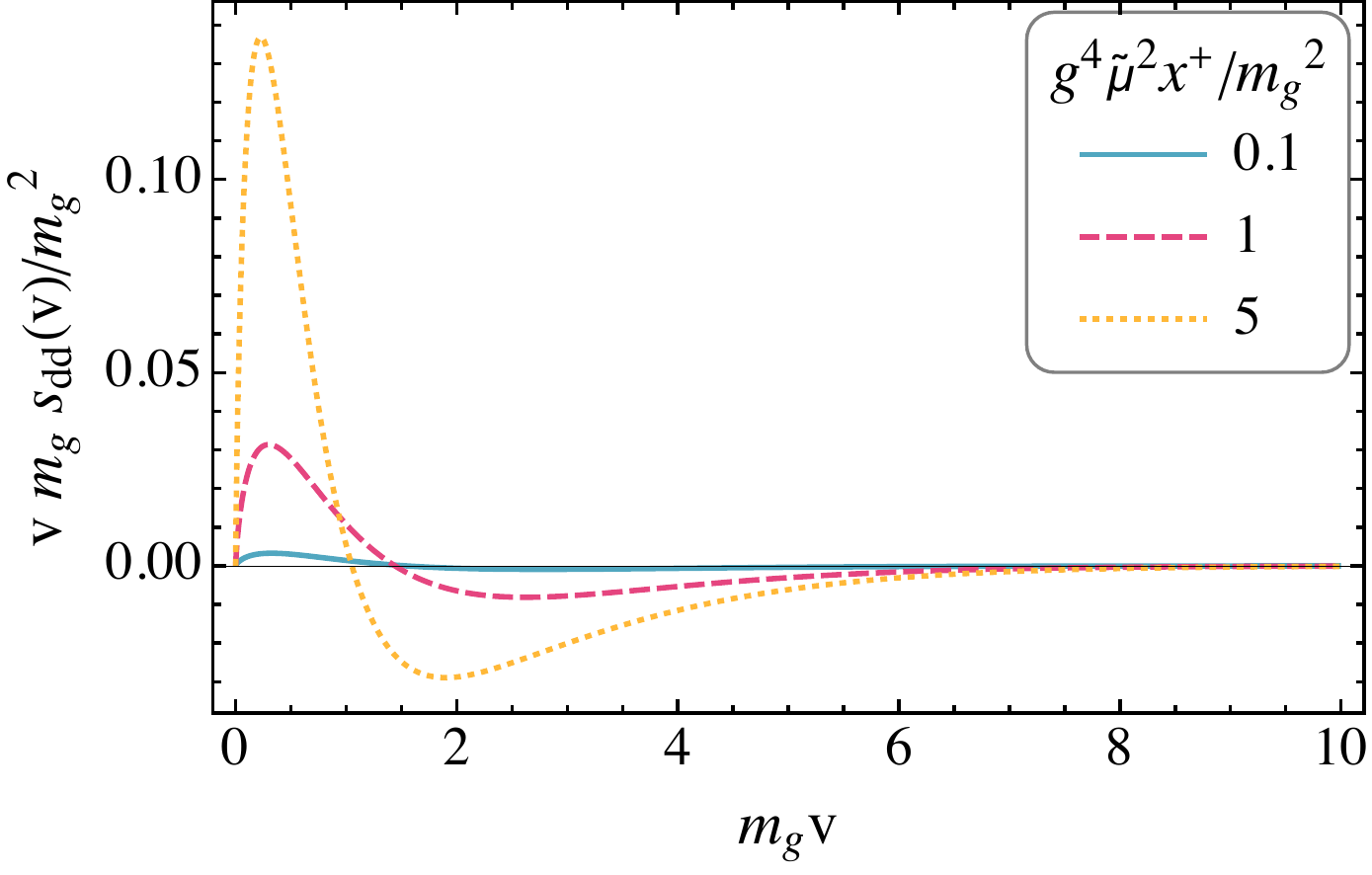}
  } 
  \subfigure[\ $s_{12}$]
  {   \includegraphics[scale=0.48]{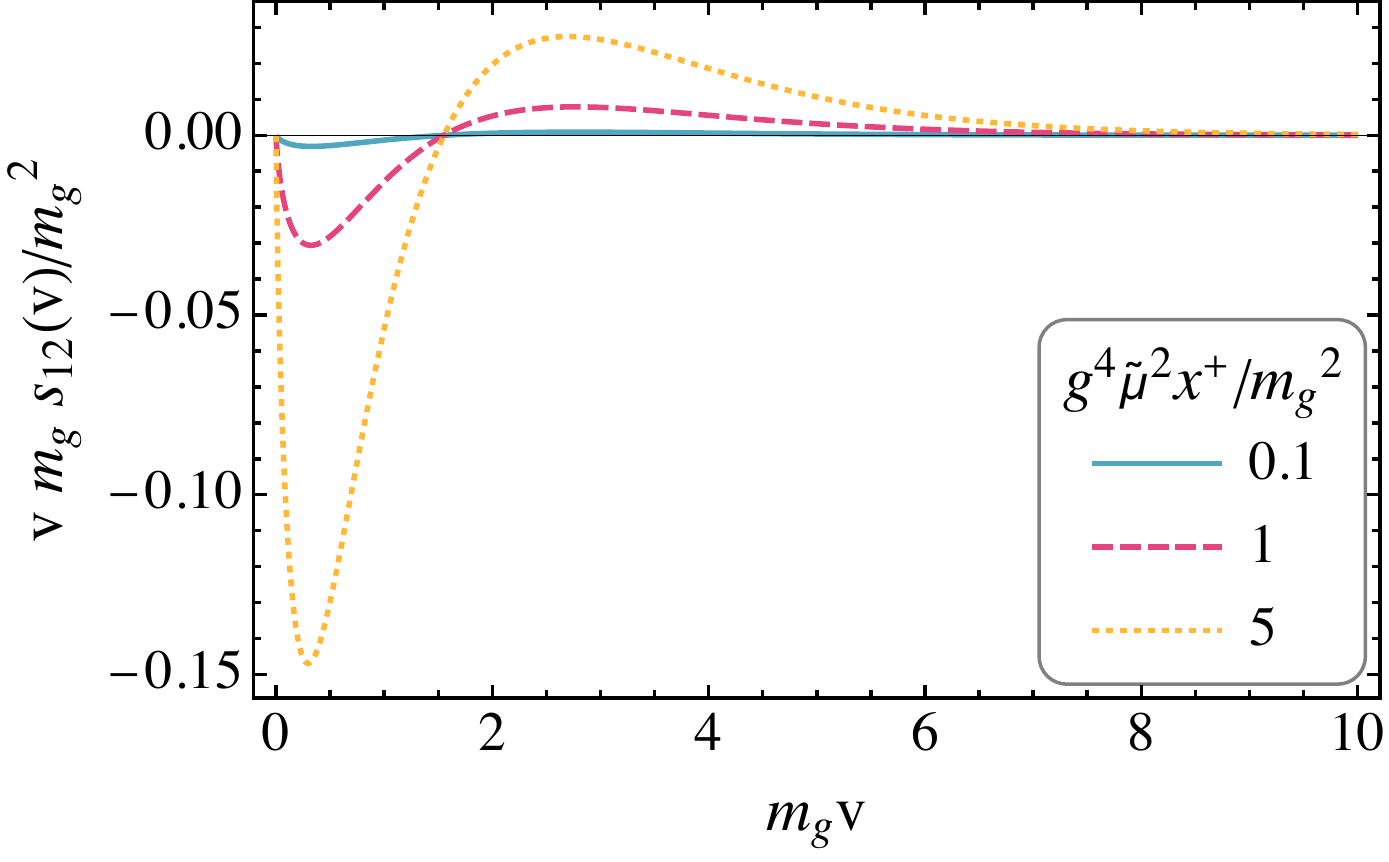}
  }  
  \subfigure[\ $s_{23}$]
  {   \includegraphics[scale=0.48]{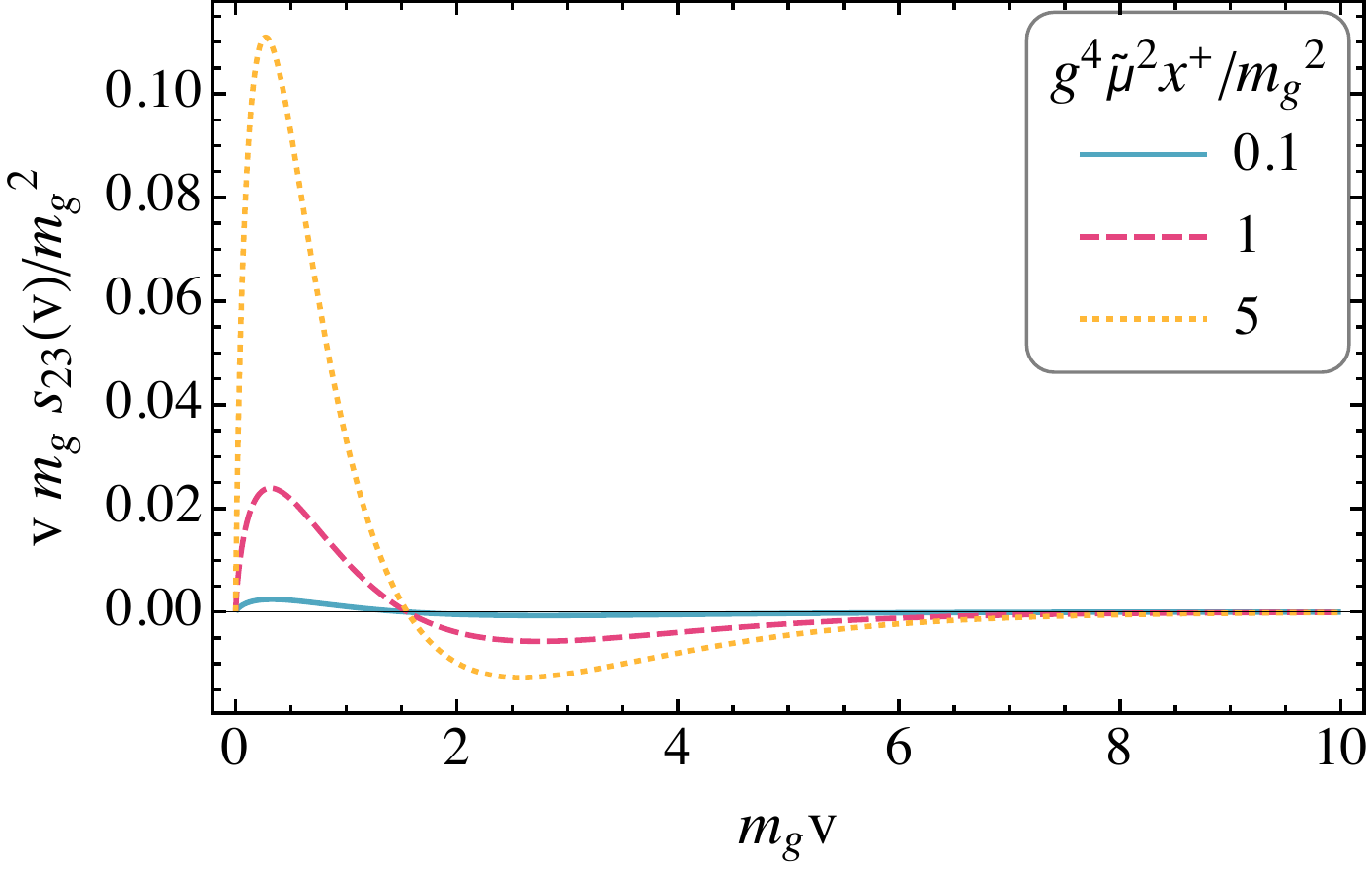}
  }
  \caption{ Plot of $v s_{mn}(v)$ ($mn=dd,12,23$) as a function of $v$ according to Eq.~\eqref{eq:DDS_res}.  }
  \label{fig:vs}
\end{figure*}

To compute the momentum broadening for a quark-gluon state in general, the next step is to integrate over $\vec v$ as shown in Eq.~\eqref{eq:pperp_qg_red},
\begin{multline}
   \int_{\v}
    f_{Rel} (\vec v_{\perp})
    \vec \nabla_{u,q} 
    \cdot 
    \vec \nabla_{u,g}
    \mathcal{P}_{qg,c}(0,x^+; \\
    \vec v_\perp+ \vec u_{q,\perp}, \vec u_{q,\perp},\vec v_\perp, \vec u_{q,\perp}-\vec u_{g,\perp})
    \big|_{\vec u_{q,\perp},\vec u_{g,\perp}=\vec 0_\perp}
    \;.
  \end{multline}
  All three nonvanishing elements exhibit a logarithmic divergence at $v=0$, as indicated by the $\mathcal G_2$ in Eq.~\eqref{eq:DDs_v0lim}. 
  We present the behavior of those elements multiplied by $v$ in Fig.~\ref{fig:vs}. In the plots, we take a dimensionless quantity $v m_g$, such that we are looking at the quark-gluon separation in  units of $1/m_g$. 

\section*{}

 \bibliography{qA}
 \end{document}